\documentclass[twocolumn]{aastex63}

\usepackage{rotating}
\usepackage[online]{threeparttablex}
\usepackage{makecell}
\newcolumntype{H}{>{\setbox0=\hbox\bgroup}c<{\egroup}@{}}
\newcommand{\hi}{H\,{\sc i}\ }

\defcitealias{LCH_2022}{Paper I}
\defcitealias{LCH_2023}{Paper II}

\newcommand{\kms}{km~s$^{-1}$}

\graphicspath{{./}{chap5/}} 

\begin{document}

\title{Observed Timescales of Stellar Feedback in Star-Forming, Low-Mass Galaxies}

\author[0000-0001-5368-3632]{Laura Congreve Hunter}
\affiliation{Department of Physics and Astronomy, Dartmouth College, 17 Fayerweather Hill Rd, Hanover, NH 03755, USA}

\author{Liese van Zee}
\affiliation{Department of Astronomy, Indiana University, 727 East 3rd Street, Bloomington, IN 47405, USA}

\author[0000-0002-2970-7435]{Roger E. Cohen}
\affiliation{Rutgers University, Department of Physics and Astronomy, 136 Frelinghuysen Road, Piscataway, NJ 08854, USA}

\author[0000-0001-5538-2614]{Kristen B. W. McQuinn}
\affiliation{Rutgers University, Department of Physics and Astronomy, 136 Frelinghuysen Road, Piscataway, NJ 08854, USA}
\affiliation{Space Telescope Science Institute, 3700 San Martin Dr, Baltimore, MD 21218, USA}

\author[0009-0000-9670-2194]{Madison Markham}
\affiliation{Department of Physics and Astronomy, Georgia State University, 25 Park Pl NE, Atlanta, GA 30303, USA}

\author[0000-0002-6650-3757]{Justin A. Kader}
\affiliation{Department of Physics and Astronomy, University of California, Irvine,  4129 Frederick Reines Hall, Irvine, CA 92697, USA}

\author[0000-0002-2492-7973]{Lexi Gault}
\affiliation{Department of Astronomy, Indiana University, 727 East 3rd Street, Bloomington, IN 47405, USA}

\author[0000-0001-8416-4093]{Andrew E. Dolphin}
\affiliation{Raytheon Company, 1151 E. Hermans Road, Tucson, AZ 85756, USA}
\affiliation{University of Arizona, Steward Observatory, 933 North Cherry Avenue, Tucson, AZ 85721, USA}

\begin{abstract}

Understanding the timescales of atomic gas turbulence is crucial to understanding the interplay between star formation and the interstellar medium (ISM).  To investigate the timescales of turbulence low-mass galaxies ($10^{6.8}<M_\odot<10^9$), this study combines temporally resolved star formation histories (SFHs)—derived from color-magnitude diagrams—with kinematic data of the atomic and ionized hydrogen in a large sample of nearby, star-forming, low-mass galaxies. To best understand the timescales involved, SFHs and gas kinematics were analyzed in 400$\times$400 parsec regions to capture the local impacts of star formation. No strong correlation was found between the ionized gas velocity dispersion and the star formation activity over the past 5–500 Myr.  In contrast, a consistent and significant correlation between the atomic hydrogen turbulence measures and the star formation activity t$\geq$100 Myr ago was identified.  This correlation suggests the star formation activity and atomic gas are coupled on this timescale.   This connection between star-formation activity $>$100 Myr ago, and the \hi turbulence properties, may be related to the time scales over which turbulence decays in the ISM. Additionally, the results demonstrate a possible difference in the global and local turbulence properties of low-mass galaxies.
\end{abstract}

\section{Introduction}
\label{chapt5:section_full_sample}

To accurately model galaxy evolution, understanding how the energy from star formation impacts the interstellar medium is essential.  Stellar feedback drives turbulence in the interstellar medium (ISM; e.g., \citealt{Spitzer78,Elmegreen04}) and plays a central role in galaxy evolution, influencing systems across a range of physical scales and timescales.

Prior observational studies using integrated light star formation rates (SFR) have found evidence of a correlation between star formation and turbulence in the ISM. For example, \citet{Moiseev15} demonstrated that increased ionized gas velocity dispersions result from energy output from current star formation activity. Other studies \citep{Joung09, Tamburro09, Stilp13c} suggest that atomic gas turbulence is linked to star formation, particularly at high SFR surface densities.  However, the spatial correlation between star formation activity and elevated energy density of the ISM is not clear. \citet{Hunter21} and \citet{Elmegreen22} cross-correlated the kinetic energy density of the atomic gas with SFR surface density for LITTLE THINGS \citep{LTHINGS} and THINGS \citep{things} galaxies, and found no significant correspondence. They concluded that stellar feedback has limited impact on large-scale \hi turbulence, primarily affecting molecular gas clouds instead.

Utilizing time-resolved star formation histories (SFHs), \cite{Stilp13b} found a correlation between globally averaged \hi turbulence and star formation activity from 30–40 Myr ago. Further work in \citet[hereafter \citetalias{LCH_2022,LCH_2023}]{LCH_2022,LCH_2023} extended this approach to local turbulence and star formation properties.  They analyzed five low-mass galaxies by dividing them into 400~pc regions and identified a local correlation between atomic gas turbulence and star formation approximately 100 Myr ago. As discussed in \citetalias{LCH_2022,LCH_2023}, the timescales associated with turbulence differ between local and global scales. These differences may reflect fundamental variations in the turbulence properties of galaxies across spatial scales.

The observed correlation between past star formation and current ISM turbulence may be related to the dissipation timescale of energy in atomic gas. The dissipation of energy in atomic gas is key to understanding the processes that sustain turbulence. This timescale can rule out energy sources based on the required energy and input efficiencies necessary to maintain the observed turbulence.  For thin galactic disks with modest velocity dispersions, the dissipation timescale is estimated at 5–15 Myr \citep[e.g.,][]{Tamburro09, Stilp13c, Utomo19}. Maintaining turbulence through only supernovae would require an efficiency energy being converted into turbulent energy of $\simeq$100\% or greater.  This implies that turbulence must be driven by additional energy sources. In contrast, observations and simulations have indicated the dissipation timescale may be $\sim$100 Myr \citep[e.g.,][]{Bacchini20a, Orr2020}. This longer timescale aligns with the few-to-ten percent energy transfer efficiencies from SNe to atomic gas found in simulations \citep[e.g.,][]{Thornton98, Martizzi16, Fierlinger16}. The local correlation timescale of roughly 100 Myr identified in \citetalias{LCH_2022,LCH_2023} support this longer dissipation timescale. These findings suggest that stellar feedback, even at modest efficiencies, could be the primary driver of turbulence in the atomic ISM.

This paper builds on the results of \citetalias{LCH_2022,LCH_2023} by expanding the sample to 26 galaxies. The sample spans two orders of magnitude in stellar mass (log(M$\odot$) = 6.92–9.05), far-UV star formation rate (log(M$\odot$ yr$^{-1}$) = --3.01 to --0.7), and \hi mass (log(M$_\odot$) = 7.00–9.21). This broader sample enables a deeper investigation into the differences between local and global turbulence properties and the impact of galactic properties on the feedback timescale. Section \ref{obs_c5} discusses the data used from the Very Large Array (VLA\footnote{The VLA is operated by the NRAO, which is a facility of the National Science Foundation operated under cooperative agreement by Associated Universities, Inc.}), \textit{Hubble Space Telescope (HST)}, and WIYN\footnote{The WIYN Observatory is a joint facility of the NSF's National Optical-Infrared Astronomy Research Laboratory, Indiana University, the University of Wisconsin-Madison, Pennsylvania State University, the University of Missouri, the University of California-Irvine, and Purdue University. } 3.5m telescope. Section~\ref{Methods_c5} summarizes the methods detailed in \citetalias{LCH_2022,LCH_2023} for determining the SFHs and measuring the turbulence of the atomic and ionized gas.  Section~\ref{halpha_res_c5} presents the H$\alpha$ results, Sections~\ref{local_time_c5} an present the results for the \hi on the 400 pc scale.  Section~\ref{global_time_c5} presents the global timescale results. Section~\ref{sec:discussion_ch5} discusses the implications of these results.   Section~\ref{conclusions_c5} summarizes the results and conclusions. The Appendix includes a complete figure set of showing the gas kinematics and stellar populations of the observed galaxies.

\section{Observational Data}\label{obs_c5}

The galaxy sample was selected to include low-mass, star forming galaxies within 5.5 Mpc with available \textit{Hubble Space Telescope (HST)} and VLA \hi data. The  HST observations imaging  were required to be deep enough to derive CMD star formation histories.  Distance cuts ensured galaxies were close enough to reliably derive CMD-based SFHs and resolve the \hi kinematics on the 400 pc scale.  Archival F814W, F606W, F555W, and F475W \textit{HST} observations of resolved stars were used to create CMDs and derive SFHs. The VLA \hi radio synthesis observations were used to determine the atomic gas surface densities and velocity dispersions. Ionized gas kinematics were derived from observations with the SparsePak IFU on the WIYN 3.5m telescope. While \textit{HST} and VLA data were required, SparsePak observations were not available for all galaxies in the final sample.

%\startlongtable
\begin{table*}
\begin{rotatetable*}
    \caption{Galaxy Sample and Observed Properties}
    \centerwidetable
    \movetableright=-1in
    \footnotesize
    \begin{tabular}{lcccccccccccc}    
        \hline
        \label{table:galax_ch5} 
        \centering  
        Galaxy & RA & Dec & Dist & m$_{FUV}$ & A$_{FUV}$ & m$_B$ & m$_{3.6}$ & R$_{3.6}$ & B/A & SFR FUV & Stellar Mass & log(sSFR) \\
         & J2000 & J2000 & Mpc & mag & mag & mag & mag & arcsec & & log(M$_\odot$ yr$^{-1}$) & log(M$_\odot$) & log(yr$^{-1}$) \\
         (1) & (2) & (3) & (4) & (5) & (6) & (7) & (8) & (9) & (10) & (11) & (12) & (13) \\
         [0.5ex] 
        \hline 
        UGC 0685 & 01:07:22.4 & 16:41:04 & 4.81$\pm$0.04 & 16.03$\pm$0.03 & 0.24 & 14.15 & 10.94 & 67.6 & 0.64 & -2.17$\pm$0.01 & 7.99$\pm0.09$ & -10.16$\pm$.09  \\
        NGC 0784 & 02:01:16.9 & 28:50:14 & 5.37$\pm$0.02 & 13.92$\pm$.05 & 0.16 & 12.26 & 9.28 & 202.6 & 0.28  & -1.26$\pm$0.02 & 8.58$\pm$0.09 & -10.01$\pm$.09 \\
        NGC 2366 & 07:26:54.7 & 69:12:57 & 3.28$\pm$0.03 & 12.43$\pm$0.05 & 0.09 & 11.53 & 8.83 & 191.2 & 0.34 & -1.12$\pm$0.02 & 8.50$\pm$0.09 & -9.62$\pm$0.09\\
        Holmberg II & 08:34:06.9 & 66:10:39 & 3.38$\pm$0.05 & 12.28$\pm$0.06 & - & 11.37 & 8.80 & 209.9 & 0.77 & -0.95$\pm$0.03 & 8.54$\pm$0.09 & -9.49$\pm$0.09 \\
        UGC 4459 & 08:34:07.2 & 66:10:54 & 3.68$\pm$0.03 & 15.27$\pm$0.05 & 0.10 & 14.41 & 11.81 & 54.5 & 0.89 & -2.15$\pm$0.02 & 7.41$\pm$0.09 & -9.56$\pm$0.09 \\
        Holmberg I & 09:40:35.1 & 71:10:46 & 4.02$\pm$0.06 & 14.59$\pm$0.09 & 0.10 & 13.25 & 10.66 & 95.9 & 0.94 & -1.80$\pm$0.04 & 7.95$\pm$0.09 & -9.75$\pm$0.09 \\
        Sextans B & 10:00:00.1 & 05:19:56 & 1.43$\pm$0.02 & 13.65$\pm$0.06 & 0.10 & 11.87 & 9.09 & 114.2 & 0.85 & -2.33$\pm$0.04 & 7.68$\pm$0.09 & -10.00$\pm$0.09 \\
        Sextans A & 10:11:00.8 & -04:41:34 & 1.45$\pm$0.05 & 12.54$\pm$0.05 & 0.10 & 11.96 & 9.24 & 169.3 & 0.86 & -1.87$\pm$0.04 & 7.63$\pm$0.09 & -9.50$\pm$0.09 \\
        IC 2574 & 10:28:23.5 & 68:24:44 & 3.93$\pm$0.04 & 12.13$\pm$0.06 & - & 10.97 & 8.12 & 411.8 & 0.37 & -0.72$\pm$0.03 & 8.94$\pm$0.09 & -9.66$\pm$0.09 \\
        NGC 3738 & 11:35:48.8 & 54:31:26 & 5.3$\pm$0.05 & 13.77$\pm$0.05 & 0.43 & 11.94 & 9.01 & 124.1 & 0.70 & -1.10$\pm$0.02 & 8.85$\pm$0.09 & -9.95$\pm$0.09\\
        NGC 3741 & 11:36:06.2 & 45:17:01 & 3.22$\pm$0.18 & 15.14$\pm$0.05 & - & 14.48 & 11.90 & 42.2 & 0.68 & -2.26$\pm$0.05 & 7.26$\pm$0.10 & -9.51$\pm$0.09 \\
        NGC 4068 & 12:04:02.4 & 52:35:27 & 4.38$\pm$0.04 & 14.29$\pm$0.05 & - & 13.09 & 10.33 & 94.4 & 0.56 & -1.64$\pm$0.02 & 8.15$\pm$0.09 & -9.8$\pm$0.09 \\
        NGC 4163 & 12:12:09.0 & 36:10:08 & 2.88$\pm$0.04 & 15.34$\pm$0.05 & - & 13.54 & 10.51 & 79.0 & 0.65 & -2.43$\pm$0.02 & 7.72$\pm$0.09 & -10.15$\pm$0.09 \\
        NGC 4190$\star$ & 12:13:44.8 & 36:38:03 & 2.83$\pm$0.05 & 14.77$\pm$0.05 & 0.12 & 13.40 & 10.5 & 49.8 & 0.91 & -2.17$\pm$0.03 & 7.71$\pm$0.09 & -9.88$\pm$0.09 \\
        UGC 7577 & 12:27:40.9 & 43:29:44 & 2.61$\pm$0.06 & 14.82$\pm$0.05 & 0.10 & 12.76 & 9.77 & 169.9 & 0.50 & -2.27$\pm$0.03 & 7.93$\pm$0.09 & -10.20$\pm$0.09 \\
        UGCA 292 & 12:38:40.1 & 32:46:01 & 3.85$\pm$0.09 & 16.21$\pm$0.05 & 0.10 & 15.70 & 12.26 & 47.7 & 0.84 & -2.49$\pm$0.03 & 7.27$\pm$0.09 & -9.76$\pm$0.09 \\
        %IC 3687$\star$ & 12:42:15 & 38:30:12 & 4.57$\pm$0.03 & 14.62$\pm$0.05 & 0.11 & 13.75 & 11.1 & 101.7 & 0.89 & -1.70$\pm$0.02 & 7.88$\pm$0.09 & -9.58$\pm$0.09 \\ 
        UGC 8024 & 12:54:05.3 & 27:08:59 & 4.04$\pm$0.06 & 14.80$\pm$0.05 & - & 14.05 & 11.61 & 73.8 & 0.54 & -1.92$\pm$0.04 & 7.57$\pm$0.09 & -9.49$\pm$0.09 \\
        GR8 & 12:58:40.4 & 14:13:03 & 2.19$\pm$0.12 & 15.20$\pm$0.05 & - & 14.67 & 12.13 & 55.0 & 0.67 & -2.61$\pm$0.05 & 6.83$\pm$0.10 & -9.45$\pm$0.09 \\
        %IC 4182$\star$ & 15:05:49.5 & 37:36:18 & 4.41$\pm$0.03 & 13.31$\pm$0.05 & 0.09 & 11.86 & 9.2 & 180.8 & 0.91 & -1.23$\pm$0.02 & 8.61$\pm$0.09 & -9.82$\pm$0.09 \\
        UGC 8201 & 13:06:24.9 & 67:42:25 & 4.83$\pm$0.04 & 14.51$\pm$0.05 & - & 13.06 & 10.65 & 122.9 & 0.50 & -1.65$\pm$0.02 & 8.11$\pm$0.09 & -9.76$\pm$0.09 \\
        NGC 5204$\star$ & 13:29:37 & 58:25:07 & 4.76$\pm$0.05 & 12.92$\pm$0.05 & 0.33 & 11.84 & 9.07 & 150.4 & 0.60 & -0.90$\pm$0.02 & 8.73$\pm$0.09 & -9.63$\pm$0.09 \\
        UGC 8638 & 13:39:19.4 & 24:46:32 & 4.29$\pm$0.04 & 15.77$\pm$0.05 & 0.02 & 14.40 & 11.43 & 62.2 & 0.55 & -2.25$\pm$0.02 & 7.70$\pm$0.09 & -9.95$\pm$0.09 \\
        UGC 8651 & 13:39:53.8 & 40:44:21 & 3.10$\pm$0.06 & 15.95$\pm$0.05& - & 14.20 & 11.58 & 65.6 & 0.53 & -2.61$\pm$0.03 & 7.35$\pm$0.09 & -9.97$\pm$0.09 \\
        NGC 5253 & 13:39:56.0 & -31:38:24 & 3.44$\pm$0.02 & 12.36$\pm$0.05 & - & 10.91 & 7.57 & 164.9 & 0.49 & -1.09$\pm$0.02 & 9.05$\pm$0.09 & -10.13$\pm$0.09 \\
        UGC 9128 & 14:15:56.9 & 23:03:23 & 2.21$\pm$0.07 & 16.21$\pm$0.05 & - & 14.43 & 11.92 & 57.7 & 0.60 & -3.01$\pm$0.03 & 6.92$\pm$0.09 & -9.93$\pm$0.09 \\
        UGC 9240 & 14:24:43.4 & 44:31:33 & 2.83$\pm$.04 & 14.82$\pm$0.05 & 0.02 & 13.22 & 10.54 & 73.1 & 0.80 & -2.24$\pm$0.02 & 7.69$\pm$0.09 & -9.92$\pm$0.09 \\
        NGC 6789 & 19:16:42.0 & 63:58:15 & 3.55$\pm$0.007 & 16.21$\pm$0.05 & - & 13.99 & 10.87 & 57.6 & 0.84 & -2.60$\pm$0.02 & 7.76$\pm$0.09 & -10.35$\pm$0.09 \\ 
        \hline 
        \end{tabular}
        \tablecomments{Column (4) CMD Distances from CMD from \cite{Tully13} Column (5) FUV magnitudes from \cite{Lee11} Column (6) FUV attenuation from \cite{Lee09}  Columns (7-10) from new WIYN 0.9m photometry Column (9) semi-major axis at 23 mag/sq arcsec in 3.6 Spitzer data $\star$ B-band magnitude, R$_{25}$ and B/A values from  B-band photometery in \cite{Cook14} and IR magnitudes from \cite{Dale09} Column (11) FUV SFR based on scaling relations in \cite{Kennicutt12} (12) masses based off 3.6 micron fluxes in \cite{Dale09}}
\end{rotatetable*}
\end{table*}

Our final sample includes twenty-six galaxies with distance between 1 and 5.5 Mpc with VLA \hi synthesis imaging and are a representative sample of low-mass, star forming galaxies within the Local Volume.  Details of the full sample are listed in Table \ref{table:galax_ch5}.  Of these galaxies, twenty-five have sufficient data for the ionized gas measures.  

\subsection{VLA Observations}

This study uses new and archival VLA B, C, and D-configuration observations.  The majority of the archival data were obtained as part of VLA-ANGST \citep{VLAANGST} and LITTLE THINGS \citep{LTHINGS} {  {with a velocity resolutions between 0.623 and 2.58 km s$^{-1}$.  Two additional galaxies were observed for this project (see Table \ref{table:hiobs_ch5}) {with a velocity resolutions 0.825 km s$^{-1}$}.  The data were reprocessed in \emph{AIPS}\footnote{The Astronomical Image Processing System (AIPS) was developed by the NRAO.} following the procedures in \citetalias{LCH_2022,LCH_2023}, and \cite{Richards18}. The individual observing blocks were reduced and combined after Doppler correction and continuum subtraction. For the new VLA data we applied Hanning Smoothing before imaging to remove ringing across the frequency channels halving the velocity resolution to 1.65 km s$^{-1}$.  For most data sets, a low-resolution data cube with a robust of 5 weighting was chosen for the final data cube with no uvtaper or uvrange limits.

For a few galaxies whose observations were strongly weighted towards longer-baselines, a uvtaper of 40 k$\lambda$ and uvrange 0 to 50 k$\lambda$ was applied to increase sensitivity at the expense of some spatial resolution.   For NGC 0784, a robust of 0.5 was used to increase spatial resolution due to the absence of B-configuration data and the galaxy’s distance. Additional details on the \hi data processing can be found in \citetalias{LCH_2022}, and \cite{Richards18}. The final data cubes are detailed in Table \ref{table:cubes_ch5} including the velocity resolution of the data cubes, total \hi fluxes and \hi masses.  The 0$^{th}$, 1$^{st}$, and 2$^{nd}$ moment maps were created in GIPSY \citep{GIPSY} following the procedure in \citetalias{LCH_2022} and are presented in the Appendix. 

\begin{table*} 
    \centering
    \small
    \caption{\centering New \hi Observations}
        \begin{tabular}{l c c c c c} 
        \hline
        \hline 
        Galaxy & Array  & Project & Dates & Time on Source  & Ch Sep \\ 
         & & & & (hrs) & (km s$^{-1}$) \\[0.5ex] 
        \hline 
        NGC 5204 & C & 20A-085 & 2020 May 19, 20 & 6.63 & 0.825  \\
        NGC 5204 & B & 20A-085 & 2020 Aug 31, Sep 21, 23, 25 20 & 2.56 & 0.825  \\
        NGC 5204 & B & 21B-058 & \makecell{2021 Oct 30, Nov 2, 16, Dec 6, \\ 2022 Jan 14, 19, 28} & 5.24 & 0.825  \\
        NGC 5204 & D & 21A-027 & 2021 Mar 26, Apr 1, 9 & 1.84 & 0.825 \\
        UGC 8638 & C & 20A-085 & 2020 Mar 12, May 6, 12, June 3 & 6.41 & 0.825 \\
        UGC 8638 & B & 20A-085 & 2020 Oct 9, 18 & 2.17 & 0.825 \\
        UGC 8638 & B & 21B-058 & \makecell{2021 Nov 7, 8, \\ 2022 Jan 4, 15, 18, 19} & 6.63 & 0.825 \\
        UGC 8638 & C & 21B-058 & 2022 Feb 22, 26 & 2.98 & 0.825 \\[1ex]
        \hline 
    \end{tabular} 
    \label{table:hiobs_ch5} 
\end{table*}

\begin{table*}[th!] 
    \centering
    \small
    \caption{\centering \hi Data Cubes and Properties}
        \begin{tabular}{c c c c c c c c c} 
        \hline
        \hline 
        Galaxy & $\Delta$v & Beam & P.A. & RMS  & \hi flux & \hi Mass \\
         & km s$^{-1}$ & arcsec$\times$arcsec & deg & mJy bm$^{-1}$ & Jy km s$^{-1}$  & log(M$_\odot$) \\ [0.5ex] 
        \hline 
        UGC 0685 & 2.58 & 19.25$\times$16.53 & -52.2 & 0.669 & 13.41$\pm$1.3 & 7.86$\pm$0.04  \\
        NGC 0784 & 2.58 & 22.20$\times$17.62 & 89.8 & 0.673 & 82.1$\pm$8.2 & 8.75$\pm$0.04  \\
        NGC 2366 & 2.58 & 22.89$\times$21.25 & -3.6 & 0.521 & 198$\pm$20 & 8.70$\pm$0.04  \\
        Holmberg II & 2.57 & 10.73$\times$10.40 & -0.9 & 1.047 & 220$\pm$22 & 8.77$\pm$0.05 \\
        UGC 4459 & 2.57 & 19.65$\times$19.39 & 35.1 & 0.519 & 22.3$\pm$2.2 & 7.85$\pm$0.04 \\
        Holmberg I & 2.58 & 9.77$\times$7.50 & -72.0 & 1.116 & 91.9$\pm$9.2 & 8.54$\pm$0.05 \\
        Sextans B & 1.29 & 21.25$\times$20.11 & 6.8 & 0.580 & 98.3$\pm$9.8 & 7.68$\pm$0.04  \\
        Sextans A & 2.58 & 13.67$\times$10.68 & -11.1 & 0.469 & 156$\pm$16 & 7.89$\pm$0.05  \\
        IC 2574 & 1.65 & 13.18$\times$12.45 & 30.4 & 0.563 & 444$\pm$44 & 9.21$\pm$0.04 \\
        NGC 3738 & 2.58 & 15.05$\times$8.29 & 82.4 & 0.484 & 20.9$\pm$2.1 & 8.14$\pm$0.04 \\
        NGC 3741 & 2.58 & 9.47$\times$6.66 & 79.5 & 0.943 & 42$\pm$4 & 8.01$\pm$0.07 \\
        NGC 4068 & 2.47 & 11.83$\times$11.29 & 14.2 & 0.7095 & 40.1$\pm$4.0 & 8.30$\pm$0.04 \\
        NGC 4163$\star$ & 1.28 & 15.911$\times$13.941 & -89.1 & 0.971 & 8.73$\pm$0.9 & 7.22$\pm$0.05\\
        NGC 4190 & 1.29 & 11.54$\times$9.62 & -8.7 & 0.997 & 24.3$\pm$2.4 & 7.66$\pm$0.05 \\
        UGC 7577 & 1.28 & 11.84$\times$11.30 & -88.4 & 0.969 & 21.8$\pm$2.2 & 7.54$\pm$0.05 \\
        NGC 4449 & 5.15 & 12.39$\times$10.01 & -85.5 & 0.633 & 340$\pm$34 & 9.16$\pm$0.04  \\
        UGCA 292 & 1.29 & 17.40$\times$16.03 & -63.1 & 0.604 & 16.0$\pm$1.6 & 7.75$\pm$0.05 \\
        UGC 8024 & 2.58 & 15.89$\times$15.39 & -25.1 & 0.618 & 100.$\pm$10 & 8.58$\pm$0.05  \\
        GR8 & 1.29 & 17.43$\times$16.50 & -57.4 & 0.680 & 8.8$\pm$.9 & 7.00$\pm$0.06  \\
        UGC 8201 & 1.29 & 13.48$\times$13.12 & 30.3 & 0.404 & 33.5$\pm$3.4 & 8.27$\pm$0.05 \\
        NGC 5204 & 1.65 & 17.62$\times$12.49 & 87.9 & 0.509 & 146$\pm$14 & 8.89$\pm$0.04 \\
        UGC 8638 & 1.65 & 13.16$\times$10.79 & 53.5 & 0.446 & 4.6$\pm$0.5 & 7.30$\pm$0.04 \\
        UGC 8651 & 2.58 & 13.95$\times$11.28 & -71.9 & 0.699 & 13.4$\pm$1.3 & 7.48$\pm$0.05 \\
        NGC 5253 & 2.58 & 17.60$\times$10.11 & -1.1 & 0.917 & 43$\pm$4 & 8.08$\pm$0.04 \\
        UGC 9128$\star$ & 1.29 & 13.279$\times$10.326 & 86.2 & 1.05 & 13.0$\pm$1.3 & 7.18$\pm$0.05 \\
        UGC 9240 & 2.58 & 16.11$\times$14.29 & 89.9 & 0.611 & 41$\pm$4 & 7.88$\pm$0.05  \\
        NGC 6789 & 2.47 & 11.92$\times$10.52 & -64.8 & 0.536 & 4.9$\pm$.5 & 7.17$\pm$0.05 &\\[1ex] 
        \hline 
    \end{tabular} 
    \tablecomments{\small $\star$ Galaxies from \citetalias{LCH_2022} that were reprocessed for the full sample included on this table }
    \label{table:cubes_ch5} 
\end{table*}

\subsection{Archival \textit{HST} Observations}

The SFHs were derived from \textit{HST} observations with either the Advanced Camera for Surveys (ACS; \citealt{ACS}) or the Wide Field Planetary Camera 2 (WFPC2; \citealt{WFPC2}). Observation details are provided in Table \ref{table:HST_obs_ch5} \footnote{All of the data presented in this article were obtained from the Mikulski Archive for Space Telescopes (MAST) at the Space Telescope Science Institute. The specific observations analyzed can be accessed via \dataset[doi:10.17909/ykf7-pz26]{https://doi.org/10.17909/ykf7-pz26}.}. Observations were taken with the F814W (I-band) filter and at least one of the following: F606W (V), F555W (V), or F475W (g).  ACS has a 202"$\times$202" field of view,  pixel scale of 0.05" pixel$^{-1}$) and the WFPC2 instrument has three 800$\times$800 pixel wide field CCDs, with a 0.1" pixel $^{-1}$ pixel scale, and a 800$\times$800 pixel planetary camera CCD with a 0.05" pixel $^{-1}$ pixel scale. 

\begin{table*}[ht!] 
    \centering
    \footnotesize
    \caption{\centering \textit{HST} Observations}
        \begin{tabular}{c c c c c c c c c} 
        \hline
        \hline 
        Galaxy & \textit{HST}  & Inst. & No. of & F475W & F555W & F606W & F814W \\
         & PID & & {Fields} & {sec} & {sec} & {sec} & {sec} \\ [0.5ex] 
        \hline 
        UGC 0685 & 10210 & ACS & 1 & -- & -- & 934 & 1226 \\
        NGC 0784 & 10210 & ACS & 1 & -- & -- & 933 & 1226 \\
        NGC 2366 & 10605 & ACS & 2 & -- & {4780} & -- & {4780} \\
        Holmberg II & 10605 & ACS & 2 & -- & 4660 & -- & 4660 \\
        UGC 4459 & 10605 & ACS & 1 & -- & {4768} & -- & {4768} \\
        Holmberg I & 10605 & ACS & 1 & -- & {5829} & -- & {5936} \\
        Sextans B & 10915 & WFPC2 & 1 & -- & -- & 2700 & 3900 \\
        Sextans A & 5915 & WFPC2 & 1 & -- & 1800 & -- & 1800 \\
        Sextans A & 7496 & WFPC2 & 1 & -- & 19200 & -- & 38400 \\
        IC 2574 & 10605 & ACS & 2 & -- & 4784 & -- & 4784 \\
        IC 2574 & 9755 & ACS & 1 & -- & 6400 & -- & 6400 \\
        NGC 3738 & 12546 & ACS & 1 & -- & -- & 900 & 900 \\
        NGC 3741 & 10915 & ACS & 1 & 2262 & -- & -- & 2331 \\
        NGC 4068 & 9771 & ACS & 1 & -- & -- & 1200 & 900 \\
        NGC 4163 & 9771 & ACS & 1 & -- & -- & 1200 & 900 \\
        {NGC 4163} & {10915 }& { ACS} & {1} & -- & -- & {2292} & {2250 } \\
        NGC 4190 & 10905 & ACS & 1 & -- & -- & 2200 & 2200 \\
        UGC 7577 & 11986 & WFPC2 & 1 & -- & -- & 2400 & 4800 \\
        %NGC 4449 & 10585 & A. Aloisi & ACS & 2 & -- & 2460 & -- & 2060 \\
        UGCA 292 & 10905 & ACS & 1 & -- & -- & 926 & -- \\
        UGCA 292 & 10915 & ACS & 1 & -- & -- & -- & 2274 \\
        UGC 8024 & 10905 & ACS & 1 & -- & -- & 924 & 1128 \\
        GR8 & 10915 & ACS & 1 & 2244 & -- & -- & 2259 \\
        UGC 8201 & 10605 & ACS & 1 & -- & 4768 & -- & 4768 \\
        NGC 5204 & 8601 & WFPC2 & 1 & -- & -- & 600 & 600 \\
        UGC 8638 & 9771 & ACS & 1 & -- & -- & 1200 & 900 \\
        UGC 8651 & 10210 & ACS & 1 & -- & -- & 1016 & 1209 \\
        NGC 5253 & 10765 & ACS & 2 & -- & 2400 & -- & 2360 \\
        UGC 9128 & 10210 & ACS & 1 & -- & -- & 985 & 1174 \\
        UGC 9240 & 10915  & ACS & 1 & -- & -- & 2301 & 2265 \\
        NGC 6789 & 8122 & WFPC2 & 1 -- & 8200 & -- & 8200 \\ [1ex] 
        \hline 
    \end{tabular} 
    \label{table:HST_obs_ch5} 
\end{table*}

The optical imaging were processed identically to STARBIRDS \citep{McQuinn15a} matching the approach in \citetalias{LCH_2022,LCH_2023}. A brief summary is provided here; for a detailed description, see \citet{McQuinn10a}. Photometry was performed using DOLPHOT \citep{Dolphin00,Dolphot} on the pipeline processed, charge transfer efficiency corrected images.  The photometry was filtered to include well-recovered point sources with the same quality cuts on signal-to-noise-ratios, crowding conditions, and sharpness parameters as applied in STARBIRDS. Artificial star tests were run on the individual images to measure the completeness of the stellar catalogs. To ensure accurate completeness functions for the SFH derivation, approximately 5–6 million artificial stars were injected per field of view—sufficient to sample individual 400$\times$400~pc regions robustly.

\subsection{SparsePak Observations} \label{sec:c5_sp}

Spatially resolved spectroscopy of the ionized gas were taken with the SparsePak IFU \citep{Sparspak} on the WIYN 3.5m telescope between December of 2015 and February of 2023. SparsePak fields were selected to cover the majority of the high-surface-brightness areas and much of the diffuse ionized gas in each galaxy (see Table \ref{table:sparse_ch5}). The observations followed the same setup used in \citetalias{LCH_2022,LCH_2023}, summarized here.  All observations were taken with the Bench Spectrograph in the same set up using the 316@63.4 grating, the X19 blocking filter, and observing at order 8.  This configuration results in a wavelength range of 6480--6890 {\AA}, centered on 6683.933 {\AA} with a velocity resolution of 13.9 km s$^{-1}$ pixel$^{-1}$.  Most pointings used a three-point dither pattern to fill gaps between fibers.  In some cases, only one or two pointings in the dither pattern were observed. Similarly, while most pointings consisted of three equal-length exposures, a few had only one or two (see Table~\ref{table:sparse_ch5}). For galaxies larger than the SparsePak field of view, the sky fibers fell within the galaxy and additional sky frames were taken to correct for telluric line contamination.

The  SparsePak data were processed as described in \citetalias{LCH_2023} using the \emph{IRAF}\footnote{IRAF is distributed by NOAO, which is operated by the Association of Universities for Research in Astronomy, Inc., under cooperative agreement with the National Science Foundation \citealt{IRAF86,IRAF93}} HYDRA package.  After sky subtraction, a custom Python routine was applied to remove sky line residuals and exposures from the same pointing were averaged.  To reduce noise, the averaged spectra were smoothed with a Gaussian kernel ($\sigma =$ 1 pixel or 0.306\AA). Emission lines were fit with a Gaussian using the IDL-based Peak Analysis software (PAN; \citealt{PAN}), providing line fluxes and widths.  Recessional velocities were measured with the FXCOR task in IRAF.  The measured H$\alpha$ line Full-Width at Half Maximums (FWHM) were corrected for the instrumental broadening of 48.5$\pm$1.6 km s$^{-1}$, measured from the equivalently smoothed ThAr spectra. Given the instrumental broadening, we estimate the minimum robustly measured observed FWHM to be 48.5$\pm2\sigma$. Since the true velocity FWHM is a convolution of the instrumental broadening and the observed velocity width, the minimum robustly measured true velocity FWHM is $\simeq$17.9~\kms. The minimum measured true velocity FWHM in the data is 19~\kms, meaning all reported velocity widths are robustly measured. The FWHM values are then converted to velocity dispersions ($\sigma_{H\alpha}$) for the analysis and the minimum robustly measured $\sigma_{H\alpha}$ is 7.8~\kms. The PAN fits and H$\alpha$ line profiles where visually inspected and the H$\alpha$ lines that passed were mapped to their SparsePak fiber placements. The PAN measured line fluxes, and velocity dispersions and FXCOR-derived velocity fields are in the Appendix. 

\begin{table*} 
    \centering
    \small
    \caption{\centering SparsePak Observations}
        \begin{tabular}{l c c c c} 
        \hline
        \hline 
        Galaxy  & No. of  & No. of  Filled & Date & ToS   \\
         & Fields & Fields & of Obs & sec \\ [0.5ex] 
        \hline 
        UGC 0685 & 1 & 1 & 2015, Dec 14 & 1200 \\
        NGC 0784 & 1 & 1 & 2015, Dec 14 & 1200 \\
        NGC 0784 & 2 & 0 & 2015, Dec 14 & 1800 \\
        NGC 0784 & 3 & 2 & 2021, Dec 7, 8 & 2340 \\
        NGC 2366 & 9 & 9 & 2019, Jan 7, 8, 9, 10 11, & 2340 \\
        NGC 2366 & 2 & 2 & 2019, Jan 8 & 780 \\
        NGC 2366 & 1 & 1 & 2019, Jan 12 & 1560 \\
        Holmberg II & 13 & 8 & 2021, Dec 7, 8, 11, 12; 2023 Feb 12, 13 & 2340 \\
        Holmberg II & 5 & 2 & 2021 Dec 11, 12; 2023 Jan 22, Feb 12 & 1560\\
        UGC 4459 & 1 & 1 & 2019, Nov 1 & 2340 \\
        UGC 4459 & 1 & 0 & 2021, May 4 & 1560 \\
        Holmberg I & 4 & 3 & 2019, Jan 11, 12 & 2340 \\
        Sextans B & 6 & 3 & 2019, Jan 7, 8, 11 & 2340 \\
        Sextans B & 1 & 1 & 2019, Jan 8 & 780 \\
        Sextans A & 7 & 5 & 2023, Jan 22, 23, Feb 12 & 2340 \\
        Sextans A & 1 & 0 & 2023, Jan 23, Feb 12 & 1560  \\
        IC 2574 & 11 & 8 & 2019, Jan 11, 12, 13 & 2340 \\
        IC 2574 & 2 & 0 & 2019, Jan 12, 13 & 780 \\
        NGC 3738 & 1 & 1 & 2021, May 2 & 2700 \\
        NGC 3741 & 1 & 1 & 2017, Apr 22 & 2700 \\
        NGC 4068 & 2 & 2 & 2016, Apr 3 & 1800 \\
        NGC 4163 & 1 & 1 & 2017, Apr 23 & 2700 \\
        NGC 4190 & 2 & 1 & 2021, May 3 & 2700 \\
        UGC 7577 & 2 & 1 & 2023, Feb 12 & 2340 \\
        UGCA 292 & 2 & 1 & 2017, Apr 21 & 2700 \\
        %IC 3687 & 3 & 2 & 2021, May 4 & 2340 \\
        UGC 8024 & 1 & 1 & 2017, Apr 23 & 2700 \\
        GR8 & 1 & 1 & 2021, May 7 & 2700 \\
        %IC 4182 & 5 & 4 & 2022, Apr 28, 30 & 2700 \\
        %IC 4182 & 1 & 0 & 2022, Apr 30 & 1800 \\
        UGC 8201 & 2 & 1 & 2021, May 6, 8 & 2700 \\
        UGC 8201 & 1 & 1 & 2021, May 6, 8 & 2340 \\
        NGC 5204 & 5 & 4 & 2022, Apr 24, 25, 28 & 2700 \\
        UGC 8638 & 1 & 1 & 2021, May 8 & 2700 \\
        UGC 8651 & 1 & 1 & 2017, Apr 23 & 2160 \\
        UGC 9128 & 1 & 1 & 2017, Apr 22 & 2700\\
        UGC 9240 & 1 & 1 & 2016, Apr 3, 5 & 2520 \\
        NGC 6789 & 1 & 1 & 2016, Apr 2 & 1800 \\ [1ex]
        \hline 
    \end{tabular} 

    \tablecomments{Difference between No. Fields and No. Filled Fields is the number of Fields with one or two points in the dither pattern completed.}
    \label{table:sparse_ch5} 
\end{table*}
\vspace{-3mm}

\section{Methods}\label{Methods_c5}

Each galaxy was divided into square regions of $\sim$400 pc per side as shown in Figures \ref{06853_frame} and \ref{56663_frame}.  This spatial scale enables us to investigate the localized impact of star formation on the ISM and compare our results to large-scale measurements from studies such as \cite{Stilp13b, Hunter21, Elmegreen22}. For each region, we independently measure the SFH, ionized gas velocity dispersion, and atomic gas velocity dispersions and energy surface density. The 400~pc region size balances the relevant observational constraints and theoretical expectations. It is large enough to contain sufficient star counts to reliably derive SFHs with acceptable time resolution ($\simeq$ 25 Myrs in the most recent time bins), and small enough to preserve the signatures of local turbulence effects.  This scale is also physically motivated: it is comparable to the scale heights of dwarf galaxies and to the spatial extent over which supernovae and superbubbles are predicted to deposit energy into the ISM \citep[e.g.,][]{Kimetal17, Gentry17, Bacchini20a}.

A large portion of the regions in our sample are located towards the outskirts of their galaxies. These regions have lower \hi column densities and historically have low SFRs. \citetalias{LCH_2022,LCH_2023} did not include these outer regions.  For \citetalias{LCH_2022}, all regions were required to have sufficient ($>$50) young stars to reconstruct the recent SFH.  This cut excludes {the} galaxies' outer areas. In \citetalias{LCH_2023}, the HST footprint for Holmberg II does not cover the outer \hi gas disk.  We analyze the full sample of regions in Section~\ref{full_anal} and then reanalyze the sample with the same restrictions as \citetalias{LCH_2022} in Section \ref{blue_anal}. 

\begin{figure*}[!tb]
    \centering
    \includegraphics[width=\textwidth]{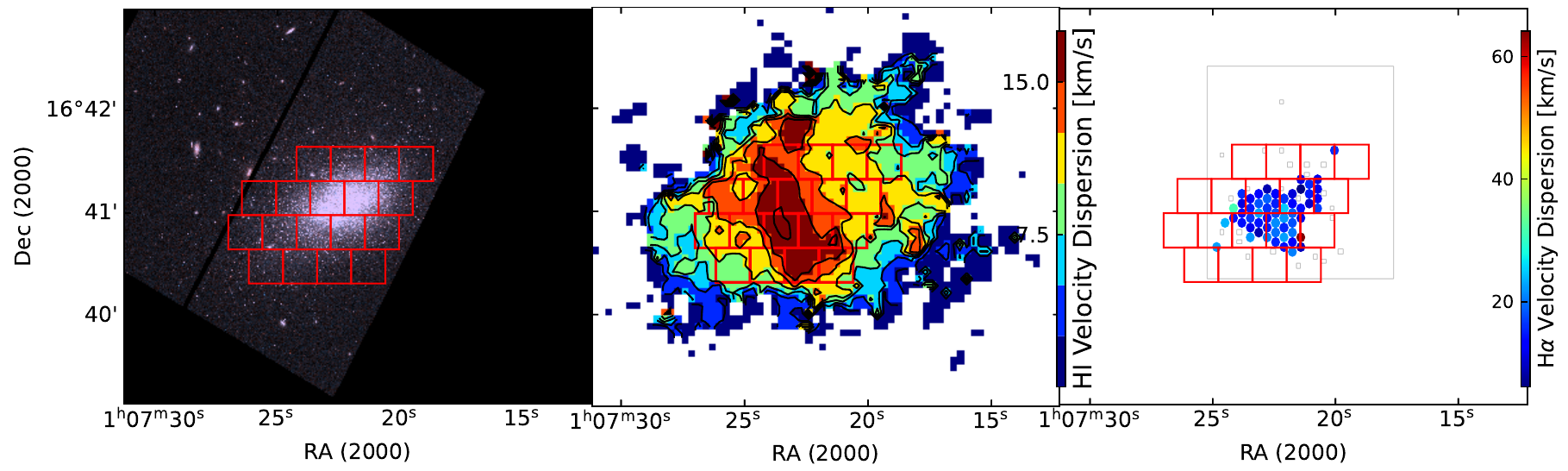}
        \caption{{UGC 0685 Regions Maps} Left: Two color image from \textit{HST} F814W (red) and F606W (blue) observations with ACS, Center: \hi dispersion map from VLA observations with isovelocity contours in 2.5 km s$^{-1}$ step size, Right: $\sigma_{H\alpha}$ map from the SparsePak IFU on the WIYN 3.5m telescope, with each filled circle corresponding to a fiber's size and position on the sky. Overlaid on all three panels are the outlines of the regions used for the analysis.}
    \label{06853_frame}
\end{figure*}

\begin{figure*}[!tb]
    \centering
    \includegraphics[width=\textwidth]{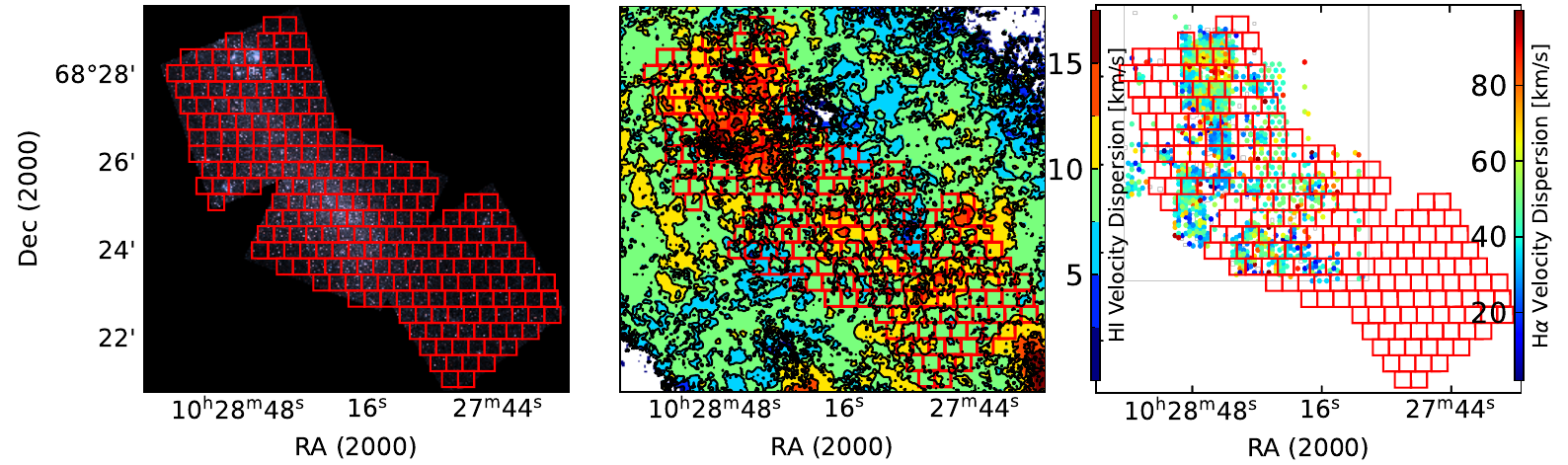}
        \caption{{IC 2574 Regions Maps} Left: Two color image from \textit{HST} F814W (red) and F555W (blue) observations with ACS, Center: \hi dispersion map from VLA observations with isovelocity contours in 2.5 km s$^{-1}$ step size, Right: $\sigma_{H\alpha}$ map from the SparsePak IFU on the WIYN 3.5m telescope, with each filled circle corresponding to a fiber's size and position on the sky. Overlaid on all three panels are the outlines of the regions used for the analysis.}
    \label{56663_frame}
\end{figure*}

\subsection{Local Star Formation Histories}\label{sec:SFH_c5}

The numerical CMD-fitting code MATCH was used to reconstruct SFHs for individual 400$\times$400 pc regions from resolved stellar populations \citep{Dolphin02}.  For detailed description of the methods, see \cite{McQuinn10a} and the references therein. In summary, MATCH generates synthetic simple stellar populations (SSPs) assuming a Kroupa initial mass function (IMF; \citealt{Kroupa01}), a binary fraction of 35\% with a flat binary mass ratio distribution, and the PARSEC stellar library \citep{Bressan12}. For the SFH fits, we constrained metallicity to be non-decreasing with time. This serves to guard against severe age-metallicity degeneracies in cases where available photometry doesn't reach faint-ward of the ancient main sequence turnoff, and has been applied in such cases in many other SFH studies employing a similar fitting approach for Local Volume dwarfs (e.g., \citealt{McQuinn10a, Weisz11, Weisz14, Williams17}). No internal differential extinction was assumed; based on the low-masses of galaxies in the sample, internal extinction is expected to be low (i.e., the mass-metallicity relation; \citealt{Berg12}). For the foreground extinction, the \cite{Schlafly11} recalibration of the \cite{Schlegel98} dust emission maps was used.  All regions in the same galaxy were assumed to have the same foreground extinction correction.  Observational errors are simulated using the completeness, photometric bias, and photometric scatter measured in the artificial star tests. The synthetic CMDs were combined linearly (along with simulated CMDs of foreground stars) to calculate the expected distribution of stars on the CMD for any SFH.  With the synthetic and observed V vs (V-I) CMDs, a maximum likelihood algorithm was used to determine the SFH most likely to have produced the observed data for each region.  Random uncertainties were estimated by applying a hybrid Markov Chain Monte Carlo \mbox{simulation \citep{Dolphin13}.} The systematic uncertainties were calculated following the prescriptions in \cite{Dolphin12}. A time binning for the SFH of $\Delta$log(t/{Myr})$\simeq$0.3 was adopted over the most recent 500 Myr, which covers the timescales of interest for star-formation driven turbulence. The resulting time intervals are 4-10~Myr, 10-25~Myr, 25-50~Myr, 50-100~Myr, 100-200~Myr, 200-500~Myr, and 500~Myr-14~Gyr.  Example CMDs and SFHs for selected regions are shown in Figure \ref{CMD_c5}.

\begin{figure*}[!tb]
    \centering
    \includegraphics[width=.85\textwidth]{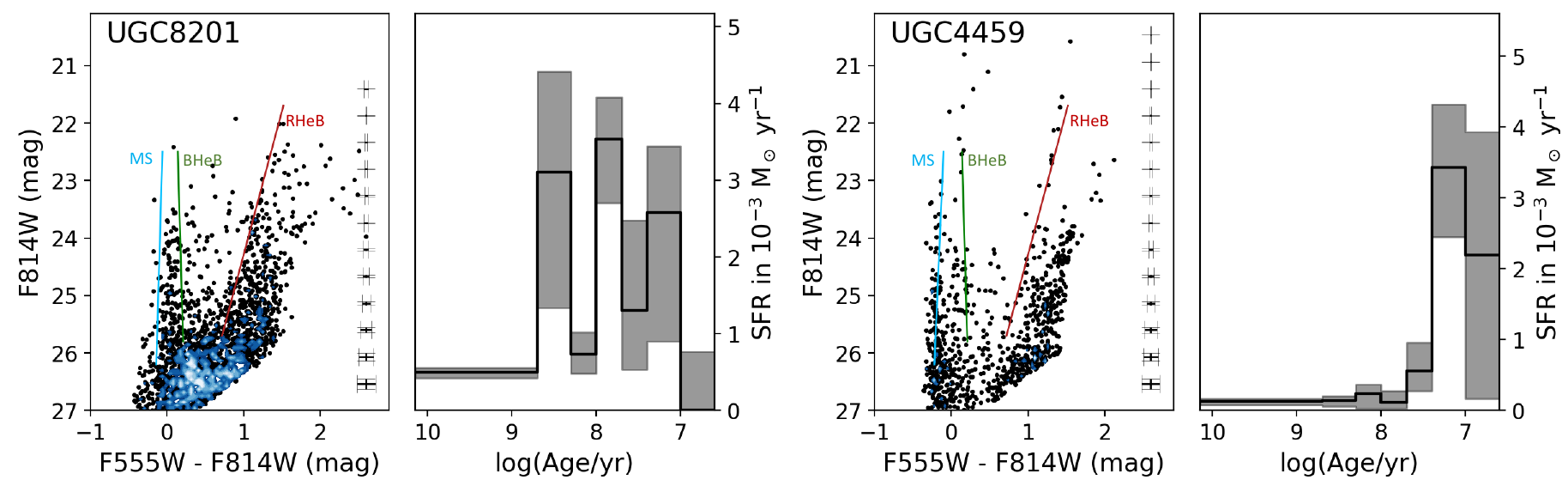}
        \caption{\textbf{Example CMD's} Representative CMDs and SFHs for 400$\times$400~pc regions in UGC 8201 and UGC 4459 . For both regions, the MS (blue), blue Helium burning stars (HeB) (green), and red HeB (red) sequences are traced.  The SFHs have $\leq$25 Myr time resolution in the two most recent time bins with $\Delta$log(t/yr)=0.3 time steps covering the 500 Myr baseline necessary for this projects scientific goals.  The gray shading is the combined systematic and random uncertainties of the SFH. The CMD-derived SFHs are compared with regional measurements of the \hi and H$\alpha$ turbulence to determine the time over which stellar feedback impacts multiple phases of the ISM. }
    \label{CMD_c5}
\end{figure*}

\subsection{Ionized Gas Turbulence Measurements}

We followed the same methodology as \citetalias{LCH_2023} to determine the turbulence in the ionized gas for each region. For each region, the SparsePak spectra were visually inspected and those with well-defined H$\alpha$ line profiles were centered to remove bulk motions and then co-added. After stacking, the continuum was subtracted, and the resulting line profile was corrected for instrumental broadening. From the stacked line profile, the FWHM and velocity dispersion $\sigma_{H\alpha}$ were measured.  The S/N of the stacked profiles was calculated with the peak of the stacked line as the signal, and the noise as the noise of individual fiber spectrum (Noise$_{spec}$, 1.8 $\times 10^{-17}$ erg s$^{-1}$ cm$^{-2}$ \AA$^{-1}$) divided by the square root of the number of fibers ($Num_{fiber}$) contributing to the line profile.
\begin{equation}
    \frac{S}{N}=Peak/ \Bigl( \frac{Noise_{spec}}{\sqrt{Num_{fiber}}} \Bigr)
\end{equation}

For the uncertainty of the line, a S/N of 10 was set to correspond to a 10\% uncertainty of the peak's strength and a S/N of 100 corresponding to a 1\% uncertainty. For lines with S/N greater than 100, the uncertainty was set to $\frac{Noise_{spec}}{\sqrt{Num_{fiber}}}$.  For the line width, the uncertainty was set to 10\% the instrumental broadening.  For the final uncertainty, the two uncertainties were added in quadrature, with the uncertainty from instrumental broadening dominating.

\subsection{HI Turbulence Measures} \label{turb_hi_c5}

The \hi turbulence measures follow the methods detailed in \citetalias{LCH_2022,LCH_2023} and outlined in \cite{Stilp13b} and \cite{Ianjama12}.  The two independent methods are summarized here. Each region's velocity was determined from the second moment maps and from the co-added line-of-sight profiles corrected for bulk motions such as rotation. For the moment maps, the flux weighted average of the second moment map was measured for each region: 

\begin{equation}
    \sigma_{m2}=\frac{\Sigma_i \sigma_i N_{HI,i}}{\Sigma_i N_{HI,i}}
\end{equation}
where N$_{HI,i}$ is the \hi column density per pixel, and $\sigma_i$ is the second moment velocity dispersion of each pixel.  Representative values for the velocity dispersions derived from the moment maps are listed in Table \ref{table:turbulence_c5}. For the uncertainty of the second moment velocity dispersion, we use the standard deviation of the weighted mean.

\begin{figure}
    \centering
    \includegraphics[width=0.45\textwidth]{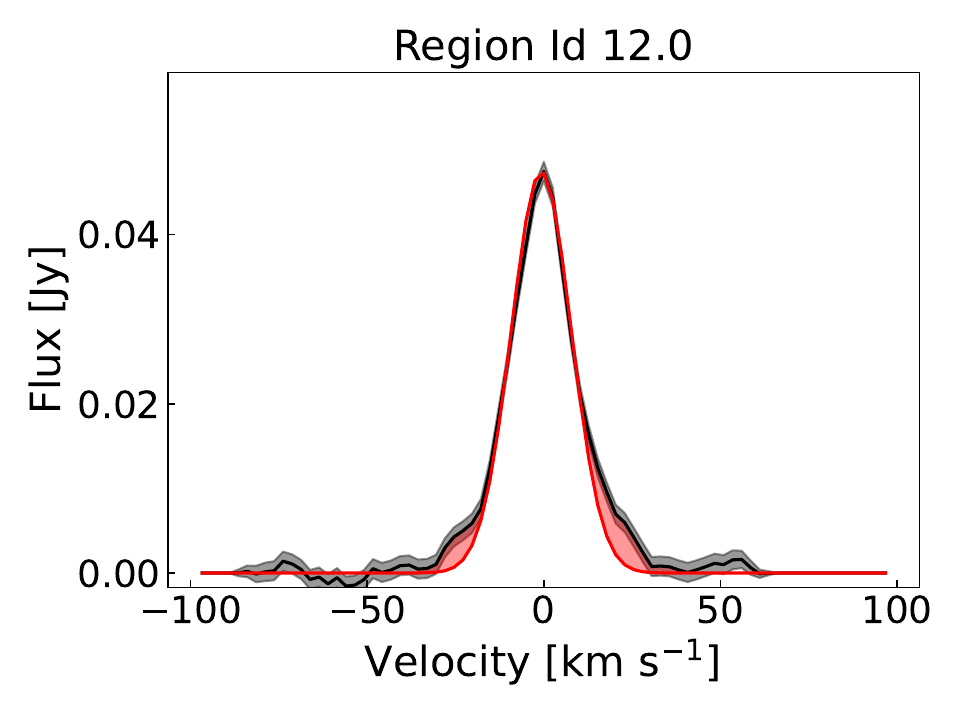}
    \caption{\textbf{UGC 9240 Superprofile:} The superprofile of a selected region in UGC 9240.  The black line is the bulk-motion corrected \hi flux from the region and the red line is the Gaussian fit for the data.  The shaded gray region is the error on the data, while the shaded red region is the wings of the \hi flux.  The wings are the high velocity, low density gas that is poorly fit by a single Gaussian.}
    \label{super_pro_9240}
\end{figure}

Superprofiles are co-added, line-of-sight \hi flux profiles corrected for rotational velocities.  We construct a superprofile for each region to determine the \hi velocity dispersion (Example: Figure \ref{super_pro_9240}). 

To create the superprofiles, we begin by fitting a Gaussian-Hermite function to the \hi line profile of each pixel of the \hi data cubes to determine the pixel-by-pixel line centers.  We then shift the line centers to 0 km s$^{-1}$ and remove the gas' bulk motions.  In \citetalias{LCH_2023}, regions were excluded if less than one-third of the pixels in the region had a \hi line detection above 3$\sigma$. This cut ensured only regions with reliable \hi detections are included in the superprofile. For this sample, additional cuts are made.  We also required the resulting superprofile to have a peak greater than 3 times the uncertainty of the peak.  The uncertainty of each point in the superprofile is:

\begin{equation}
    \sigma = \sigma_{ch,rms} \times \sqrt{N_{pix}/N_{pix/beam}}
    \label{uncereq_c5}
\end{equation}
where $\sigma_{ch,rms}$ is the rms noise per channel, N$_{pix}$ is the number of pixels contributing to a given point in the superprofile, and N$_{pix/beam}$ is %the number of profiles within one resolution element or 
pixels per beam.  

For each superprofile, a Gaussian was scaled to the amplitude and the FWHM of the line profile.  The observed \hi line profiles contain higher velocity and lower density gas which are poorly fit by the Gaussian.  This gas, which is above the Gaussian fit is described as the wings of the superprofile (Figure \ref{super_pro_9240}).  We determine the velocity dispersion of the low-density gas from these wings. To measure the \hi flux in the wings of the superprofiles, we require the peak to be at least 6 times the uncertainty of the peak. From the scaled Gaussian fits we measure three parameters:

\begin{enumerate}
    \item $\sigma_{central}$: the $\sigma$ of the scaled-Gaussian profile fit to the FWHM and amplitude of the observed \hi superprofile
    \item $f_{wings}$: the fraction of \hi flux in the wings of the superprofile
    \item $\sigma^2_{wings}$:  the rms velocity of the \hi flux in the profile wings
\end{enumerate}

We estimate the errors on these parameters following the methods of \citetalias{LCH_2022,LCH_2023}. The superprofile were refit 2000 times adding Gaussian noise to each point based off Equation \ref{uncereq_c5}. The ranges of $\sigma_{cen}$ and $\sigma_{wing}$ per galaxy are listed in Table \ref{table:turbulence_c5}.  Additionally, the \hi energy surface density ($\Sigma_{HI}$) was determined for each region.   The $\Sigma_{HI}$ per region was estimated from the average \hi surface density (M$_{HI}$/A$_{HI}$) of the region, where M$_{HI}$ is the \hi mass within the region and A$_{HI}$ is the area of the region. (M$_{HI}$/A$_{HI}$) is multiplied by 3/2 to account for the motion in all three directions, assuming isotropic velocity dispersion. The equations for $\Sigma_{HI}$ are:

\begin{enumerate}
    \item $\Sigma_{E,m2}$ is the \hi energy surface density from the second moment averages ($\sigma_{m2}$)
    \begin{equation}
        \Sigma_{\text{E,m2}} = \frac{3 M_{HI}}{2 A_{HI}}\sigma_{\text{m2}}^2
    \end{equation}
    \item $\Sigma_{E,central}$ is the \hi energy surface density derived from the superprofiles ($\sigma_{central}$):
    \begin{equation}
        \Sigma_{\text{E,central}} = \frac{3 M_{HI}}{2 A_{HI}}(1-f_{\text{wing}})(1-f_{\text{cold}})\sigma_{\text{central}}^2
    \end{equation}
    M$_{HI}$ is the total \hi mass within the region, M$_{HI}$(1-\textit{f}$_{\text{wings}}$)(1-f$_{\text{cold}}$) is the total \hi mass contained within the central peak corrected for the dynamically cold HI, and the fraction of \hi within the wings of the superprofile.  As in \citetalias{LCH_2022,LCH_2023}, $f_{\text{cold}}$=0.15 was chosen to be consistent with \cite{Stilp13a} and previous estimates for dwarf galaxies \citep{Young03,Bolatto11,Warren12}.  
    \item $\Sigma_{\text{E,wing}}$ is the \hi energy surface density derived for the wings of the superprofiles:
    \begin{equation}
        \Sigma_{\text{E,wing}} = \frac{3 M_{HI}}{2 A_{HI}}f_{\text{wings}}\sigma_{\text{wings}}^2
    \end{equation}
\end{enumerate}

We assumed 10\% uncertainty for the \hi surface density (M$_{HI}$/A$_{HI}$) based-off uncertainties in \hi fluxes in \cite{vanzee97}, and differences between single dish observations and the VLA \hi fluxes. 

\begin{table*}
\begin{rotatetable*}
\centering
\footnotesize
\caption{Median and Range Turbulence Measures} 
    \begin{tabular}{l c c c c c c c c c c c c c c c }
        \hline
        Galaxy & M$_{HI}$/A$_{HI}$ & 25\% & 75\% & $\sigma_{H\alpha}$ & 25\% & 75\% & $\sigma_{\text{m2}}$ & 25\% & 75\% & $\sigma_{\text{central}}$ & 25\% & 75\% & $\sigma_{\text{wings}}$ & 25\% & 75\% \\
        & M$_\odot$/pc$^2$ & \multicolumn{2}{c}{Range} & km/s & \multicolumn{2}{c}{Range} & km/s & \multicolumn{2}{c}{Range} & km/s & \multicolumn{2}{c}{Range} & km/s & \multicolumn{2}{c}{Range}\\
        \hline 
        UGC 0685 & 10.4 & 5.9 & 14.9 & 17.0 & 15.1 & 19.0 & 13.2 & 10.8 & 15.1 & 13.1 & 10.5 & 15.0 & 33.7 & 29.6 & 37.3 \\
        NGC 0784 & 23.2 & 16.7 & 30.9 & 17.5 & 15.4 & 18.8 & 13.2 & 12.6 & 14.1 & 12.0 & 11.0 & 13.3 & 31.9 & 28.7 & 34.1 \\
        NGC 2366 & 12.3 & 8.4 & 17.9 & 17.7 & 16.2 & 19.7 & 14.2 & 13.1 & 15.4 & 10.9 & 9.7 & 12.3 & 30.2 & 26.8 & 34.8 \\
        Holmberg II & 17.5 & 12.3 & 20.8 & 18.1 & 16.2 & 19.9 & 10.0 & 9.2 & 11.2 & 11.3 & 10.0 & 13.7 & 29.7 & 25.3 & 33.9 \\
        UGC 4459 & 7.8 & 4.7 & 12.5 & 17.1 & 15.5 & 18.3 & 10.4 & 9.8 & 11.2 & 10.1 & 8.9 & 12.0 & 26.4 & 24.7 & 28.5 \\
        Holmberg I & 17.7 & 12.3 & 20.5 & 15.8 & 14.4 & 18.2 & 7.9 & 7.1 & 8.5 & 11.3 & 9.9 & 12.4 & 29.2 & 26.3 & 31.9 \\
        Sextans B & 6.5 & 6.1 & 7.2 & 16.8 & 15.8 & 18.7 & 8.5 & 8.4 & 8.8 & 7.0 & 7.0 & 7.6 & 20.6 & 20.6 & 21.7\\
        Sextans A & 6.4 & 6.1 & 7.2 & 19.4 & 17.4 & 19.7 & 10.0 & 8.7 & 10.6 & 10.5 & 8.4 & 12.4 & 28.9 & 23.7 & 31.4\\
        IC 2574 & 11.0 & 7.9 & 14.0 & 18.0 & 15.9 & 19.7 & 9.5 & 8.7 & 10.7 & 9.3 & 8.0 & 11.0 & 23.3 & 21.3 & 27.4 \\
        NGC 3738 & 11.6 & 9.5 & 17.8 & 21.5 & 19.0 & 24.4 & 17.6 & 15.1 & 19.8 & 17.7 & 16.2 & 22.1 & 47.3 & 41.2 & 53.0\\
        NGC 3741 & 17.8 & 16.5 & 27.0 & 16.9 & 16.7 & 17.9 & 8.0 & 6.9 & 8.6 & 11.1 & 10.6 & 11.7 & 32.8 & 22.5 & 42.1 \\
        NGC 4068 & 14.3 & 12.2 & 17.7 & 17.5 & 16.4 & 19.2 & 8.9 & 8.5 & 9.4 & 10.3 & 9.4 & 11.8 & 27.3 & 24.3 & 30.2 \\
        NGC 4163 & 4.8 & 4.1 & 8.8 & 16.7 & 16.2 & 17.7 & 7.7 & 7.6 & 8.7 & 8.9 & 8.7 & 9.9 & 20.9 & 20.5 & 22.9 \\
        NGC 4190 & 16.7 & 15.4 & 18.3 & 19.9 & 17.7 & 20.4 & 8.6 & 8.2 & 9.2 & 10.9 & 9.9 & 11.5 & 27.1 & 25.9 & 29.2\\
        UGC 7577 & 5.8 & 5.4 & 6.4 & 18.0 & 17.7 & 18.5 & 4.8 & 4.7 & 5.2 & 6.8 & 5.9 & 6.9 & 14.5 & 12.1 & 14.7 \\
        UGCA 292 & 18.9 & 12.9 & 20.7 & 16.0 & 16.0 & 16.3 & 8.7 & 8.7 & 8.9 & 8.5 & 8.2 & 9.2 & 19.6 & 19.0 & 19.6\\
        UGC 8024 & 13.3 & 12.0 & 14.3 & 16.6 & 15.7 & 17.9 & 9.9 & 9.6 & 10.4 & 9.7 & 8.9 & 10.4 & 25.1 & 24.5 & 27.7\\
        GR8 & 5.2 & 2.7 & 7.0 & 18.6 & 18.0 & 19.4 & 8.2 & 7.9 & 8.4 & 8.7 & 7.7 & 10.3 & 21.9 & 19.9 & 22.6 \\
        UGC 8201 & 7.3 & 4.2 & 10.3 & 17.3 & 15.9 & 20.6 & 12.5 & 10.1 & 14.0 & 12.9 & 10.7 & 15.4 & 32.8 & 28.2 & 38.3\\
        NGC 5204 & 16.4 & 15.7 & 16.8 & 18.8 & 18.4 & 19.2 & 14.1 & 13.5 & 14.5 & 12.2 & 12.0 & 12.3 & 37.1 & 36.9 & 37.6 \\
        UGC 8638 & 6.8 & 5.2 & 10.0 & 18.6 & 17.5 & 20.3 & 11.2 & 10.9 & 11.7 & 11.9 & 11.4 & 14.1 & 29.2 & 25.7 & 32.8 \\
        UGC 8651 & 5.6 & 4.2 & 7.2 & 15.0 & 8.9 & 18.1 & 6.3 & 5.8 & 6.6 & 8.2 & 7.4 & 8.7 & 22.9 & 18.6 & 25.07 \\
        NGC 5253 & 10.1 & 6.4 & 15.1 & N/a & N/a & N/a & 14.4 & 11.5 & 16.7 & 19.1 & 16.3 & 20.3 & 46.9 & 44.5 & 56.2 \\
        UGC 9128 & 6.5 & 5.2 & 9.0 & 15.2 & 15.0 & 15.5 & 8.7 & 7.2 & 9.6 & 9.3 & 9.0 & 13.0 & 21.3 & 20.6 & 22.4 \\
        UGC 9240 & 7.3 & 5.0 & 14.2 & 15.8 & 15.4 & 17.0 & 10.6 & 9.9 & 11.5 & 11.5 & 10.4 & 13.4 & 29.7 & 27.6 & 32.9 \\
        NGC 6789 & 5.4 & 4.8 & 13.1 & 17.2 & 16.1 & 17.7 & 8.9 & 8.0 & 9.9 & 12.2 & 10.3 & 12.5 & 25.8 & 23.6 & 40.4 \\[1ex] 
        \hline 
        \end{tabular}
        \tablecomments{\small \textbf{{M$_{HI}$/A$_{HI}$ is the average \hi surface density per regions used to derive the \hi energy surface density\\
    $\sigma{_H\alpha}$ is the median H$\alpha$ velocity dispersion and the upper and lower quartiles for each galaxy \\
    $\sigma{_{m2}}$ is the 2nd moment map velocity dispersion and the upper and lower quartiles for each galaxy \\
    $\sigma{_{central}}$ is the superprofile fit velocity dispersion and the upper and lower quartiles for each galaxy \\
    $\sigma{_{wings}}$ is the wings of the superprofile velocity dispersion and the upper and lower quartiles for each galaxy \\ }}}
        \label{table:turbulence_c5} 
\end{rotatetable*}
\end{table*}

\subsection{Spearman Rank Correlation Coefficient}

In our analysis, we compare the different turbulence measures to the SFHs in the past 500~Myrs. A strong correlation between the current turbulence measures and the SFR in a given time bin would suggest that the current turbulence is influenced by the star formation activity at that time.  We quantify these correlations using the Spearman rank correlation coefficient, $\rho$, which tests  for a monotonic relationship between two variables. Values of $0 < \rho \leq 1$ indicate a positive correlation, $-1 \leq \rho < 0$ indicate an anti-correlation, and $\rho = 0$ reflects no correlation. We use standard thresholds of 0$<|\rho|\leq$0.2 is no correlation, 0.2$<|\rho|\leq$0.4 indicates a weak correlation, 0.4$<|\rho|\leq$0.7 indicates a strong correlation, and 0.7$<|\rho|<$1 indicates a very strong correlation. Along with $\rho$, the Spearman correlation test reports a corresponding \textit{P-value}, which gives the probability of obtaining a $\rho$ as extreme as the observed one under the null hypothesis of no correlation. A \textit{P}-value less than or equal to 0.05 is considered statistically significant.  For the majority of our tests the \textit{P}-values are $\leq$0.001, and we draw attention to \textit{P}-values $\geq$0.001

To test whether the regions analyzed sufficiently sample the underlying parameter space, we applied a bootstrap resampling technique. We randomly resampled the regions 3000 times, with replacement, drawing the same number of regions as in the observed sample for each iteration. This provided a distribution of possible Spearman $\rho$ values, from which we took the central 68\% interval to represent the uncertainty on $\rho$.

\section{Ionized Gas Results} \label{halpha_res_c5}

\begin{figure}[!tb]
    \centering
    \includegraphics[width=.48\textwidth]{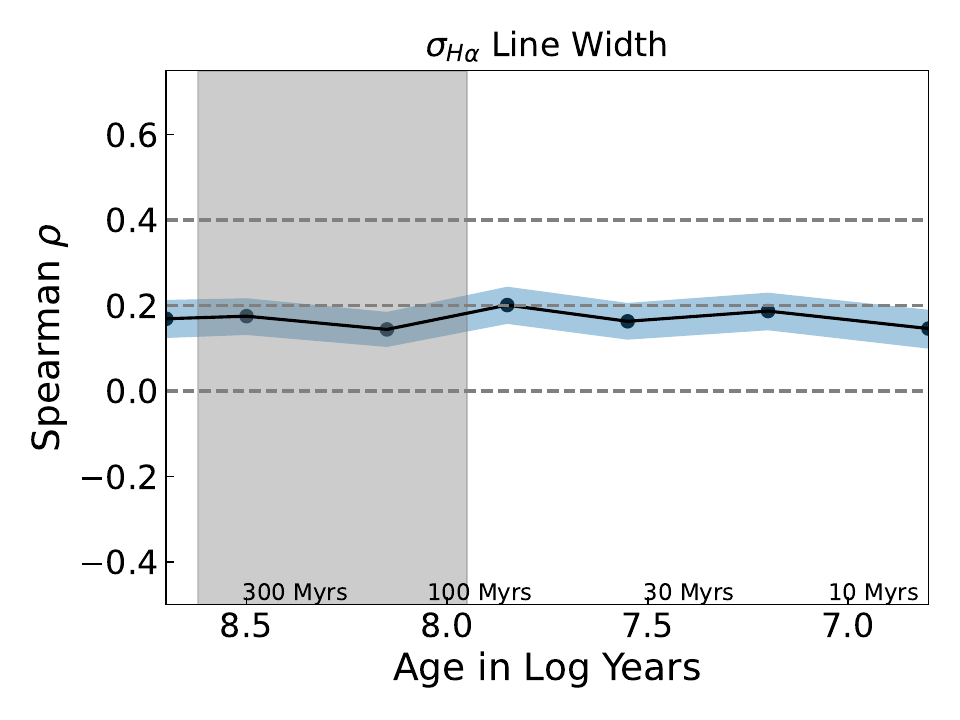}
        \caption{\small \textbf{Comparison of  H$\alpha$ derived ionized gas velocity dispersions and SFH results.} Spearman $\rho$ coefficient versus log time, which demonstrates how correlated the SFR of a given time bin is with the H$\alpha$ FWHM for the whole sample. The light blue shaded region represents the 1$\sigma$ bootstrapping error and under each point is the relevant \textit{P}value for \textit{P$\geq0.001$} . For the $\sigma_{H\alpha}$, there is no evidence of a  correlation between the velocity dispersion and the SFR in any time bin. The largest $\rho$ values is 0.20, which does not meet our $\rho>$0.20 requirement.}
    \label{boot_rs_ha_full}
\end{figure} 

The results of Spearman's rank correlation tests between the SFR at each time bin and the current H$\alpha$ velocity dispersion are shown in Figure \ref{boot_rs_ha_full}.  Each point represents the degree of correlation between the current $\sigma_{H\alpha}$ and the SFR at a specific lookback time with the 1$\sigma$ uncertainties shaded blue. {No clear increase in the strength of the correlation} is observed as at any timescale and all $\rho$ values fall between 0.13 and 0.20, indicating no statistically significant correlation. The possible indications of a correlation seen in \citetalias{LCH_2022,LCH_2023} were likely artifacts of the small sample sizes, where a few regions strongly influenced the results.  A few regions with higher SFRs 10-25~Myr ago and high current velocity dispersions appear to be responsible for the observed potential correlation in \citetalias{LCH_2022,LCH_2023}.  The larger number of regions here reduces the influence of individual regions and we observe no correlation between $\sigma_{H\alpha}$ and the SFR in the past 5-500~Myrs.

The previously observed correlation between the ionized gas kinematics and star formation activity is between the H$\alpha$ derived SFRs (sensitive to t$<$10~Myr) and $\sigma_{H\alpha}$ in IFU studies (e.g., \citealt{Green10, Moiseev15, Zhou17, Law22}). However, our CMD-derived SFHs do not provide the SFR for t$<$5~Myr \citep{McQuinn10a}, which limits our ability to test this correlation. For individual regions, H$\alpha$ fluxes are too low to reliably derive SFRs, making the data insensitive to correlations on very short timescales (see e.g., \citealt{Lee09} for a discussion of the effectiveness of star formation tracers at low SFRs).  
 
Interpreting the H$\alpha$ timescales is further complicated by uneven fiber coverage across the galaxies. Observations are biased toward regions with high-surface-brightness H$\alpha$ emission. This results in some of the 488 total regions being fully covered, while others contain a single fiber with sufficient signal-to-noise to be included. In such under-sampled regions,  the measured velocity dispersion represents only a small fraction of the area.  To robustly test for correlations on timescale longer than 5 Myr, a large sample of regions with H$\alpha$ detection across the whole region is necessary.  This analysis would be inherently biased by the requirement of current star formation to produce observable H$\alpha$ emission.

Any correlations may also be washed out by the uncertainties on the observed data. To test the impact of the uncertainties on the measured correlation strength, we simulated a data set with a sample size of 500 (similar to the number of regions analyzed) with an underlying correlation strength of $\rho=$0.2. For each simulated data point we randomly select an observed $\sigma_{H\alpha}$ and SFR along with the corresponding uncertainties. We then perturbed the simulated data point with a simulated uncertainties drawn from normal distributions centered at zero with a spread equal to the observed uncertainties. We repeat the simulation 2000 times. Applying the uncertainties on the simulated data modestly decreases the correlation strength by $\rho$ is 0.04$\pm0.025$. This offset in $\rho$ is similar to the uncertainties on $\rho$ we find from bootstrapping the data (see figure~\ref{halpha_res_c5}). The uncertainties on the data appear to mildly depress the correlation strength and may be partially responsible for the lack of correlation seen between the SFH and $\sigma_{H\alpha}$. 

\section{HI Local Timescale Results} \label{local_time_c5}

This section presents our results for the Spearman's rank correlation tests between the SFHs and \hi turbulence measures for different region and galaxy samples. In \citetalias{LCH_2022,LCH_2023}, we analyzed the correlation timescales between star formation activity and \hi turbulence in five low-mass galaxies. \citetalias{LCH_2022} focused on NGC~4068, NGC~4163, NGC~6789, and UGC~09128, identifying a clear correlation between the 3 velocity dispersion measures and 3 \hi energy surface density measures and star formation that occurred 100–200 Myr ago.   \citetalias{LCH_2023} examined Holmberg II (UGC~04305) and found a similar correlation, at a slightly earlier timescale of 70–140 Myr. 

In our analysis we consider a number of different factors that could influence our results. First, we explore the impact of including the full spatial extent of the galaxies. In \citetalias{LCH_2022}, we required regions to have recent star formation activity and contain more than 50 young stars.  These cuts effectively excluded the outer regions with little recent star formation. In \citetalias{LCH_2023}, the HST footprint of Holmberg II did not cover the outer \hi disk, excluding the outer areas from our analysis. To determine the impact of this cut,  Section~\ref{full_anal} analyzes the full set of regions, and Section~\ref{blue_anal} repeats the analysis for regions with the same \citetalias{LCH_2022} cuts based on young star counts. 

Second, we also explore differences as a function of mass and specific star formation rate (sSFR). The galaxies in the sample cover two orders in magnitude in stellar mass and roughly an order of magnitude in star formation. To measure the impact of stellar mass and sSFR on the feedback timescales we divide our sample into 3 mass bins and 2 sSFR bins. We examine these sub-samples in Section~\ref{sec:subsample}. Because of the large range of stellar masses, the physical size of the galaxies differs greatly. So, in Section~\ref{sec:equal} we test whether the largest galaxies dominate the full sample by using a more equal number of regions per galaxy.

\subsection{Full Spatial Sample Regional Results}\label{full_anal}

\begin{figure}[!tb]
    \gridline{\fig{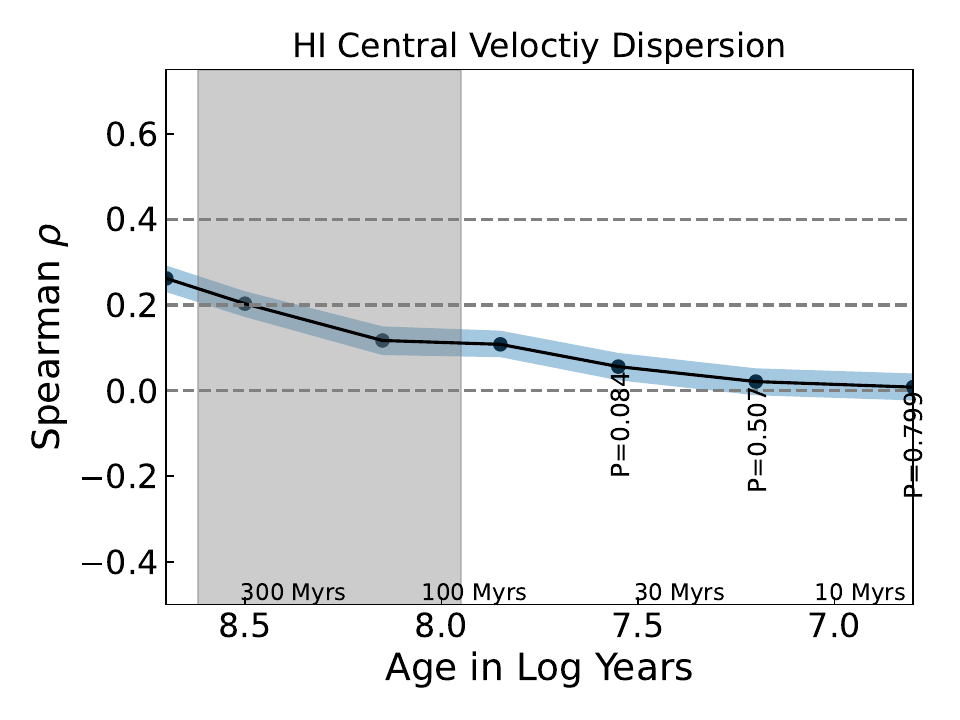}{.45\textwidth}{}}
    \gridline{\fig{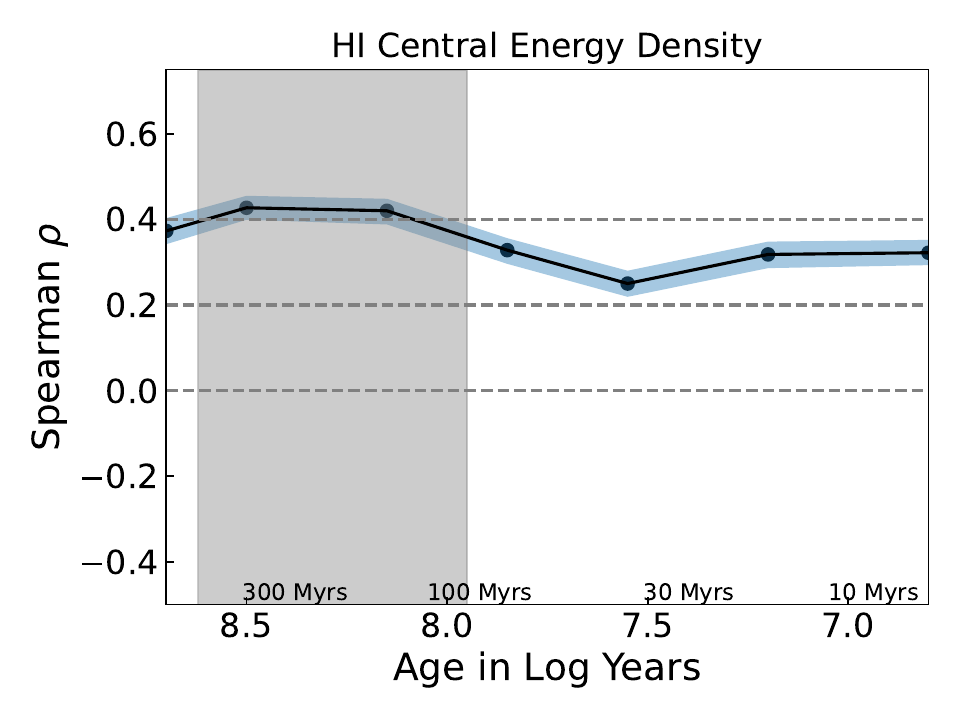}{.45\textwidth}{}}
    \caption{\small \textbf{Comparison of \hi Superprofile Measures and SFH Results: Full Spatial Coverage Sample.} 
    The Spearman $\rho$ coefficient as a function of log time, illustrating how strongly the star formation rate in each time bin correlates with (a) the \hi velocity dispersion ($\sigma_{cen}$) and (b) the \hi energy surface density ($\Sigma_{cen}$), derived from Gaussian superprofile fits, for the full galaxy sample. The light blue shaded region indicates the 1$\sigma$ bootstrapping error. Corresponding P-values are shown above for\textit{P}$\geq$0.001. For $\sigma_{cen}$ (panel a), the strongest correlation appears in the oldest time bin ($\rho=$0.262). In contrast, $\Sigma_{cen}$ (panel b) shows its strongest correlations in the second and third-to-last time bins ($\rho=$0.42 and $\rho=$0.43 respectively), corresponding to star formation activity from approximately 100–500 Myr ago. This period is highlighted by the gray shaded region. }
    \label{boot_rs_cen_full}
\end{figure}  

Figure~\ref{boot_rs_cen_full} shows the correlation for all regions between the SFR in a each time bin and the velocity dispersion and \hi energy surface density measured from the Gaussian superprofile fits. The light blue shaded region indicates the 1$\sigma$ bootstrapping error from resampling the regions 1000 times, and the corresponding P-value is listed beside each point. In Figure~\ref{boot_rs_cen_full}, the time range from 100-500~Myr is highlighted in gray as the \hi energy surface density measured from the superprofile fits has a notable increase in the correlation strength in this time range {with $\rho>$0.4}.  The other two \hi energy surface density measures not shown demonstrate a similar increase in the correlation strength at this time bin.  Although broader than the correlation timescales found in \citetalias{LCH_2022,LCH_2023}, it is in agreement with those findings.  The broader timescales appears to be driven, in part, by the diversity of the galaxy sample (see Section~\ref{sec:subsample}).  The galaxies analyzed span 2 orders of magnitude in stellar mass and \hi mass and nearly an order of magnitude in sSFR (see Table \ref{table:galax_ch5}). 

As in \citetalias{LCH_2022,LCH_2023}, the correlation between the HI energy surface density and the shorter time bins is weaker than the correlation with SFR at t$\simeq$~100 Myr.  This consistent feature appears to be driven by regions with recent star formation (t$\leq$50~Myr) having a diverse range of HI energy surface densities.  Regions with relatively high star formation rates in the first three SFH bins (SFR$\geq0.0075~M_\odot yr^{-1}$) consistently have HI energy surface densities similar to regions without star formation in the same bin, many having HI energy surface densities well below the mean of $\sim$50$\times10^{51}$ erg kpc$^{-2}$. This may imply the HI energy has not yet been significantly increased by the recent star formation event. This may be due to the delay between the formation of stars and SNe releasing large amount of energy into the ISM heating the atomic gas.

Conversely, the correlation with the velocity dispersion deviates significantly from expectations. For the velocity dispersion, the correlation with past SFR increases with lookback time and is strongest with $\rho=$0.26 at t~$>$~500 Myr.  This is seen in Figure \ref{boot_rs_cen_full} for the velocity dispersion from the superprofile fits.  The correlation with the velocity dispersions measured from the wings and 2$^{nd}$ moment maps follow the same trend. This time bin is the average SFR for t$>$500 Myr and lacks time resolution. Additionally, this is longer than the $\leq$500 Myr for stellar clusters to dissolve \citep{Gieles08,Bastian10,Bastian11}.

This weak correlation between the current velocity dispersion and average past SFR was not present in \citetalias{LCH_2022,LCH_2023}. Figure \ref{sfrvhi_8.7} demonstrates the SFR in the oldest time bin plotted against the current \hi velocity dispersion for the full sample of regions. It appears to be driven by a combination of a large number of regions with historically low SFRs and low current velocity dispersions and a small set of regions with historically higher SFRs and higher current velocity dispersions. A clear clustering of regions with low SFRs and velocity dispersions $\leq$15 km/s is evident.  In contrast, \citetalias{LCH_2022,LCH_2023} excluded some of these outer regions by requiring regions to have a minimum of 50 young stars to be included in the sample.  In Section~\ref{blue_anal}, we reanalyze the sample of regions using the same young star selection criteria as in \citetalias{LCH_2022} to assess the impact of these cuts on the correlation results.

The higher historical SFRs are associated with two of the most massive galaxies ($log(M_\star)>8.7$) in the sample NGC~3738 and NGC~5253. Both NGC~3738 and NGC~5253 have \hi gas disks with signs of disturbance, with NGC~5253 in particular being clearly disrupted with an asymmetric \hi disk. Both of these galaxies have significantly higher velocity dispersion compared to the majority of the sample (see Table~\ref{table:turbulence_c5}).  In Sections~\ref{sec:subsample}, only the high-mass sample has evidence of this correlation between the historical SFR and current velocity dispersion. Excluding NGC~3738 and NGC~5253 from the full sample decreases the correlation between the velocity dispersion and the SFRs in the oldest bins. Only the $\sigma_{mom2}$ shows a significant correlation with the SFR in the oldest bin.

Additionally, there is the impact of inclination on the results to consider. The inclinations of the galaxies in the sample are diverse ranging from face-on to nearly edge on as seen in the Appendix. The galaxies' inclination impacts the 3-dimensional volume of a given region. As projection effects stretch a region, stars and gas further apart than 400~pc are included in the same region. This results in our analysis combining regions of different effective sizes together. The inclination strongly impacts the HI kinematics measurements in ways that cannot be easily resolved with a geometric correction. Looking through an inclined galaxy results in more rotationally broadened \hi line profiles and often double-peaked line profiles ---the two peaks corresponding to the near and far side of the disk. This combination of more rotational broadening and altered line profile shape adds additional uncertainty to the \hi velocity dispersion and ${\Sigma_{\text \hi}}$. Combined with the impact to the region volume, galaxy inclinations likely weaken correlation strengths we observe and add additional uncertainty to the correlation timescales derived here. 

%  %due to some of the characteristics of the regions in the sample.

\begin{figure}[!tb]
    \gridline{\fig{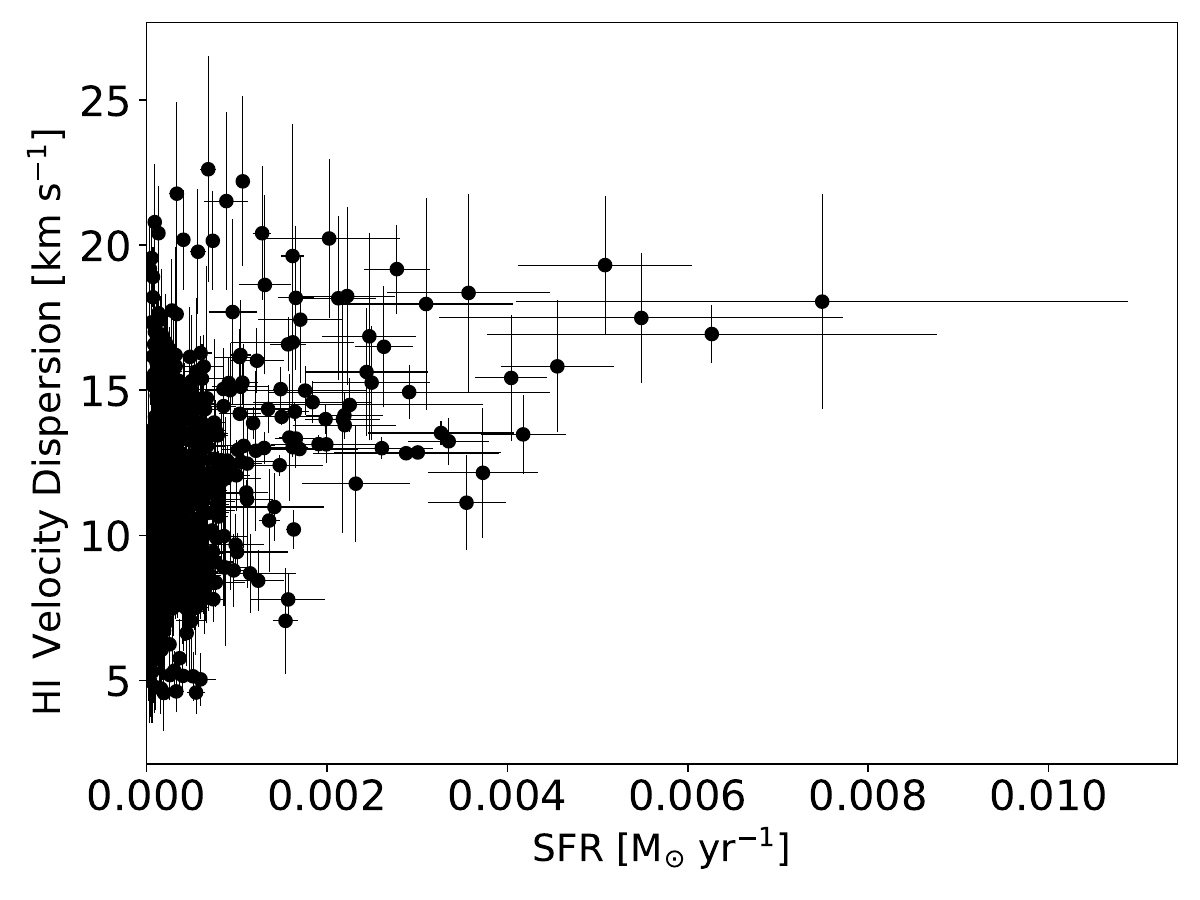}{.45\textwidth}{}}
    \gridline{\fig{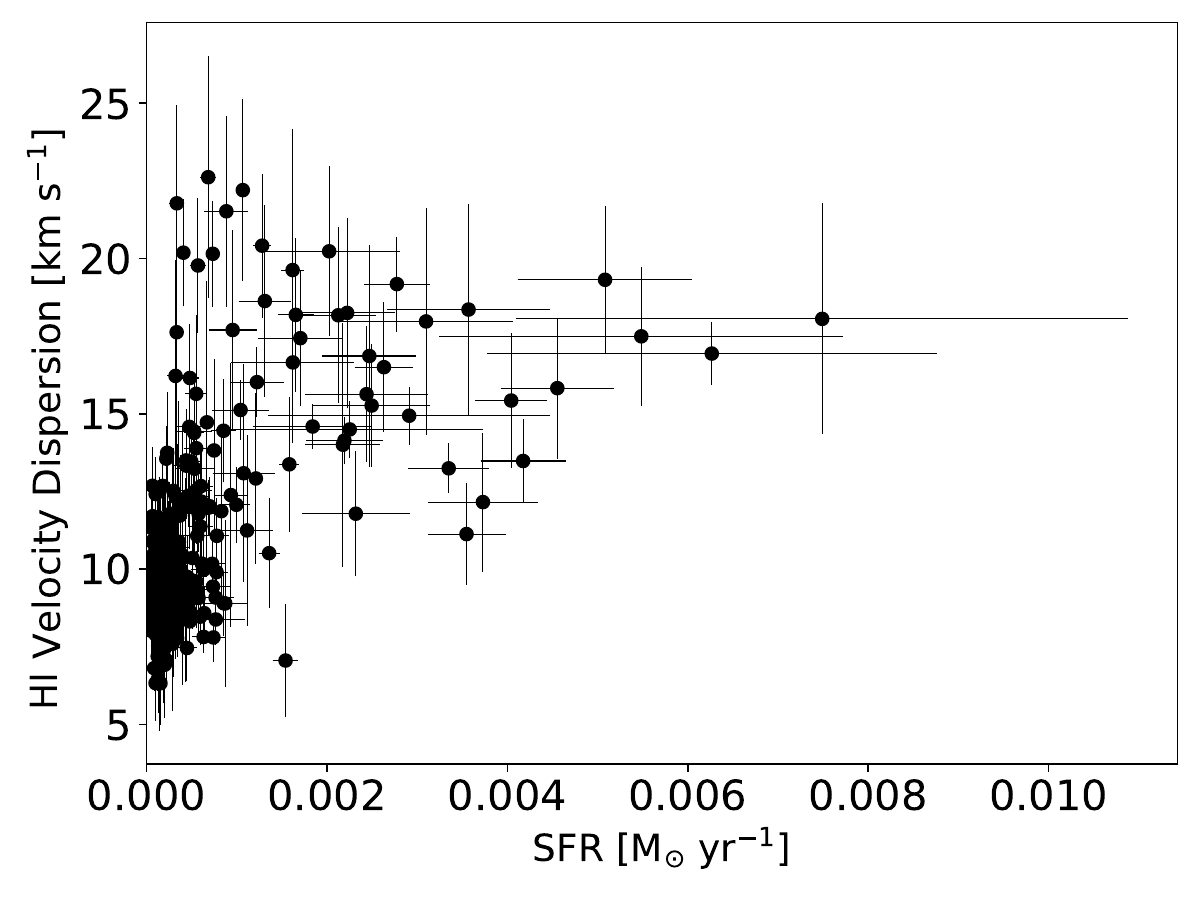}{.45\textwidth}{}}
        \caption{\small \textbf{Comparison of $\sigma_{HI}$ and SFR 500 Myr to 10 Gyr Ago: Full Spatial Coverage Sample and High-Mass Sample.} The \hi velocity dispersion ($\sigma_{HI}$) from the moment maps is plotted against the SFR averaged over 500 Myr to 10 Gyr for the Full Spatial Coverage Sample (Panel A) and for galaxies with $log(M_\star)>8.7$. Error bars on the SFR values represent the 68\% confidence intervals. A prominent clustering of regions is evident at low SFRs and velocity dispersions between 5 and 15 km/s.  This concentration of regions with historically low star formation and low current velocity dispersions is driving the correlation seen in Figure \ref{boot_rs_cen_full}a between the velocity dispersion and star formation in the oldest time bin. }
    \label{sfrvhi_8.7}
\end{figure}

\subsection{Recently Star Forming Regions} \label{blue_anal}

\begin{figure}[!tb]
    \gridline{\fig{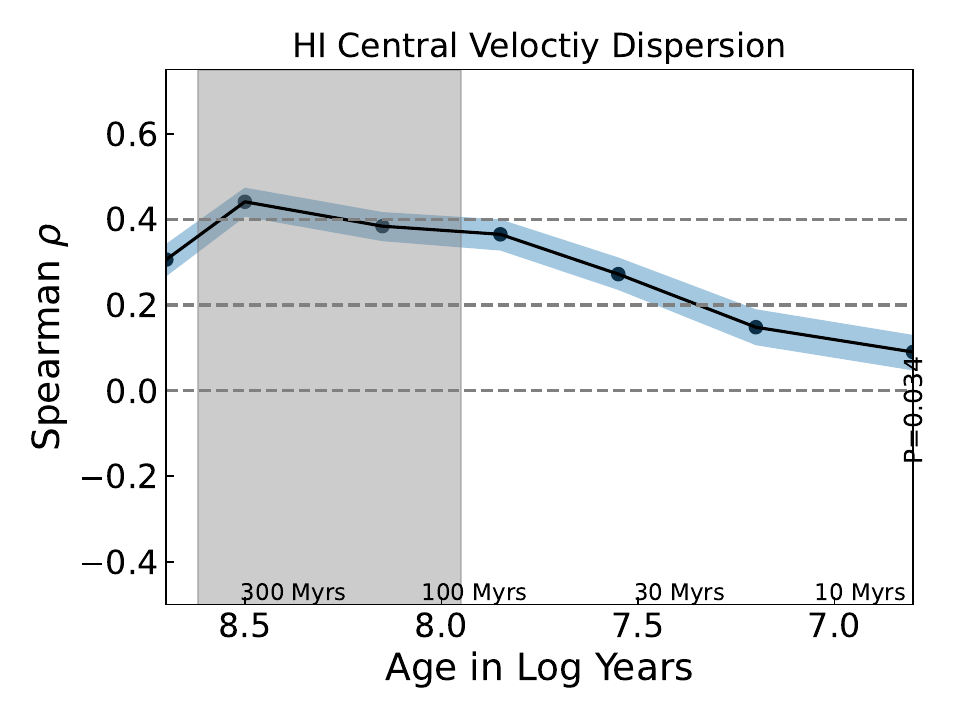}{.45\textwidth}{}}
    \gridline{\fig{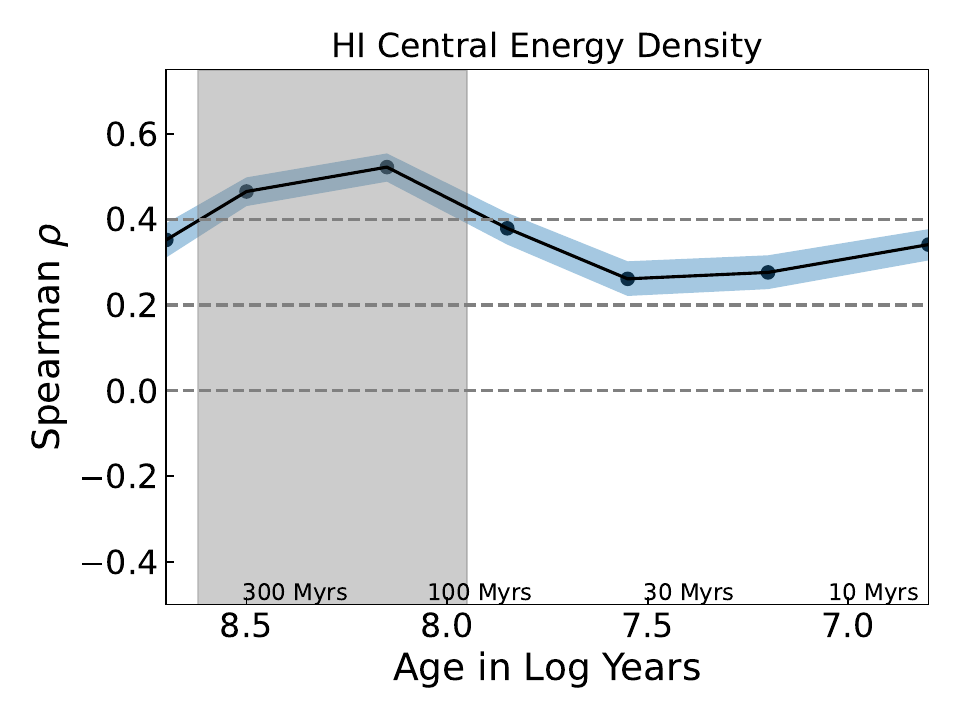}{.45\textwidth}{}}
    \caption{\small \textbf{Comparison of \hi Superprofile Measures and SFH Results: Recent Star Formation Sample.} Comparing \hi turbulence measures and SFH for the regions with indications of recent star formation activity. Spearman $\rho$ coefficient versus log time, indicating the strength of the correlation between the SFR in each time bin and (a) the \hi velocity dispersion ($\sigma_{cen}$), and (b) the \hi energy surface density ($\Sigma_{cen}$), both derived from Gaussian superprofile fits.  This figure only includes regions with evidence of recent star formation activity using the same selection criteria as \citetalias{LCH_2022}. The light blue shaded region represents the 1$\sigma$ bootstrapping error and the corresponding \textit{P}--value is shown beside points with $P\leq0.001$. For the $\sigma_{cen}$ (Panel a), the strongest correlation is with the 200-500 Myr time bin ($\rho=$0.44). For $\Sigma_{cen}$ (Panel b) the strongest correlation is seen at in the second and third to last time bins ($\rho$=0.47 and $\rho=$0.52 respectively), corresponding to 100-500 Myr. The gray region highlights the time range from 100-500 Myr ago.} 
    \label{boot_rs_cen_blue}
\end{figure} 

To test the impact of the young star cuts used in \citetalias{LCH_2022}, the full sample of galaxies was re-analyzed. Following the methods of \citetalias{LCH_2022}, we focused on regions with evidence of recent star formation activity.  As described in \citetalias{LCH_2022}, the initial proof-of-concept study included regions only if they contained at least 50 young stars, ensuring well-constrained recent SFHs. However, this criterion was found to exclude many regions with star formation activity over the past 500 Myr. As a result, in \citetalias{LCH_2023} (Holmberg II) we did not apply this cut. 

In the full sample, this cut leads to the exclusion of regions with low stellar mass, low \hi mass, and low velocity dispersion. For most galaxies, applying the young star criteria removes only $\simeq$10\% of regions and has minimal effect on the results. In Holmberg II, for example, the cut excludes just 13 out of 125 regions. Including the cut in \citetalias{LCH_2023} would have had no impact on the results. However a few galaxies are strongly impacted such as IC~2574. For IC~2574 nearly half its regions are removed, dropping from 236 to 126. For the full sample, 556 of 961 regions include the minimum number of young stars.  

Re-analyzing the sample with the young star requirement removes some regions with poorly constrained recent SFHs and lower signal-to-noise \hi kinematics. As shown in Figure \ref{boot_rs_cen_blue}, the strongest correlation is between the velocity dispersion and the SFR 200-500 Myr ago $\rho=$0.44 and a modest correlation with star formation 50-200 Myr ago ($\rho=$0.365 and 0.381). The more significant correlation  implies the inclusion of the low star-formation regions drove the lack of a correlation seen with the velocity dispersion for the full sample. The correlation between the star formation and \hi energy surface density appears to be a more robust signal which is less impacted by the inclusion of regions with little star formation. 

There remains some correlation ($\rho=$0.31) between the oldest star formation time bin and the \hi velocity dispersion which was not seen in \citetalias{LCH_2022,LCH_2023}. Excluding NGC~3738 and NGC~5253 from this re-analysis reduces the correlation with the oldest bin to $\rho=0.2$, which is our threshold for a weak correlation.  

Notably, the correlation between \hi energy surface density and SFR from 100–500 Myr ago is stronger for the sample with the young star cuts than in the full sample ($\rho=$0.52 for the 100--200~Myr bin and $\rho=$0.47  for the 200--500~Myr bin). In particular, the \hi energy surface density shows a {stronger correlation at the} 100-200 Myr timescale compared to the 200-500 Myr timescale in Figure \ref{boot_rs_cen_blue} Panel b. This preference for the 100-200 Myr timescale aligns precisely with \citetalias{LCH_2022}, which applied the same young star cut.    This suggests that applying the young star cuts modestly impacts the results, shifting the correlation with the \hi energy surface density correlation to a slighter shorter timescale.

\subsection{Sample Division by Mass and sSFR}\label{sec:subsample}

To understand which galactic properties influence the correlation timescale, we analyzed sub-samples of galaxies divided by stellar mass and FUV sSFR. For the mass-based divisions, galaxies were grouped into three bins:
\begin{itemize}
    \item M$_\star \leq 7.7$ log(M$_\odot$)
    \item $7.7 <$ M$_\star < 8.7$ log(M$_\odot$)
    \item M$_\star > 8.7$ log(M$_\odot$)
\end{itemize}
For the sSFR-based division, galaxies were split into two sub-sample:
\begin{itemize}
    \item sSFR $<$ -9.7 log(yr$^{-1}$)
    \item sSFR $>$ -9.7 log(yr$^{-1}$)
\end{itemize}

For the low-mass and medium-mass samples, there is no significant correlation between the \hi velocity dispersion and the star formation activity in any time bin.  The high-mass sample is similar to the full spatial sample with a correlation between the current velocity dispersion and the star formation in the oldest time-bin. As seen in Figure~\ref{sfrvhi_8.7}, the high-mass sub-sample contains both a large number of regions with velocity dispersions below 12 km s$^{-1}$ and historically low SFR below 0.002 M$_\odot$ yr$^{-1}$ and a significant number of regions with high historical SFRs and high current velocity dispersions.  The correlation is much stronger for the high-mass sub-sample as NGC~3738 and NGC~5253 are two of the four galaxies in the sub-sample.  

\begin{figure}
    \gridline{\fig{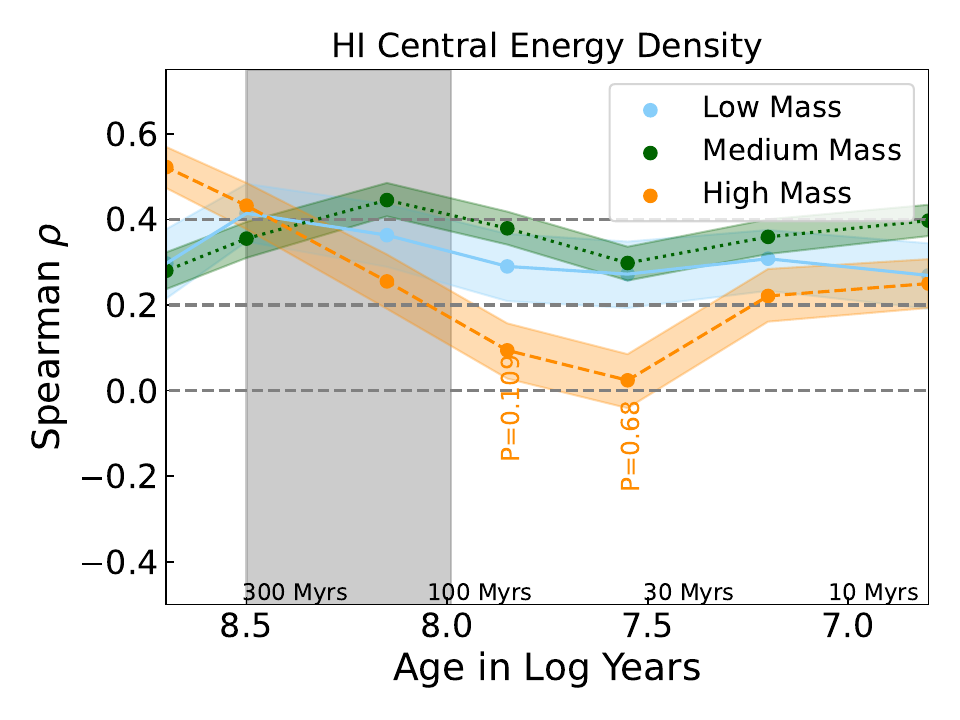}{.45\textwidth}{}}
    \vspace{-1.875cm}
    \gridline{\fig{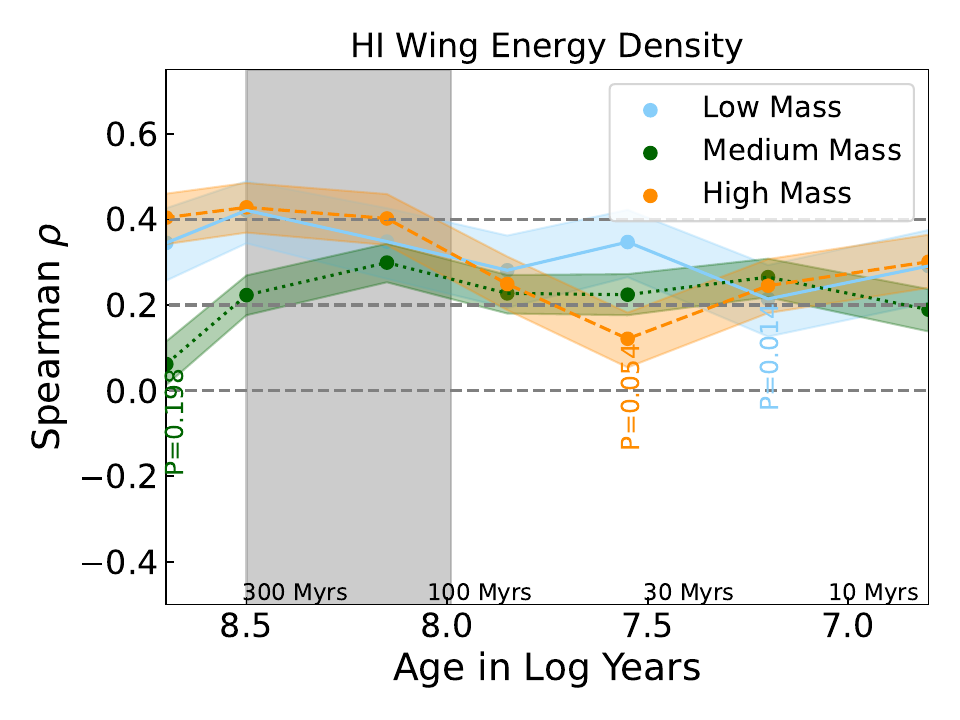}{.45\textwidth}{}}
    \vspace{-1.875cm}
    \gridline{\fig{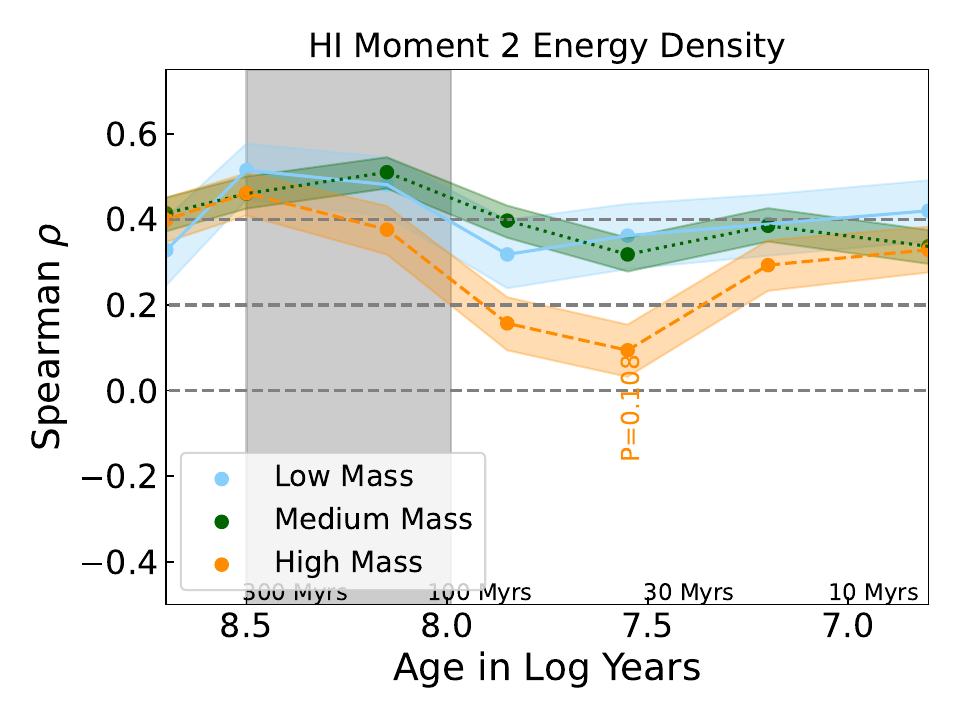}{.45\textwidth}{}}
    \vspace{-5mm}
    
     \caption{\small \textbf{Comparison of $\Sigma_{HI}$ and SFH results for different mass samples.} Comparing \hi turbulence measures and SFH for the galaxies in different mass bins; galaxies with stellar masses $6.7 < $M$_\star< 7.7$ log(M$_\odot$) in blue solid line, galaxies with stellar masses $7.7 < $M$_\star< 8.7$ log(M$_\odot$) in green dotted line, and galaxies with stellar masses M$_\star> 8.7$ log(M$_\odot$) in orange dashed line. Panel (a) shows the correlation with energy surface density from the Gaussian superprofiles, panel (b) with the energy surface density of the superprofile wings, and panel (c) with energy surface density from the second moment maps. For the low mass sample, all three plots show the strongest correlation at 200–500 Myr with $\rho\geq0.4$. For the medium mass sample, the strongest correlation is seen at at the SFR 100-200 Myr ago in panels a and c. For the high mass sample, $\rho\geq0.4$ in the last two time bins (SFR more than 200~Myr ago) in panels b and c.  The time range between 100 and 500~Myr is highlighted in gray.}
    \label{boot_rs_mass}
\end{figure}

Across all three mass sub-samples, the \hi energy surface density shows signs of the 100–500 Myr correlation timescale identified in the full spatial sample.  For the low-mass sub-sample, indications of a specific correlation timescale are the weakest with the $\rho$ ranging between 0.21 and 0.41 across all time bins for both $\Sigma_{HI}$ measured from Gaussian superprofile fit and from the \hi wings. While the correlation coefficient is the highest at 200-500~Myr time bin, the lower bound of the correlation is well below 0.4 with $\rho\simeq$0.34 and accounting for the uncertainties the correlation is nearly flat. The correlation is stronger ($\rho=$0.52) for the $\Sigma_{HI}$ measured from the \hi moment maps (figure \ref{boot_rs_mass}, in blue). For the medium-mass sub-sample, the strongest correlation is at 100-200 Myr (see Figure \ref{boot_rs_mass}, in green ) for $\Sigma_{HI}$ measured from Gaussian superprofile ($\rho=$0.45) and the moment maps ($\rho=$0.51). This is consistent with \citetalias{LCH_2022,LCH_2023}, as all the galaxies (except UGC 9128) analyzed previously fall within this mass range.  For the high-mass sub-sample,  the correlation timescale broadens and shifts to longer timescales.  The strongest correlation timescale is with the last two time bins (figure \ref{boot_rs_mass}, in orange). Similar to the velocity dispersion correlation, the \hi energy surface density correlations are impacted by the inclusion of NGC~3738 and NGC~5253 in the high-mass sample because of the small number of galaxies in this sub-sample. 

For the sSFR sub-samples, The correlation plots in  Figure~\ref{boot_rs_ssfr} are mostly flat. For the low sSFR sub-sample velocity dispersion correlations follow the same trends in  as seen for the full sample (see Figure \ref{boot_rs_cen_full}a). In Figure~\ref{boot_rs_ssfr} (blue solid line) there is a stronger correlation between the SFR and \hi energy surface density at t$\geq$100 Myr compared to the shorter timescales {and there a dip in the correlation at 25-50~Myr ago}. Noticeably, for the low-sSFR sample, like the high-mass sample, there is a strong correlation with the oldest time bin representing the majority of the galaxies' lifetime. Excluding NGC~3738 and NGC~5253 from this sample has a small impact on the \hi energy surface density flattening out the correlation even further.

The more interesting results is the high sSFR sub-sample (sSFR$>$-9.7). This is unsurprising as the impact of stellar feedback on the ISM is expected to be more observable for galaxies with higher sSFRs. In \cite{Tamburro09,Stilp13c}, they found a clear correlation between \hi turbulence measures and star formation activity for galaxies and regions with higher FUV SFR densities. For the high sSFR, the velocity dispersion does not have a clear timescale with the correlation increasing with lookback time. The velocity dispersion appears most related to the star formation 50-500 Myr ago than the current star formation or the star formation at t$>$500 Myr (figure \ref{boot_rs_cen_high_ssfr_velocity}). 

For the high sSFR sample in figure~\ref{boot_rs_ssfr} (orange dashed line), the strongest  correlation for all three \hi energy surface density measures is at 100-200 Myr ago with the second strongest correlation at 200-500 Myr timescale. Panel 3 of figure~\ref{boot_rs_ssfr} shows the largest difference in the correlation strength at 100-500~Myr compared to the other time bins emphasizing the importance of this timescale.  As with the low-sSFR sample, the correlation plot is quite flat with $\rho$ between 0.3 and 0.38 at most time steps and peaks at $\rho\simeq0.5$. The clearest difference between the high-sSFR sample and low-sSFR sample is the high-sSFR show a large drop in the correlation at the oldest time bin and does not have the same dip at 25-50~Myr. 

\begin{figure}
    \gridline{\fig{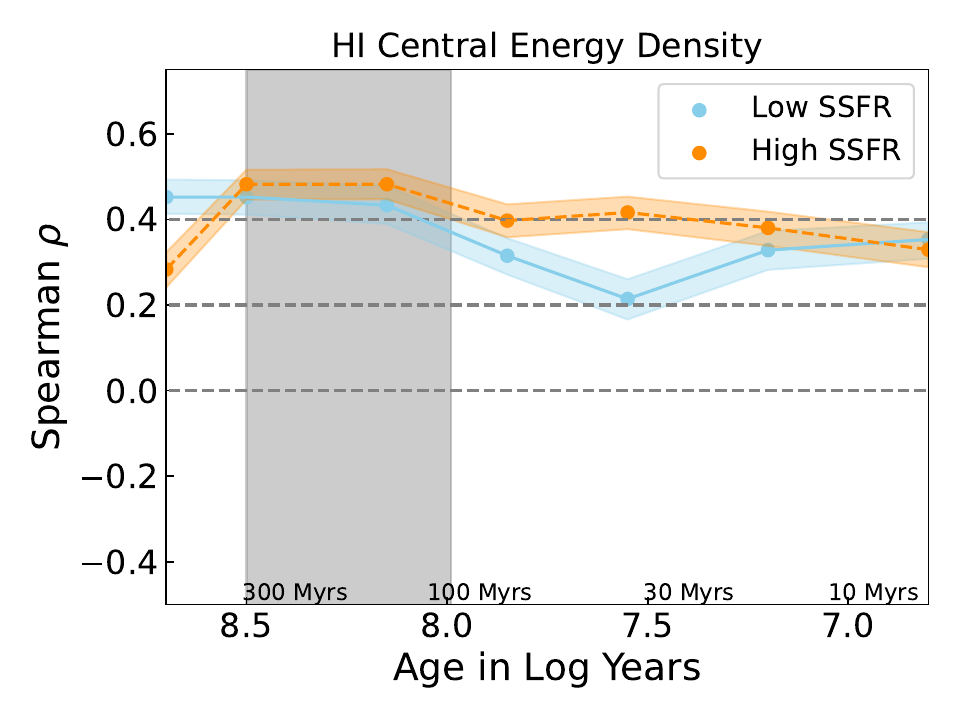}{.45\textwidth}{}}
    \vspace{-1.875cm}
    \gridline{\fig{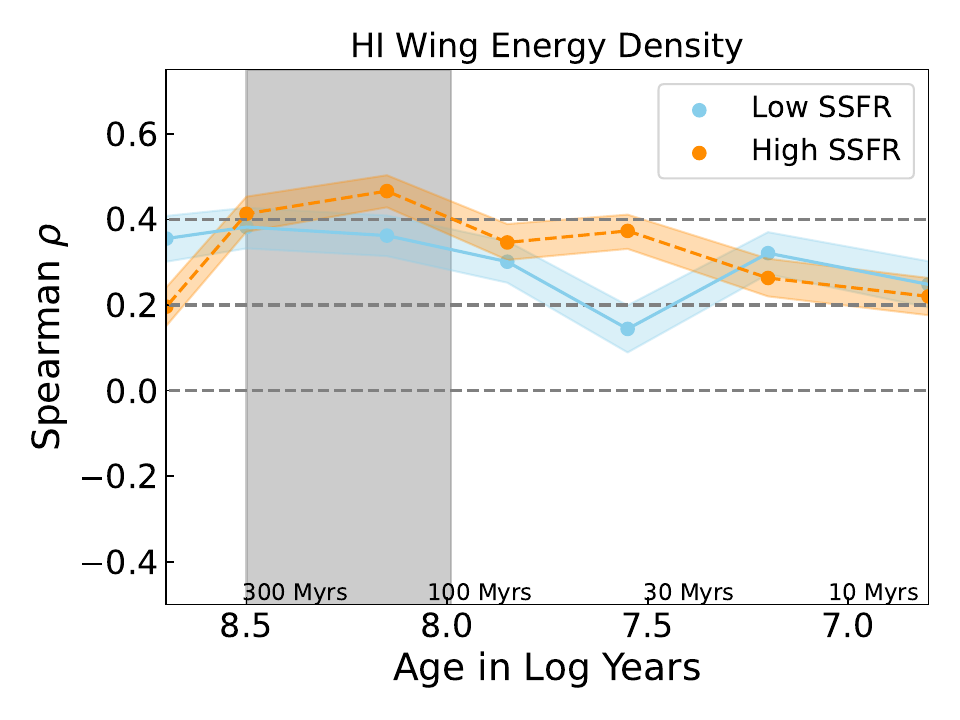}{.45\textwidth}{}}
    \vspace{-1.875cm}
    \gridline{\fig{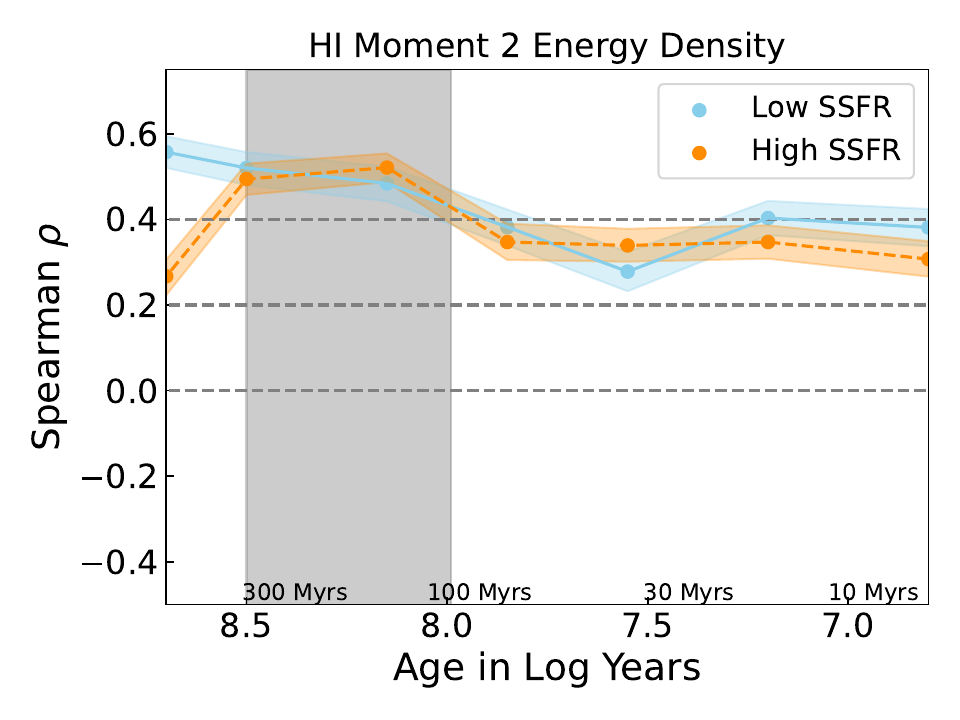}{.45\textwidth}{}} 
    \vspace{-5mm}
     \caption{\small \textbf{Comparison of $\Sigma_{HI}$ and SFH Results: Binned sSFR Galaxies} Comparing \hi turbulence measures and SFH for the galaxies with low specific star formation rates FUV sSFR$< -9.7$ log(yr$^{-1}$, orange dashed line) and galaxies with high sSFR (log(sSFR)$>$-9.7, blue solid line).  a) is the correlation of the SFH with the energy surface density of the Gaussian superprofiles, b ) is the correlation of the SFH with the energy surface density of the wings of the superprofiles, and c) is the correlation of the SFH with the energy surface density measured from the second moment maps. There is no time range with a consistently stronger correlation compared to the other time points for the low sSFR sample, the correlation is slightly stronger for t$\geq$100 as it trends from $\rho\simeq0.35$ to $\rho\simeq0.45$. There is a defined drop in the correlation at t$=$25~Myr. For the high sSFR sample, there is a increase in correlation strength for SFR 100--500 Myr compared to the other time bins especially in panel c. The 100-500 Myr time bins are highlighted by the gray shaded region.}
    \label{boot_rs_ssfr}
\end{figure}

\begin{figure}[!tb]
    \centering
    \includegraphics[width=.48\textwidth]{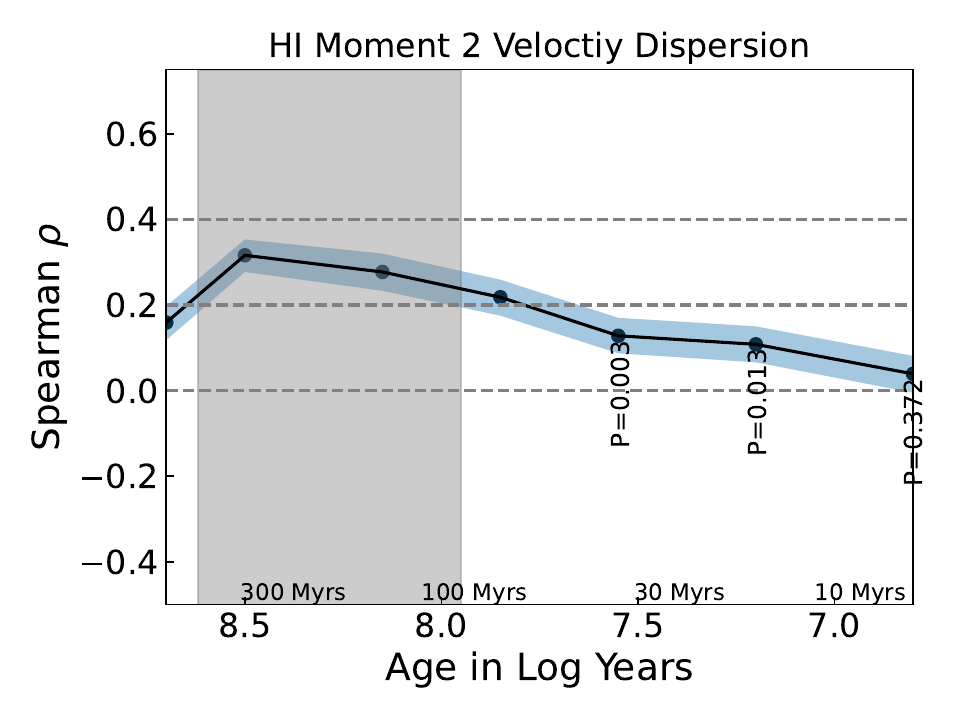}
    \vspace{-5mm}

    \caption{\small \textbf{Comparison of \hi Velocity Dispersion and SFH Results: High sSFR Galaxies} Comparing \hi velocity dispersion derived form the moment maps and SFH for the galaxies with high sSFR (log(sSFR)$>$-9.7). Spearman $\rho$ coefficient and corresponding \textit{P}--value for points where $P\geq0.001$  plotted against log time showing how correlated the SFR of a given time bin is with the \hi turbulence measures. The light blue shaded region represents the 1$\sigma$ bootstrapping error.  For the \hi velocity dispersion, there is the indication of a correlation 50-500 Myr ago as $\rho>0.2$.  The gray region highlights the time range from 100-500 Myr ago.}
    \label{boot_rs_cen_high_ssfr_velocity}
\end{figure}

The correlation timescale results of the sub-samples are in broad agreement with the results from \citetalias{LCH_2022} (100-200~Myr), \citetalias{LCH_2023} (70-140~Myr) and the full galaxy sample (100-500~Myr). All sub-samples demonstrate a correlation in 100-200~Myr time bin, with some indication of a correlation in the adjacent 200-500 Myr time bin.  The medium-mass and high sSFR sub-samples align well with \citetalias{LCH_2022} (see \citetalias{LCH_2022} Figure 16).
Figures \ref{boot_rs_cen_full} through \ref{boot_rs_ssfr} support the picture of stellar feedback impacting the local atomic gas on the timescales of hundreds of millions of years as suggested in \citetalias{LCH_2023}, \cite{Weisz09}, and \cite{Orr2020}.  The variation in the breadth of the correlation for the different sub-samples is likely due to the characteristics of the galaxies in each sub-sample. The differences among the three mass sub-samples may indicate that there is a relationship between galaxy mass and the timescales of stellar feedback (see Section \ref{sec:discussion_ch5} for further discussion). These results demonstrate the range of possible timescale for low-mass galaxies. While the exact timescale likely varies by galaxy, it appears to fall between $\sim$70 Myr and a few hundred Myr. With the exception of Holmberg II, most galaxies in the sample either lack sufficient data or are too highly inclined for a detailed analysis.

\subsection{Equal Region Sampling per Galaxy} \label{sec:equal}

The number of regions per galaxy varies widely because of the two orders of magnitude range in stellar mass and differences in the spatial coverage and depth of the \textit{HST} imaging. Sextans B has only 4 regions, due to its proximity and WFPC2's footprint, while IC 2574 has 235 regions across three ACS pointings. As a results larger galaxies, more distant galaxies, and those with multiple HST pointings contribute more regions potentially weighting the results towards these galaxies.  This uneven sampling is evident in Figures~\ref{06853_frame} and \ref{56663_frame}. To mitigate this effect, we redid the analysis by re-sampling the regions included.

For the re-sampling analysis, the number of regions per galaxy was capped at 10. Every region for galaxies with 10 or fewer regions were included, while for galaxies with 11 or more regions 10 of their regions with a SFH and turbulence measure were randomly selected. This process was repeated for each of the seven turbulence measures (six from \hi and one from H$\alpha$). Regions lacking SparsePak coverage or H$\alpha$ detection were excluded from the H$\alpha$ re-sampling, and regions falling below the \hi cutoffs were excluded from the \hi re-sampling.  The resulting re-sampled datasets include 225 regions for the \hi and 192 for the H$\alpha$, compared to the full samples of 961 and 485 regions, respectively. The re-sampling was repeated 2000 times to well-sample the combinations of regions from the larger galaxies.  

For the ionized gas, the results were identical to the whole sample with the same flat profile seen in Figure \ref{boot_rs_ha_full}.  There is no evidence of a preferred timescale between 5 and 500 Myr for the ionized gas for this data set.  The results presented in this paper, along with the previous lack of clear results, indicates  this analysis is not sensitive to any correlations between the ionized gas turbulence and SFHs. 

For the re-sampled \hi regions, the results for the full sample, low-mass, medium-mass and both sSFR sub-samples are nearly identical to the previous results in Sections~\ref{full_anal} and \ref{sec:subsample}. For the high-mass sub-sample, the correlation with the \hi energy surface density has shifted even more strongly towards t$>$500 Myr as NGC~3738 and NGC~5253 now contribute over half the regions used.  It is clear that the identified correlation timescales are not driven by primarily single large galaxy, such as Holmberg II or IC 2574, but by all the galaxies within the sample and sub-samples. 

\section{Local vs. Global Timescale Results} \label{global_time_c5}

To compare local and global correlation timescales with those found in \citet{Stilp13b}, we analyzed global SFHs and global \hi turbulence for the galaxies in this study.   To account for differences in galaxies sizes, we compared global \hi turbulence measures with global SFR surface densities ($\Sigma_{\mathrm{SFR}}$). The global SFHs were derived using CMD isochrone fitting with MATCH following the methods of Cohen et al in prep.  Each galaxy was divided into four equally-populated elliptical annuli based on structural parameters from 3.6~$\mu$m Spitzer observations. The outermost annulus extended to 4.4 disk scale lengths.  SFHs were computed independently for each ellipse, enabling artificial star tests to better reflect local crowding conditions.  Additionally, the faint magnitude limits for the SFH fitting were evaluated at the 50\% completeness limit for each ellipse, so the less crowded ellipse can make use of fainter photometry. The four SFHs were statistically combined, propagating the per-ellipse uncertainties to construct the global SFH.  A time binning of approximately $\Delta$log(t/yr) = 0.15 was used—-matching the binning applied to Holmberg II in \citetalias{LCH_2023} ---to align with the analysis of \citep{Stilp13b}. The global star formation histories were divided by the area of the galaxy used for the SFH resulting in $\Sigma_{SFR}$. This was not done for the regional SFHs as the regions have equal area.

Global \hi turbulence is measured by masking the \hi data outside the \textit{HST} footprint and outermost annulli used for the CMD fitting, ensuring a close spatial match between the \hi and SFH measurements. As with the regional analysis, the bulk motion-corrected \hi line profiles were stacked to create a global superprofile. This superprofile was fit using the same methods as the individual regions.  Additionally, the \hi mass-weighted average of the 2nd moment map velocity dispersions was determined within the \textit{HST} footprint. The \hi energy surface densities were also computed. 

Figure \ref{global_ssfr} shows the results of the global SFHs and \hi turbulence measures comparison. Due to the much smaller sample size, the uncertainties are larger than in the local analysis. The gray-shaded region in Figure \ref{global_ssfr} highlights the 25–40~Myr time range in which \citet{Stilp13b} identified the strongest correlation between the SFR and \hi turbulence. While the correlation in Figure \ref{global_ssfr} at 25–40~Myr is not the strongest (0.28$\leq\rho\leq$0.58) in our 560~Myr time frame, it is the strongest {correlation} within the past 100 Myr. This is notable, as the analysis by \citet{Stilp13b} only extended to 100~Myr and did not probe longer timescales. 

\begin{figure*}
\gridline{\fig{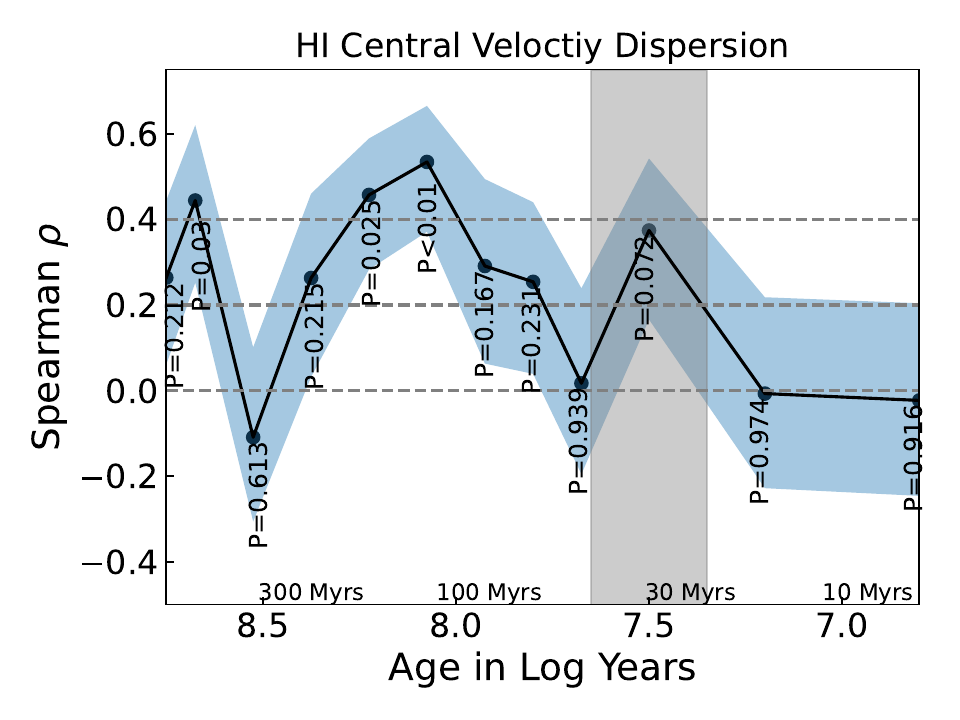}{.425\textwidth}{}
    \fig{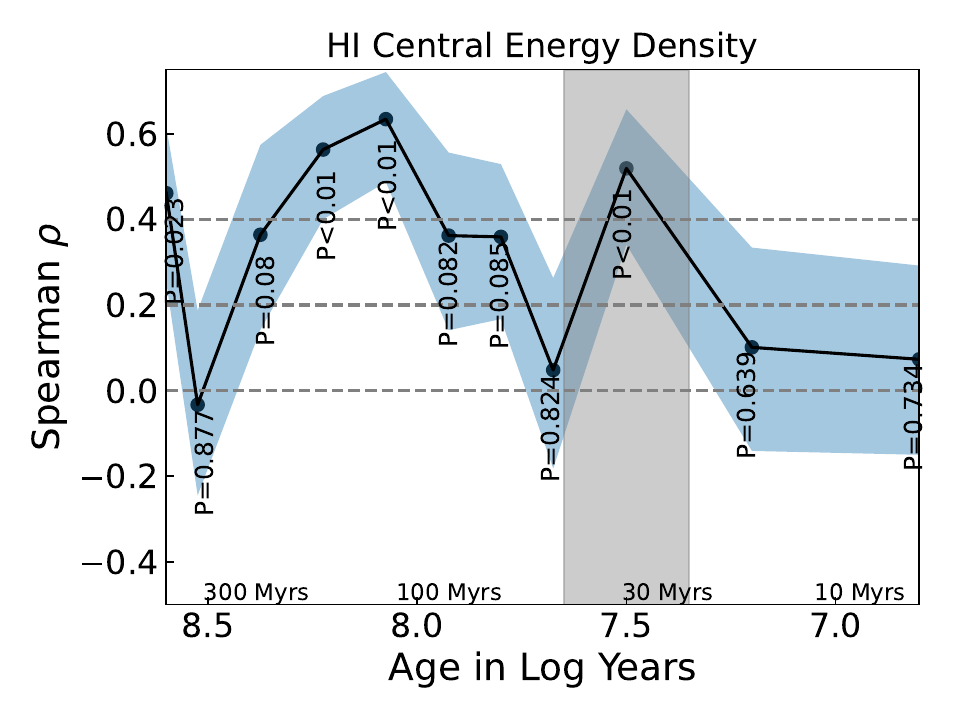}{.425\textwidth}{}}
\gridline{\fig{/vdisp_wing\_global_rs.pdf}{.425\textwidth}{}
    \fig{/Energy_wing\_global_rs.pdf}{.425\textwidth}{}}
\gridline{\fig{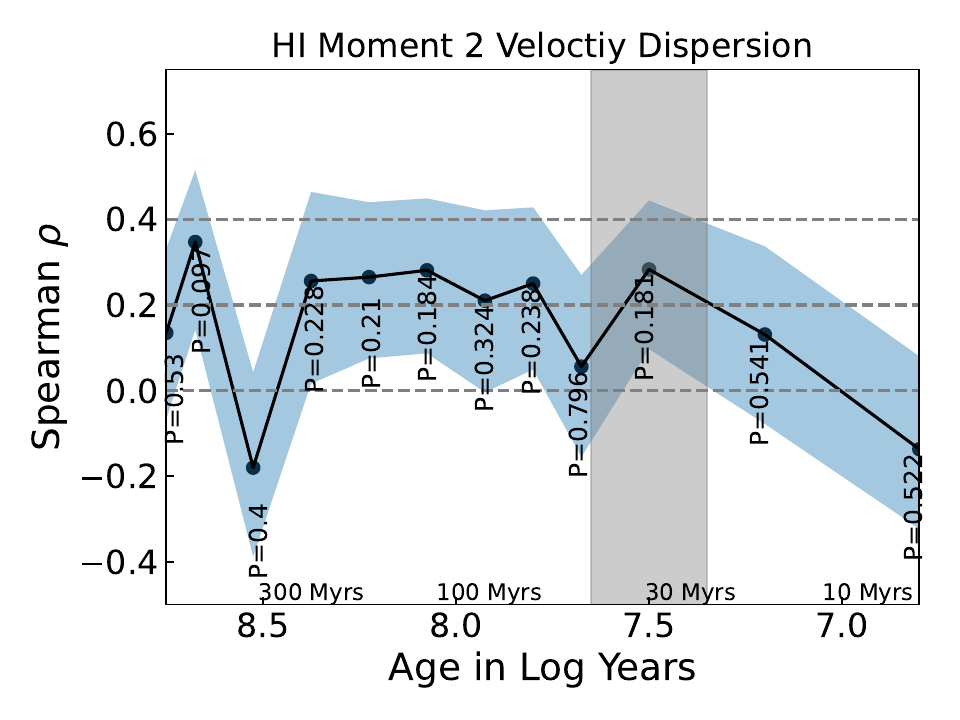}{.425\textwidth}{}
    \fig{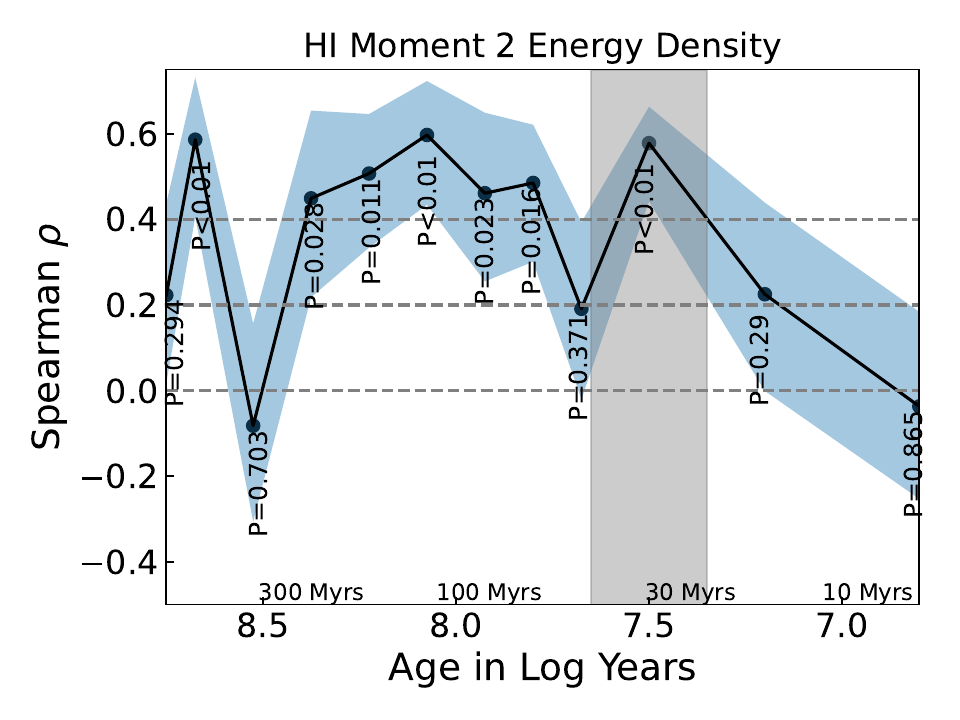}{.425\textwidth}{}}   
    \vspace{-5mm}
    
     \caption{\small \textbf{Comparison of Global \hi Turbulence and SFH Results} The median Spearman $\rho$ coefficient and corresponding \textit{P} value plotted against log time showing how correlated the SFR of a given time bin is with the \hi turbulence measures. The light blue shaded region represents the 25\% and 75\% quarterlies of the range of $\rho$ when re-sampling the regions included for each galaxy.  a) and b) are the correlation of the SFH with the velocity dispersion and energy surface density from the Gaussian superprofiles, c) and d) are the correlation of the SFH with the velocity dispersion and energy surface density from the wings of the superprofiles, and e) and f) are the correlation of the SFH with the velocity dispersion and energy surface density measured from the second moment maps. There is some correlation between the different \hi turbulence measures and the SFR rates at t=25-40 Myr and t$\geq$100 Myr. }
    \label{global_ssfr}
\end{figure*}

A clear correlation with $\rho\geq$0.4 is found between \hi turbulence and star formation  for the t$\simeq$100-300~Myr time range for \hi energy surface density, consistent with the timescales identified in both the regional analysis in this work and in \citetalias{LCH_2022,LCH_2023}.  The strength of this global correlation is comparable to that observed at regional scales but with much larger uncertainties. Unlike the regional results, however, the global correlation is evident in both the velocity dispersion and \hi energy surface density measures --- similar to the trends seen in the high SFR subsample and the initial four galaxies analyzed in \citetalias{LCH_2022}. Because of the higher uncertainties, it is unclear if there is a timescale between 100$\leq$t$\leq$300 Myr at which the correlation is the strongest; instead, the correlation appears broadly constant across this time span.

If the global correlation timescale is related to the dispersion timescales of atomic gas, a broad correlation is expected—since both the \hi turbulence and SFHs are averaged over entire galactic disks.   The inner regions of the disks, with shorter scale lengths, have shorter dissipation timescale on the order of tens of Myr, while the flared outer regions exhibit longer timescales, up to several hundred Myr \citep{Bacchini20b}. This spatial variation explains the shorter correlation timescale reported by \citet{Stilp13b} compared to the analysis here and in \citetalias{LCH_2022,LCH_2023}. \citet{Stilp13b} analysis focused on the galaxies' inner disk regions due to their \hi column density limits and available HST coverage. These regions of the disk are likely to have shorter dissipation timescales than what was found here in our regional analysis. The 30-40~Myr peak identified in \citet{Stilp13b}, agrees with the modest 25-40~Myr increase in the correlation strength seen in Figure \ref{global_ssfr}.  As discussed in \citet{Stilp13b}, this timescale aligns with the lifetimes of the lowest-mass supernova progenitors, suggesting a potential physical connection between star formation and turbulence injection mechanisms on this timescale.

\section{Discussion: What Affects Local Turbulence Timescales? } \label{sec:discussion_ch5}

Figures \ref{boot_rs_cen_full} through \ref{boot_rs_ssfr} consistently point to the 100–500 Myr time range as a characteristic timescale for turbulence on the 400 pc scale. The results presented in the previous section indicate that present-day \hi turbulence is strongly influenced by the star formation activity several hundred Myr ago. This is consistent with previous works, though the timescales identified here are longer and broader.  This supports the scenario in which energy is injected into the atomic gas by multiple star formation events and then decays slowly.  The breadth of the timescales seen in Figure \ref{boot_rs_cen_full} is driven by the diverse characteristics of the galaxies in the full sample.   

\subsection{sSFR and Mass Effects}

From the analysis in Section~\ref{sec:subsample}, it appears that galaxies with higher sSFRs (FUV sSFR$\geq$-9.7 log(yr$^{-1}$)) exhibit shorter correlation timescales, while those with lower sSFRs trend toward longer correlation timescales.  Notably, the high-sSFR sub-sample is the only sub-sample that shows any indication of a correlation between the \hi velocity dispersion and the SFR.  The effects of the mass and sSFR appear separable: the mass-selected samples contain galaxies spanning the full range of sSRF, and vice versa. However, the influence of sSFR on the observed trends appears stronger than that of stellar mass. The differences between the high- and low-sSFR sub-samples are more pronounced than those between the mass sub-samples.

As discussed in the introduction, previous observational studies have compared \hi turbulence measures to H$\alpha$ and UV SFRs (e.g., \citealt{vanzee99,Tamburro09,Stilp13c,Hunter21,Elmegreen22}).  In the context of the long correlation timescales found here, the lack of correlation between the \hi turbulence and recent SFRs (averaged over the past 10-100 Myr) in past work is understandable (\citealt{vanzee99,Hunter21,Elmegreen22}). Our analysis reveals that the correlation between the \hi energy density and star formation rate in the past 100 Myr is notably weaker than the correlation with the star formation rate 100-500 Myr ago. 

Previously, \cite{Tamburro09} and \cite{Stilp13c} reported a correlation between the \hi turbulence and SFRs at high SFR surface densities derived from FUV and 24 micron imaging. Notably, \cite{Stilp13c} utilized similar methods to construct \hi superprofiles from stacked \hi line profiles. In the present work, only the high sSFR sub-sample demonstrates a correlation between the \hi velocity dispersion and star formation less than 100 Myr ago.  Figure \ref{boot_rs_cen_high_ssfr_velocity}a shows a weak indication of a correlation between superprofile-derived \hi velocity dispersion and the star formation 25-100 Myr ago.  The correlation between the \hi velocity dispersion and SFR for this sub-sample extends from 25-500 Myr and contributes to the shorter correlation timescale seen for the high-sSFR sub-sample compared to the low-sSFR sub-sample.

The longer correlation timescales seen for the high-mass sub-sample, relative to the medium-mass sub-sample, is unexpected. Based on the mass–metallicity relation \citep{Berg12}, higher-mass galaxies are expected to have higher metallicities, which should lead to more efficient cooling and, consequently, shorter turbulence dissipation timescales. However, the high-mass sub-sample includes NGC~5253 and NGC~3738 which have high velocity dispersions and many regions with historically high SFRs as well as a significant number of low velocity dispersion regions located in the outer parts of IC~2574 (similar to the full spatial coverage sample, see Figure \ref{sfrvhi_8.7}).  In the context of the flared disks observed for dwarf galaxies (e.g., \citealt{Banerjee11,Bacchini19}), these regions may have longer dissipation timescales compared to the inner regions (see Section \ref{driving} for further discussion). That said, the low- and medium-mass sub-samples also include a similar fraction of regions located in galaxy outskirts.  However, when the analysis is re-done so that each galaxy contributes equally, the trend toward longer correlation timescales for the high-mass sub-sample becomes more pronounced as the influence of NGC~5253 and NGC~3738 becomes stronger. There effect becomes more significant when the  number of regions per galaxy is equal. 

In contrast, the slightly longer correlation timescale observed for the low-mass sample compared the medium-mass sub-sample agrees with expectations.  The lower-mass galaxies have lower velocity dispersions on average (avg($\sigma_{mom2}=9.2$ for the low-mass sub-sample versus avg($\sigma_{mom2}=10.0$ for the medium-mass sub-sample).  As explained in Section~\ref{driving}, the dissipation timescale correlates inversely with the velocity dispersion. Thus, the lower-mass galaxies may be expected to have slightly longer correlation timescales. Additionally, the lower metallicities of the lower-mass galaxies likely result in less efficient cooling and therefore longer turbulence dissipation timescales. 

\subsection{Driving Scale and Dissipation Timescale} \label{driving}

As discussed in \cite{Bacchini20a}, the turbulence dissipation timescale ($\tau_d$) is fundamentally linked to the driving scale of turbulence ($L_D$), which represents the characteristic scale at which energy is injected into the ISM. The dissipation timescale can be expressed as:
\begin{equation}
    \tau_d=\frac{E_{turb}}{\dot{E}_{turb}}=\frac{L_D}{v_{rms}}
\end{equation}
Turbulence is envisioned as “eddies” that develop at a variety of different spatial scales. The driving scale ($L_D$) corresponds to the scale at which turbulent energy is injected at and the largest of these eddies form. These large-scale motions subsequently cascade to smaller scales, where the energy is eventually dissipated. As such, $L_D$ plays a critical role in setting the overall timescale over which turbulent energy in the ISM is lost. 

Estimates of $L_D$ for \hi turbulence vary widely across observations and simulations. For dwarf galaxies, $L_D$ have been inferred to range from approximately twice the disk scale height -- roughly 200–600 pc, based on measurements by \cite{Patra20,Bacchini20b}, and \cite{Pina2025} -- to as large as 2.3 kpc \citep{Chepurnov15}. \cite{Chepurnov15} determined a 2.3~kpc turbulence driving scale from the velocity power spectrum of the SMC. Even larger $L_D$, up to 6 kpc, have been predicted in simulations of dwarf galaxies \citep{Dib05}. The 400~pc region size used in this analysis is at lower end of the range of predicted driving scale for dwarf galaxies.  As discussed in \citetalias{LCH_2023}, 400~pc regions appear to be more sensitive to the local correlation timescale than larger size regions which fall within the range of predicted driving scales. Furthermore, the 400~pc scale is similar in size to the size of large \hi holes observed in nearby galaxies (100$-$1000~pc e.g., \citealt{Kamphuis91, Puche92, Boomsma08, Pokhrel20}).   

\cite{Bacchini19, Bacchini20b} measured the disk scale heights of several galaxies in our sample. For Holmberg II, analyzed in \citetalias{LCH_2023}, \cite{Bacchini20b}'s measured a disk thickness ranging from 200 to 600 pc, with an average of $\sim$400~pc. Assuming $L_D=$800~pc (twice the scale height), we can estimate the turbulence dissipation timescale using Equation 4 from \cite{Bacchini20a}, based on the formulation in \cite{Maclow99}:
\begin{equation}
    \tau_d=\frac{L_D}{v_{rms}}=(10\text{Myr})\Big(\frac{L_D}{\text{100pc}}\Big)\Big(\frac{v_{rms}}{10\text{km s$^{-1}$}}\Big)^{-1}
\end{equation}
Using the median $\sigma_{m2}$ value from Table~\ref{table:turbulence_c5} (10 km s$^{-1}$), this yields a dissipation timescale of approximately 80 Myr. This estimate is in excellent agreement with the measured correlation timescale of 70–140~Myr reported in \citetalias{LCH_2023}.

Combining the disk scale heights from \cite{Bacchini19,Bacchini20b} with the velocity dispersions measured in this study yields dissipation timescales in the range of 60–225 Myr for the galaxies in both samples. This range is somewhat shorter than the 100–500 Myr correlation timescale found for the full sample. However, all the overlapping galaxies are in the high sSFR sub-sample, for which the measured correlation timescale is 100–200 Myr. The predicted 60-225~Myr timescale is in excellent agreement with the observed 100-200 Myr correlation timescale. Modeled low-mass galaxies have a wide range disk thickness (150-3000~pc) and velocity dispersions ($\sim$5-20~km s$^{-1}$).  These values correspond to dissipation timescales ranging from roughly 30 to 1000 Myr.  This large range of physically motivated dissipation timescales aligns with the broad 100-500 Myr timescale seen in this paper.

If the atomic gas scale height is correlated with the turbulence driving scale, then the broad range of correlation timescales observed in this study is likely a consequence of variations in galaxy geometry. Specifically, thinner disk galaxies—or the thinner inner regions of disks—would be expected to have shorter dissipation (and thus correlation) timescales. However, the current analysis combines both inner and outer disk regions, averaging over a range of physical parameters. This contributes to the broad range of correlation timescales observed. A more targeted analysis of radial trends in individual, face-on disk galaxies would better isolate the impact of disk thickness on the dissipation scale.  Such an approach would also help minimize confounding variables such as stellar mass, global star formation history, and environmental factors. Holmberg II is not ideal for such a radial study as the \hi scale height is $\simeq$400 pc for much of the disk \citep{Bacchini20b, Patra20}. A galaxy with a steeper disk thickness gradient would be ideal.

\subsection{Supernovae Energy Injection}

Long dissipation timescales allow low efficiencies of energy injection by stellar feedback (such as by SNe) to drive the observed turbulence.  If the correlation timescale we measure is indeed closely tied to the turbulence dissipation timescale, then SNe could maintain \hi turbulence with energy transfer efficiencies as low as a few percent \citep{Bacchini20a}. Such low efficiencies are consistent with simulation predictions which estimate that only a few to ten percent of SN energy is converted into ISM kinetic energy (e.g., \citealt{Thornton98, Dib05, Martizzi16, Fierlinger16, Chamandy20}). In contrast, observational work has found that the observed \hi kinetic energy require efficiencies $\geq$100\% as they assume dissipation timescales at least an order of magnitude shorter than the timescale found here (e.g., \citealt{Tamburro09, Stilp13c, Utomo19}). The long correlation timescale found here supports SNe and stellar feedback as primary drivers of atomic gas turbulence. With long dissipation timescales, the energy SNe input, is sufficient to maintain the observed atomic gas kinetic energy with low efficiencies.

\subsection{Star Formation Efficiencies}

It is well established that galaxy formation models lacking stellar feedback—or incorporating overly efficient gas cooling—tend to overproduce stars, rapidly depleting the available gas reservoir (e.g., \citealt{Katz96, Keres09, Krumholz11}). In large-scale cosmological simulations such as EAGLE \citep{EAGLE} and IllustrisTNG \citep{Illustris}, the spatial and mass resolution is insufficient to resolve the detailed processes by which individual supernovae (SNe) inject energy into the interstellar medium (ISM). As a result, these simulations rely on sub-grid models to implement stellar feedback, injecting thermal and kinetic energy into the surrounding gas.  These simulations must calibrate their gas cooling parameters to match observed ISM and star formation efficiency observations. 

As the turbulence dissipation timescale is related to the gas cooling timescale, it can be considered a proxy of gas cooling efficiency.  The long dissipation timescale found here implies that the atomic gas of dwarf galaxies cools slowly. If $L_D$ correlates to disk thickness, then thicker gas disks would have longer dissipation timescales, resulting in slower cooling and reduced star formation efficiency in those regions. Dwarf galaxies in particular are expected to have thick gas disks which flare on the outskirts due to their shallow gravitational potentials (e.g., \citealt{Roychowdhury10, Banerjee11, Iorio17, Pina2025} ). This should lower their star formation efficiencies compared to spiral galaxies. Indeed, \cite{Leroy08} found that in spiral galaxies, star formation efficiency decreases with radius, and dwarf galaxies have lower efficiencies compared to spiral galaxies.  This implies that longer dissipation timescales, driven by disk structure, may play a role in decreasing star formation efficiency. As noted previously, a radial analysis of a face-on star-forming galaxy would be informative in determining if the correlation timescale is related to star formation efficiencies.

\subsection{Complications from Multiple Driving Scales}

In \cite{Kolmogorov41} turbulence model, energy is injected at large spatial scales and cascades down to smaller scales, where it dissipates.  The 400~pc scale of this study is well above the dissipation scale for atomic gas.  \cite{Chepurnov15} noted that at a resolution of 30~pc, the dissipation scale in the SMC remained unresolved. This suggests that 400~pc lies within the inertial range—between the driving and dissipation scales—where turbulence is fully developed, as described in \cite{Kolmogorov41} framework (e.g., \citealt{Lesieur08}). 

However, as noted in \cite{Elmegreen04}, this explanation of turbulent energy cascade may not hold for the ISM.  For the ISM, energy maybe injected on multiple physical scales: from galactic rotation and shear ($\sim$10~kpc) to winds from individual stars ($\sim$10~pc).  These multiple possible drivers would remove the key concept of a direct cascade from larger to smaller scales. As \cite{Elmegreen04} discusses, energy in the ISM may cascade in either direction or both simultaneously. The determination of the correlation timescale at significantly smaller resolutions and on global scales is key to understanding how energy propagates between scales in the ISM.  Additionally, analysis of the molecular gas may be a crucial to understanding the cascade of energy and turbulence in the ISM as the molecular gas must dissipate energy to collapse and cool to form stars.

\section{Conclusions} \label{conclusions_c5}

This paper presents an analysis of 26 low-mass galaxies, using the methods established in \citetalias{LCH_2022,LCH_2023} on both 400 pc and global spatial scales. The study utilizes a multi-wavelength datasets from \textit{HST}, VLA, and SparsePak (WIYN 3.5m). We compare time-resolved star formation histories to local \hi energy surface density ($\Sigma_{HI}$) and velocity dispersion ($\sigma_{HI}$), measured from Gaussian superprofiles and second moment maps assessing correlations using Spearman's rank correlation coefficient. Additionally, we examine the $H\alpha$ velocity dispersion.

From the ionized gas,  there is no evidence of a correlation timescale.  The possible correlation between the ionized gas velocity dispersion and the SFR 10-25 Myr ago seen in \citetalias{LCH_2022,LCH_2023} work is not evident in the larger sample.  It appears in \citetalias{LCH_2022,LCH_2023} the possible correlation was driven by a small number of outliers.  The lack of a timescale is unsurprising, as this analysis was not sensitive to the shortest timescale ($\leq$5 Myr).  Previous work has indicated that the ionized gas turbulence measured from H$\alpha$ line widths is correlated with the H$\alpha$ derived star formation rate (e.g., \citealt{Moiseev15,Zhou17,Law22}) which is sensitive to star formation on a timescale shorter than 10 Myr. %For future analysis, a sample of all the regions that are well covered by SparsePak will be considered.  

For the atomic gas of the full sample of galaxies, the clearest correlation timescale is between the SF 100-500 Myr ago and the current \hi energy surface density. The combination of galaxies, with a wide mix of diverse properties, appears to broaden the correlation timescale and impact the correlation strength.  The 100-500 Myr timescale found here is in good agreement with the 100-200 Myr timescale found in \citetalias{LCH_2022}, and the 70-140 Myr timescale found in \citetalias{LCH_2023}.

Considering galaxy sample with varying masses and sSFRs, the correlation timescale shifts within the 100–500~Myr range, depending on the sample characteristics. The high sSFR sub-sample is the only one to show a correlation between the past star formation activity and the current velocity dispersion and \hi energy surface density. Further, the 100–500~Myr correlation timescale identified remains significant after correcting for a few galaxies dominating the total number of regions, indicating it is driven by contributions from the full sample. 

Additionally, we analyzed the global correlation timescale  of the sample. While no single definitive timescale was identified, there is evidence supporting the 30–40~Myr timescale found by \cite{Stilp13b} when considering star formation activity over the past 100~Myr.  We further find a correlation between star formation over the past 100–300~Myr and current \hi turbulence measures. 

Taken together, the results support a picture in which stellar feedback influences the local atomic gas over a few hundred million years. The observed correlation timescale appears to be driven by the dissipation timescales of dwarf galaxies.  The broad 100–500~Myr range we identified likely reflects the diversity of the galaxies within the sample. Future analyses of local correlation timescales should focus on individual galaxies, as combining multiple systems appears to broaden the correlation timescale signal. While this study identifies a range of plausible correlation timescales, it does not pinpoint the specific galactic characteristics driving the differences in correlation timescale. By comparing the timescales of individual galaxies with a variety of properties, such as mass, sSFR, and environment, future work will can explore what causes the range in local correlation timescales. Such insights are essential for advancing our understanding of how turbulence and feedback shape galaxy evolution.

\begin{acknowledgments}     
     This work is published in memorial of Prof. Liese van Zee. 
     
     The authors thank the anonymous reviewer for their valuable insights which improved the quality of this paper.
     
     This work is financially supported through NSF Grant Nos. AST-1806522 and AST-1940800. Any opinions, findings, and conclusions or recommendations expressed in this material are those of the authors and do not necessarily reflect the views of the National Science Foundation. Based on observations with the NASA/ESA Hubble Space Telescope obtained from the Data Archive at the Space Telescope Science Institute, which is operated by the Associations of Universities for Research in Astronomy, Incorporated, under NASA contract NAS5-26555. These observations are associated with program No. 10605. Support for program  HST-AR-16144 was provided by NASA through a grant from the Space Telescope Science Institute, which is operated by the Associations of Universities for Research in Astronomy, Incorporated, under NASA contract NAS5- 26555. Work in this paper was partially supported by NSF REU grant PHY-2150234. Additional support from this work comes from the Indiana Space Grant Consortium Fellowship program and the Indiana University College of Arts and Sciences. The authors acknowledge the observational and technical support from the National Radio Astronomy Observatory (NRAO), and from Kitt Peak National Observatory (KPNO). Observations reported here were obtained with WIYN 3.5 telescope which is a joint partnership of the NSF's National Optical-Infrared Astronomy Research Laboratory, Indiana University, the University of Wisconsin-Madison, Pennsylvania State University, the University of Missouri, the University of California-Irvine, and Purdue University.  This research made use of the NASA Astrophysics Data System Bibliographic Services and the NASA/IPAC Extragalactic Database (NED), which is operated by the Jet Propulsion Laboratory, California Institute of Technology, under contract with the National Aeronautics and Space Administration.

    \textit{Facilities:} Hubble Space Telescope; the Very Large Array; the WIYN Observatory\\
    \textit{Software:} Astropy \citep{astropy13,astropy18}; GIPSY \citep{GIPSY}; Peak Analysis \citep{PAN}; IRAF \citep{IRAF86,IRAF93}; 
\end{acknowledgments}

\section{Appendix: Galaxy Maps}\label{appendix_maps}

The complete set of \hi and H$\alpha$ maps for all the galaxies in the original sample.  The atomic and ionized gas maps for Holmberg II, NGC 4068, NGC 4163, NGC 6789, and UGC 9128 were previously presented in \citetalias{LCH_2022,LCH_2023}. The bottom row of each figure includes the 400$\times$400~pc regions used for our analysis. 

\figsetstart
\figsetnum{13}
\figsettitle{Ionized and Neutral Gas Moment Maps}

\figsetgrpstart
\figsetgrpnum{13.1}
\figsetgrptitle{Maps for UGC0685}
\figsetplot{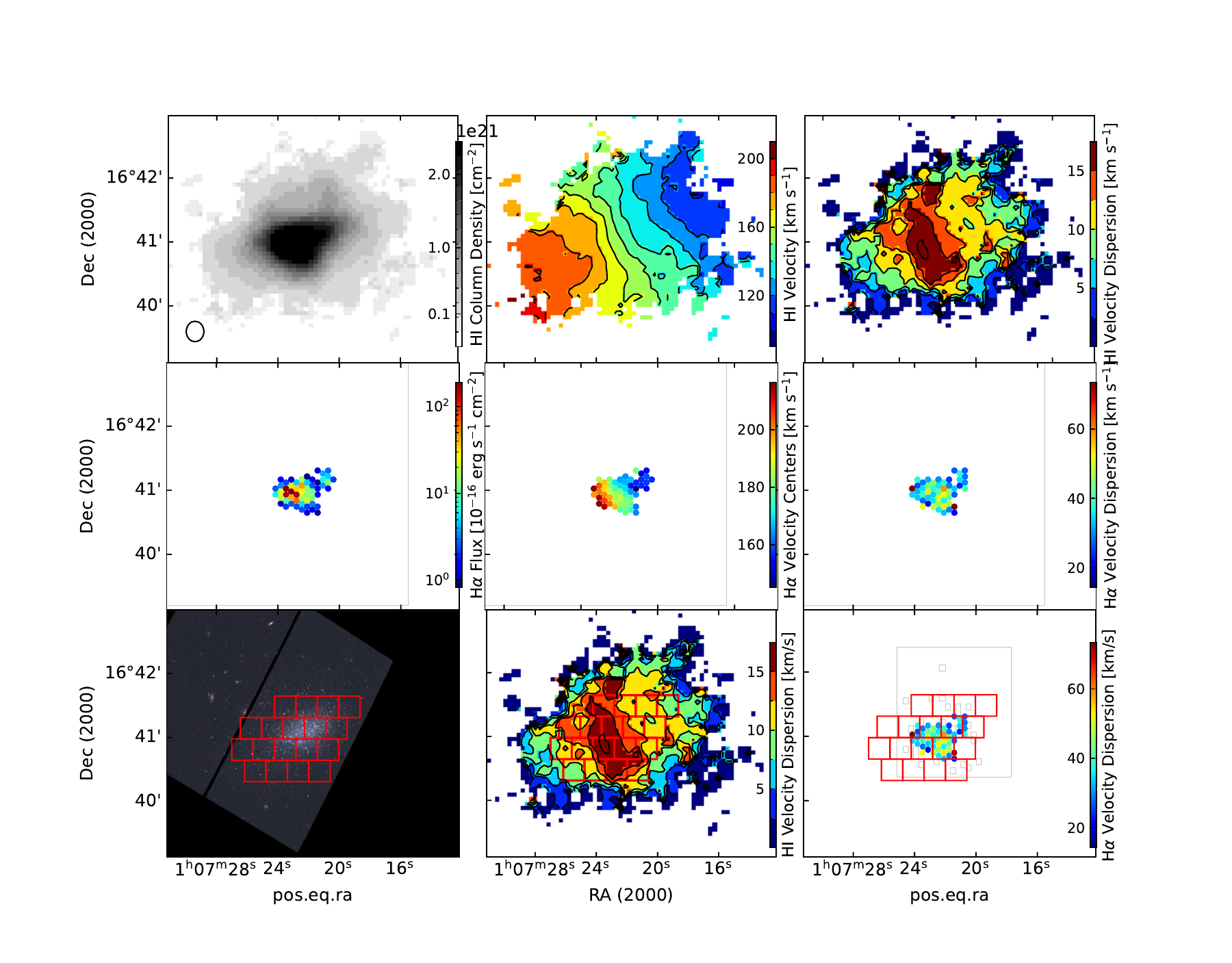}
\figsetgrpnote{\textbf{UGC 0685 Maps Top Row}  UGC 0685 \hi moment maps from VLA observations. Left: \hi column density in 10$^{21}$ hydrogen atoms cm$^{-2}$, Center: \hi velocity map with isovelocity contours spaced every 10 km s$^{-1}$, Right: \hi velocity dispersion map with isovelocity contours at 2.5 km s$^{-1}$ spacing. The beam size (19.25''$\times$16.53'') of the \hi data cube used is shown in the bottom left of the left panel. \textbf{Middle Row} UGC 0685 maps from observations with the SparsePak IFU on the WIYN 3.5m telescope, with H$\alpha$ line measurements from PAN.  Left: H$\alpha$ line flux on a log scale in units of 10$^{-16}$ erg s$^{-1}$ cm$^{-1}$, Center: H$\alpha$ line centers map, Right: H$\alpha$ velocity dispersion ($\sigma_{H\alpha}$) map. Each filled circle corresponds to a fiber's size and position on the sky. \textbf{Bottom Row} Left: Thre color image from \textit{HST} F814W (red), average of F814W and F606W (green), and F606W (blue) observations with ACS, Center: \hi dispersion map from VLA observations with isovelocity contours in 2.5 km s$^{-1}$ step size, Right: $\sigma_{H\alpha}$ map from the SparsePak IFU on the WIYN 3.5m telescope. Overlaid on all three panels are the outlines of the regions used for the analysis. }
\figsetgrpend

\figsetgrpstart
\figsetgrpnum{13.2}
\figsetgrptitle{Maps of NGC0784}
\figsetplot{f16_2.png}
\figsetgrpnote{\textbf{NGC 0784 Top Row} NGC0784 \hi moment maps from VLA observations. Left: \hi column density in 10$^{21}$ hydrogen atoms cm$^{-2}$, Center: \hi velocity map with isovelocity contours spaced every 10 km s$^{-1}$, Right: \hi velocity dispersion map with isovelocity contours at 2.5 km s$^{-1}$ spacing. The beam size (22.20''$\times$17.62'') of the \hi data cube used is shown in the bottom left of the left panel.  \textbf{ Middle Row} NGC0784 maps from observations with the SparsePak IFU on the WIYN 3.5m telescope, with H$\alpha$ line measurements from PAN.  Left: H$\alpha$ line flux on a log scale in units of 10$^{-16}$ erg s$^{-1}$ cm$^{-1}$, Center: H$\alpha$ line centers map, Right: H$\alpha$ velocity dispersion ($\sigma_{H\alpha}$) map. Each filled circle corresponds to a fiber's size and position on the sky.  \textbf{Bottom Row} Left: Three color image from \textit{HST} F814W (red), average of F814W and F606W (green), and F606W (blue) observations with ACS, Center: \hi dispersion map from VLA observations with isovelocity contours in 2.5 km s$^{-1}$ step size, Right: $\sigma_{H\alpha}$ map from the SparsePak IFU on the WIYN 3.5m telescope. Overlaid on all three panels are the outlines of the regions used for the analysis.  }
\figsetgrpend

\figsetgrpstart
\figsetgrpnum{13.3}
\figsetgrptitle{Maps of NGC2366}
\figsetplot{f16_3.png}
\figsetgrpnote{ \textbf{NGC 2366 Top Row} \hi moment maps from VLA observations Left: \hi column density in 10$^{21}$ hydrogen atoms cm$^{-2}$, Center: \hi velocity map with isovelocity contours spaces every 10 km s$^{-1}$, Right: \hi velocity dispersion map with isovelocity contours at 2.5 km s$^{-1}$ spacing. The beam size (22.89''$\times$21.25'') of the \hi data cube used is shown in the bottom left of the left panel. \textbf{Middle Row} maps from observations with the SparsePak IFU on the WIYN 3.5m telescope, with H$\alpha$ line measurements from PAN.  Left: H$\alpha$ line flux on a log scale in units of 10$^{-16}$ erg s$^{-1}$ cm$^{-1}$, Center: H$\alpha$ line centers map, Right: H$\alpha$ velocity dispersion ($\sigma_{H\alpha}$) map. Each filled circle corresponds to a fiber's size and position on the sky.  \textbf{Bottom Row} Left: Three color image from \textit{HST} F814W (red), average of F814W and F555W (green), and F555W (blue) observations with ACS, Center: \hi dispersion map from VLA observations with isovelocity contours in 2.5 km s$^{-1}$ step size, Right: $\sigma_{H\alpha}$ map from the SparsePak IFU on the WIYN 3.5m telescope. Overlaid on all three panels are the outlines of the regions used for the analysis. }
\figsetgrpend

\figsetgrpstart
\figsetgrpnum{13.4}
\figsetgrptitle{Maps of Holmberg II}
\figsetplot{f16_4.png}
\figsetgrpnote{\textbf{Holmberg II Top Row} \hi moment maps from VLA observations. Left: \hi column density in 10$^{21}$ hydrogen atoms cm$^{-2}$, Center: \hi velocity map with isovelocity contours spaced every 10 km s$^{-1}$, Right: \hi velocity dispersion map with isovelocity contours at 1.5 km s$^{-1}$ spacing. The beam size (10.73''$\times$10.40'') of the \hi data cube used is shown in the bottom left of the left panel. \textbf{Middle Row} maps from observations with the SparsePak IFU on the WIYN 3.5m telescope, with H$\alpha$ line measurements from PAN and FXCOR.  Left: H$\alpha$ flux on a log scale in units of 10$^{-16}$ erg s$^{-1}$ cm$^{-1}$, Center: H$\alpha$ recessional velocities, Right: H$\alpha$ velocity dispersion ($\sigma_{H\alpha}$) map. Each filled circle corresponds to a fiber's size and position on the sky.  \textbf{Bottom Row} Left: Three color image from \textit{HST} F814W (red), average of F814W and F555W (green), and F555W (blue) observations with ACS, Center: the VLA \hi dispersion map with isovelocity contours in 2.5 km s$^{-1}$ step size, Right: the WIYn 3.5m SparsePak IFU $\sigma_{H\alpha}$ map. Overlaid in red are the outlines of regions used for the analysis. }
\figsetgrpend

\figsetgrpstart
\figsetgrpnum{13.5}
\figsetgrptitle{Maps of UGC04459}
\figsetplot{f16_5.png}
\figsetgrpnote{\textbf{UGC 04459 Top Row} \hi moment maps from VLA observations. Left: \hi column density in 10$^{21}$ hydrogen atoms cm$^{-2}$, Center: \hi velocity map with isovelocity contours spaced every 5 km s$^{-1}$, Right: \hi velocity dispersion map with isovelocity contours at 2.5 km s$^{-1}$ spacing. The beam size (19.65''$\times$19.39'') of the \hi data cube used is shown in the bottom left of the left panel. \textbf{Middle Row} maps from observations with the SparsePak IFU on the WIYN 3.5m telescope, with H$\alpha$ line measurements from PAN.  Left: H$\alpha$ line flux on a log scale in units of 10$^{-16}$ erg s$^{-1}$ cm$^{-1}$, Center: H$\alpha$ line centers map, Right: H$\alpha$ velocity dispersion ($\sigma_{H\alpha}$) map. Each filled circle corresponds to a fiber's size and position on the sky. \textbf{Bottom Row} Left: Three color image from \textit{HST} F814W (red), average of F814W and F555W (green), and F555W (blue) observations with ACS, Center: \hi dispersion map from VLA observations with isovelocity contours in 2.5 km s$^{-1}$ step size, Right: $\sigma_{H\alpha}$ map from the SparsePak IFU on the WIYN 3.5m telescope. Overlaid on all three panels are the outlines of the regions used for the analysis. }
\figsetgrpend

\figsetgrpstart
\figsetgrpnum{13.6}
\figsetgrptitle{Maps off Holmberg I}
\figsetplot{f16_6.png}
\figsetgrpnote{\textbf{Holmberg I Top Row} \textbf{Holmberg I} \hi moment maps from VLA observations. Left: \hi column density in 10$^{21}$ hydrogen atoms cm$^{-2}$, Center: \hi velocity map with isovelocity contours spaced every 5 km s$^{-1}$, Right: \hi velocity dispersion map with isovelocity contours at 2.5 km s$^{-1}$ spacing. The beam size (9.77''$\times$7.50'') of the \hi data cube used is shown in the bottom left of the left panel. \textbf{Middle Row} maps from observations with the SparsePak IFU on the WIYN 3.5m telescope, with H$\alpha$ line measurements from PAN.  Left: H$\alpha$ line flux on a log scale in units of 10$^{-16}$ erg s$^{-1}$ cm$^{-1}$, Center: H$\alpha$ line centers map, Right: H$\alpha$ velocity dispersion ($\sigma_{H\alpha}$) map. Each filled circle corresponds to a fiber's size and position on the sky. \textbf{Bottom Row} Left: Three color image from \textit{HST} F814W (red), average of F814W and F555W (green), and F555W (blue) observations with ACS, Center: \hi dispersion map from VLA observations with isovelocity contours in 2.5 km s$^{-1}$ step size, Right: $\sigma_{H\alpha}$ map from the SparsePak IFU on the WIYN 3.5m telescope. Overlaid on all three panels are the outlines of the regions used for the analysis. }
\figsetgrpend

\figsetgrpstart
\figsetgrpnum{13.7}
\figsetgrptitle{Maps of Sextans B}
\figsetplot{f16_7.png}
\figsetgrpnote{\textbf{Sextans B Top Row} \hi moment maps from VLA observations. Left: \hi column density in 10$^{21}$ hydrogen atoms cm$^{-2}$, Center: \hi velocity map with isovelocity contours spaced every 5 km s$^{-1}$, Right: \hi velocity dispersion map with isovelocity contours at 1.5 km s$^{-1}$ spacing. The beam size (21.26''$\times$20.11'') of the \hi data cube used is shown in the bottom left of the left panel. \textbf{Middle Row} maps from observations with the SparsePak IFU on the WIYN 3.5m telescope, with H$\alpha$ line measurements from PAN.  Left: H$\alpha$ line flux on a log scale in units of 10$^{-16}$ erg s$^{-1}$ cm$^{-1}$, Center: H$\alpha$ line centers map, Right: H$\alpha$ velocity dispersion ($\sigma_{H\alpha}$) map. Each filled circle corresponds to a fiber's size and position on the sky.  \textbf{Top Row} Left: Two color image from \textit{HST} F814W (red), average of F814W and F606W (green), and F606W (blue) observations with WFPC2, Center: \hi dispersion map from VLA observations with isovelocity contours in 3 km s$^{-1}$ step size, Right: $\sigma_{H\alpha}$ map from the SparsePak IFU on the WIYN 3.5m telescope. Overlaid on all three panels are the outlines of the regions used for the analysis. }
\figsetgrpend

\figsetgrpstart
\figsetgrpnum{13.8}
\figsetgrptitle{Maps of Sextans A}
\figsetplot{f16_8.png}
\figsetgrpnote{ \textbf{Sextans A Top Row} \hi moment maps from VLA observations. Left: \hi column density in 10$^{21}$ hydrogen atoms cm$^{-2}$, Center: \hi velocity map with isovelocity contours spaced every 5 km s$^{-1}$, Right: \hi velocity dispersion map with isovelocity contours at 2.5 km s$^{-1}$ spacing. The beam size (13.67''$\times$10.68'') of the \hi data cube used is shown in the bottom left of the left panel. \textbf{Middle Row} maps from observations with the SparsePak IFU on the WIYN 3.5m telescope, with H$\alpha$ line measurements from PAN.  Left: H$\alpha$ line flux on a log scale in units of 10$^{-16}$ erg s$^{-1}$ cm$^{-1}$, Center: H$\alpha$ line centers map, Right: H$\alpha$ velocity dispersion ($\sigma_{H\alpha}$) map. Each filled circle corresponds to a fiber's size and position on the sky. \textbf{Bottom Row} Left: Three color image from \textit{HST} F814W (red), average of F814W and F555W (green), and F555W (blue) observations with WFPC2, Center: \hi dispersion map from VLA observations with isovelocity contours in 2.5 km s$^{-1}$ step size, Right: $\sigma_{H\alpha}$ map from the SparsePak IFU on the WIYN 3.5m telescope. Overlaid on all three panels are the outlines of the regions used for the analysis. }
\figsetgrpend

\figsetgrpstart
\figsetgrpnum{13.9}
\figsetgrptitle{Maps of IC 2574}
\figsetplot{f16_9.png}
\figsetgrpnote{ \textbf{IC 2574 Top Row} \hi moment maps from VLA observations Left: \hi column density in 10$^{21}$ hydrogen atoms cm$^{-2}$, Center: \hi velocity map with isovelocity contours spaces every 10 km s$^{-1}$, Right: \hi velocity dispersion map with isovelocity contours at 2.5 km s$^{-1}$ spacing. The beam size (13.18''$\times$12.45'') of the \hi data cube used is shown in the bottom left of the left panel. \textbf{Middle Row} maps from observations with the SparsePak IFU on the WIYN 3.5m telescope, with H$\alpha$ line measurements from PAN.  Left: H$\alpha$ line flux on a log scale in units of 10$^{-16}$ erg s$^{-1}$ cm$^{-1}$, Center: H$\alpha$ line centers map, Right: H$\alpha$ velocity dispersion ($\sigma_{H\alpha}$) map. Each filled circle corresponds to a fiber's size and position on the sky. \textbf{Bottom Row} Left: Three color image from \textit{HST} F814W (red), average of F814W and F555W (green), and F555W (blue) observations with ACS, Center: \hi dispersion map from VLA observations with isovelocity contours in 2.5 km s$^{-1}$ step size, Right: $\sigma_{H\alpha}$ map from the SparsePak IFU on the WIYN 3.5m telescope. Overlaid on all three panels are the outlines of the regions used for the analysis. }
\figsetgrpend

\figsetgrpstart
\figsetgrpnum{13.10}
\figsetgrptitle{Maps of NGC3738}
\figsetplot{f16_10.png}
\figsetgrpnote{ \textbf{NGC 3738 Top Row} \hi moment maps from VLA observations. Left: \hi column density in 10$^{21}$ hydrogen atoms cm$^{-2}$, Center: \hi velocity map with isovelocity contours spaced every 10 km s$^{-1}$, Right: \hi velocity dispersion map with isovelocity contours at 5 km s$^{-1}$ spacing. The beam size (10.73''$\times$10.40'') of the \hi data cube used is shown in the bottom left of the left panel. \textbf{Middle Row} maps from observations with the SparsePak IFU on the WIYN 3.5m telescope, with H$\alpha$ line measurements from PAN.  Left: H$\alpha$ line flux on a log scale in units of 10$^{-16}$ erg s$^{-1}$ cm$^{-1}$, Center: H$\alpha$ line centers map, Right: H$\alpha$ velocity dispersion ($\sigma_{H\alpha}$) map. Each filled circle corresponds to a fiber's size and position on the sky. \textbf{Bottom Row} Left: Three color image from \textit{HST} 814W (red), average of F814W and F606W (green), and F606W (blue) observations with ACS, Center: \hi dispersion map from VLA observations with isovelocity contours in 5 km s$^{-1}$ step size, Right: $\sigma_{H\alpha}$ map from the SparsePak IFU on the WIYN 3.5m telescope. Overlaid on all three panels are the outlines of the regions used for the analysis. }
\figsetgrpend

\figsetgrpstart
\figsetgrpnum{13.11}
\figsetgrptitle{Maps of NGC3741}
\figsetplot{f16_11.png}
\figsetgrpnote{ \textbf{NGC3741 Top Row} \hi moment maps from VLA observations. Left: \hi column density in 10$^{21}$ hydrogen atoms cm$^{-2}$, Center: \hi velocity map with isovelocity contours spaced every 10 km s$^{-1}$, Right: \hi velocity dispersion map with isovelocity contours at 1.5 km s$^{-1}$ spacing. The beam size (9.47''$\times$6.66'') of the \hi data cube used is shown in the bottom left of the left panel.  \textbf{Middle Row} maps from observations with the SparsePak IFU on the WIYN 3.5m telescope, with H$\alpha$ line measurements from PAN.  Left: H$\alpha$ line flux on a log scale in units of 10$^{-16}$ erg s$^{-1}$ cm$^{-1}$, Center: H$\alpha$ line centers map, Right: H$\alpha$ velocity dispersion ($\sigma_{H\alpha}$) map. Each filled circle corresponds to a fiber's size and position on the sky. \textbf{Bottom Row} Left: Three color image from \textit{HST} F814W (red), average of F814W and F475W (green), and F475W (blue) observations with ACS, Center: \hi dispersion map from VLA observations with isovelocity contours in 1.5 km s$^{-1}$ step size, Right: $\sigma_{H\alpha}$ map from the SparsePak IFU on the WIYN 3.5m telescope. Overlaid on all three panels are the outlines of the regions used for the analysis. }
\figsetgrpend

\figsetgrpstart
\figsetgrpnum{13.12}
\figsetgrptitle{Maps of NGC4068}
\figsetplot{f16_12.png}
\figsetgrpnote{ \textbf{NGC 4068 Top Row} \hi moment maps from VLA observations Left: \hi column density in 10$^{21}$ hydrogen atoms cm$^{-2}$, Center: \hi velocity map with isovelocity contours spaces every 10 km s$^{-1}$, Right: \hi velocity dispersion map with isovelocity contours at 2.5 km s$^{-1}$ spacing. The beam size (11.83''$\times$11.29'') of the \hi data cube used is shown in the bottom left of the left panel. \textbf{Middle Row} maps from observations with the SparsePak IFU on the WIYN 3.5m telescope, with H$\alpha$ line measurements from PAN.  Left: H$\alpha$ line flux on a log scale in units of 10$^{-16}$ erg s$^{-1}$ cm$^{-1}$, Center: H$\alpha$ line centers map, Right: H$\alpha$ velocity dispersion ($\sigma_{H\alpha}$) map. Each filled circle corresponds to a fiber's size and position on the sky. \textbf{Bottom Row} Left: Three color image from \textit{HST} F814W (red), average of F814W and F606W (green), and F606W (blue) observations with ACS, Center: \hi dispersion map from VLA observations with isovelocity contours in 2.5 km s$^{-1}$ step size, Right: $H\alpha$ FWHM map from the SparsePak IFU on the WIYN 3.5m telescope. Overlaid on all three panels are the outlines of the regions used for the analysis.}
\figsetgrpend

\figsetgrpstart
\figsetgrpnum{13.13}
\figsetgrptitle{Maps of NGC4163}
\figsetplot{f16_13.png}
\figsetgrpnote{ \textbf{NGC 4163 Top Row} \hi moment maps from VLA observations Left: \hi column density in 10$^{21}$ hydrogen atoms cm$^{-2}$, Center: \hi velocity map with isovelocity contours spaces every 5 km s$^{-1}$, Right: \hi velocity dispersion map with isovelocity contours at 2 km s$^{-1}$ spacing. The beam size (15.911''$\times$12.941'') of the \hi data cube used is shown in the bottom left of the left panel. \textbf{Middle Row} maps from observations with the SparsePak IFU on the WIYN 3.5m telescope, with H$\alpha$ line measurements from PAN.  Left: H$\alpha$ line flux on a log scale in units of 10$^{-16}$ erg s$^{-1}$ cm$^{-1}$, Center: H$\alpha$ line centers map, Right: H$\alpha$ velocity dispersion ($\sigma_{H\alpha}$) map. Each filled circle corresponds to a fiber's size and position on the sky. \textbf{Bottom Row} Left: Three color image from \textit{HST} F814W (red), average of F814W and F606W (green), and F606W (blue) observations with ACS, Center: \hi dispersion map from VLA observations with isovelocity contours in 2 km s$^{-1}$ step size, Right: $H\alpha$ FWHM map from the SparsePak IFU on the WIYN 3.5m telescope. Overlaid on all three panels are the outlines of the regions used for the analysis. }
\figsetgrpend

\figsetgrpstart
\figsetgrpnum{13.14}
\figsetgrptitle{Maps of NGC4190}
\figsetplot{f16_14.png}
\figsetgrpnote{\textbf{NGC 4190 Top Row} \hi moment maps from VLA observations. Left: \hi column density in 10$^{21}$ hydrogen atoms cm$^{-2}$, Center: \hi velocity map with isovelocity contours spaced every 10 km s$^{-1}$, Right: \hi velocity dispersion map with isovelocity contours at 2.5 km s$^{-1}$ spacing. The beam size (11.54''$\times$9.62'') of the \hi data cube used is shown in the bottom left of the left panel. \textbf{Middle Row} maps from observations with the SparsePak IFU on the WIYN 3.5m telescope, with H$\alpha$ line measurements from PAN.  Left: H$\alpha$ line flux on a log scale in units of 10$^{-16}$ erg s$^{-1}$ cm$^{-1}$, Center: H$\alpha$ line centers map, Right: H$\alpha$ velocity dispersion ($\sigma_{H\alpha}$) map. Each filled circle corresponds to a fiber's size and position on the sky. \textbf{Bottom Row} Left: Three color image from \textit{HST} F814W (red), average of F814W and F606W (green), and F606W (blue) observations with ACS, Center: \hi dispersion map from VLA observations with isovelocity contours in 2.5 km s$^{-1}$ step size, Right: $\sigma_{H\alpha}$ map from the SparsePak IFU on the WIYN 3.5m telescope. Overlaid on all three panels are the outlines of the regions used for the analysis.}
\figsetgrpend

\figsetgrpstart
\figsetgrpnum{13.15}
\figsetgrptitle{Maps of UGC07577}
\figsetplot{f16_15.png}
\figsetgrpnote{ \textbf{UGC 7577 Top Row} hi moment maps from VLA observations. Left: \hi column density in 10$^{21}$ hydrogen atoms cm$^{-2}$, Center: \hi velocity map with isovelocity contours spaced every 5 km s$^{-1}$, Right: \hi velocity dispersion map with isovelocity contours at 1 km s$^{-1}$ spacing. The beam size (11.84''$\times$11.30'') of the \hi data cube used is shown in the bottom left of the left panel. \textbf{Middle Row} maps from observations with the SparsePak IFU on the WIYN 3.5m telescope, with H$\alpha$ line measurements from PAN.  Left: H$\alpha$ line flux on a log scale in units of 10$^{-16}$ erg s$^{-1}$ cm$^{-1}$, Center: H$\alpha$ line centers map, Right: H$\alpha$ velocity dispersion ($\sigma_{H\alpha}$) map. Each filled circle corresponds to a fiber's size and position on the sky. \textbf{Bottom Row} Left: Three color image from \textit{HST} F814W (red), average of F814W and F606W (green), and F606W (blue) observations with WFPC2, Center: \hi dispersion map from VLA observations with isovelocity contours in 1 km s$^{-1}$ step size, Right: $\sigma_{H\alpha}$ map from the SparsePak IFU on the WIYN 3.5m telescope. Overlaid on all three panels are the outlines of the regions used for the analysis. }
\figsetgrpend

\figsetgrpstart
\figsetgrpnum{13.16}
\figsetgrptitle{Maps of UGCA292A}
\figsetplot{f16_13.png}
\figsetgrpnote{ \textbf{UGCA 292 Top Row} \hi moment maps from VLA observations. Left: \hi column density in 10$^{21}$ hydrogen atoms cm$^{-2}$, Center: \hi velocity map with isovelocity contours spaced every 5 km s$^{-1}$, Right: \hi velocity dispersion map with isovelocity contours at 1 km s$^{-1}$ spacing. The beam size (17.40''$\times$16.03'') of the \hi data cube used is shown in the bottom left of the left panel. \textbf{Middle Row} maps from observations with the SparsePak IFU on the WIYN 3.5m telescope, with H$\alpha$ line measurements from PAN.  Left: H$\alpha$ line flux on a log scale in units of 10$^{-16}$ erg s$^{-1}$ cm$^{-1}$, Center: H$\alpha$ line centers map, Right: H$\alpha$ velocity dispersion ($\sigma_{H\alpha}$) map. Each filled circle corresponds to a fiber's size and position on the sky. \textbf{Bottom Row} Left: Three color image from \textit{HST} F814W (red), average of F814W and F475W (green), and F475W (blue) observations with ACS, Center: \hi dispersion map from VLA observations with isovelocity contours in 1 km s$^{-1}$ step size, Right: $\sigma_{H\alpha}$ map from the SparsePak IFU on the WIYN 3.5m telescope. Overlaid on all three panels are the outlines of the regions used for the analysis. }
\figsetgrpend

\figsetgrpstart
\figsetgrpnum{13.17}
\figsetgrptitle{Maps of UGC08024}
\figsetplot{f16_17.png}
\figsetgrpnote{\textbf{UGC 8024 Top Row} \hi moment maps from VLA observations. Left: \hi column density in 10$^{21}$ hydrogen atoms cm$^{-2}$, Center: \hi velocity map with isovelocity contours spaced every 10 km s$^{-1}$, Right: \hi velocity dispersion map with isovelocity contours at 1.5 km s$^{-1}$ spacing. The beam size (10.73''$\times$10.40'') of the \hi data cube used is shown in the bottom left of the left panel.  \textbf{Middle Row} maps from observations with the SparsePak IFU on the WIYN 3.5m telescope, with H$\alpha$ line measurements from PAN.  Left: H$\alpha$ line flux on a log scale in units of 10$^{-16}$ erg s$^{-1}$ cm$^{-1}$, Center: H$\alpha$ line centers map, Right: H$\alpha$ velocity dispersion ($\sigma_{H\alpha}$) map. Each filled circle corresponds to a fiber's size and position on the sky. \textbf{Bottom Row} Left: Three color image from \textit{HST} F814W (red), average of F814W and F606W (green), and F606W (blue) observations with ACS, Center: \hi dispersion map from VLA observations with isovelocity contours in 1.5 km s$^{-1}$ step size, Right: $\sigma_{H\alpha}$ map from the SparsePak IFU on the WIYN 3.5m telescope. Overlaid on all three panels are the outlines of the regions used for the analysis.}
\figsetgrpend

\figsetgrpstart
\figsetgrpnum{13.18}
\figsetgrptitle{Maps of GR8}
\figsetplot{f16_18.png}
\figsetgrpnote{\\textbf{GR8 Top Row} \hi moment maps from VLA observations. Left: \hi column density in 10$^{21}$ hydrogen atoms cm$^{-2}$, Center: \hi velocity map with isovelocity contours spaced every 5 km s$^{-1}$, Right: \hi velocity dispersion map with isovelocity contours at 1.5 km s$^{-1}$ spacing. The beam size (17.43''$\times$16.50'') of the \hi data cube used is shown in the bottom left of the left panel. \textbf{Middle Row} maps from observations with the SparsePak IFU on the WIYN 3.5m telescope, with H$\alpha$ line measurements from PAN.  Left: H$\alpha$ line flux on a log scale in units of 10$^{-16}$ erg s$^{-1}$ cm$^{-1}$, Center: H$\alpha$ line centers map, Right: H$\alpha$ velocity dispersion ($\sigma_{H\alpha}$) map. Each filled circle corresponds to a fiber's size and position on the sky. \textbf{Bottom Row} Left: Three color image from \textit{HST} F814W (red), average of F814W and F475W (green), and F475W (blue) observations with ACS, Center: \hi dispersion map from VLA observations with isovelocity contours in 1.5 km s$^{-1}$ step size, Right: $\sigma_{H\alpha}$ map from the SparsePak IFU on the WIYN 3.5m telescope. Overlaid on all three panels are the outlines of the regions used for the analysis.}
\figsetgrpend

\figsetgrpstart
\figsetgrpnum{13.19}
\figsetgrptitle{Maps of UGC08201}
\figsetplot{f16_19.png}
\figsetgrpnote{ \textbf{UGC 8201 Top Row} \hi moment maps from VLA observations. Left: \hi column density in 10$^{21}$ hydrogen atoms cm$^{-2}$, Center: \hi velocity map with isovelocity contours spaced every 10 km s$^{-1}$, Right: \hi velocity dispersion map with isovelocity contours at 2.5 km s$^{-1}$ spacing. The beam size (13.48''$\times$13.12'') of the \hi data cube used is shown in the bottom left of the left panel.  \textbf{Middle Row} maps from observations with the SparsePak IFU on the WIYN 3.5m telescope, with H$\alpha$ line measurements from PAN.  Left: H$\alpha$ line flux on a log scale in units of 10$^{-16}$ erg s$^{-1}$ cm$^{-1}$, Center: H$\alpha$ line centers map, Right: H$\alpha$ velocity dispersion ($\sigma_{H\alpha}$) map. Each filled circle corresponds to a fiber's size and position on the sky. \textbf{Bottom Row} Left: Three color image from \textit{HST} F814W (red), average of F814W and F555W (green), and F555W (blue) observations with ACS, Center: \hi dispersion map from VLA observations with isovelocity contours in 2.5 km s$^{-1}$ step size, Right: $\sigma_{H\alpha}$ map from the SparsePak IFU on the WIYN 3.5m telescope. Overlaid on all three panels are the outlines of the regions used for the analysis.}
\figsetgrpend

\figsetgrpstart
\figsetgrpnum{13.20}
\figsetgrptitle{Maps of NGC5204}
\figsetplot{f16_20.png}
\figsetgrpnote{ \textbf{NGC 5204 Top Row} \hi moment maps from VLA observations. Left: \hi column density in 10$^{21}$ hydrogen atoms cm$^{-2}$, Center: \hi velocity map with isovelocity contours spaced every 10 km s$^{-1}$, Right: \hi velocity dispersion map with isovelocity contours at 2.5 km s$^{-1}$ spacing. The beam size (17.62''$\times$12.50'') of the \hi data cube used is shown in the bottom left of the left panel. \textbf{Middle Row} maps from observations with the SparsePak IFU on the WIYN 3.5m telescope, with H$\alpha$ line measurements from PAN.  Left: H$\alpha$ line flux on a log scale in units of 10$^{-16}$ erg s$^{-1}$ cm$^{-1}$, Center: H$\alpha$ line centers map, Right: H$\alpha$ velocity dispersion ($\sigma_{H\alpha}$) map. Each filled circle corresponds to a fiber's size and position on the sky. \textbf{Bottom Row} Left: Three color image from \textit{HST} F814W (red), average of F814W and F606W (green), and F606W (blue) observations with WFPC2, Center: \hi dispersion map from VLA observations with isovelocity contours in 2.5 km s$^{-1}$ step size, Right: $\sigma_{H\alpha}$ map from the SparsePak IFU on the WIYN 3.5m telescope. Overlaid on all three panels are the outlines of the regions used for the analysis.}
\figsetgrpend

\figsetgrpstart
\figsetgrpnum{13.21}
\figsetgrptitle{Maps of UGC8638}
\figsetplot{f16_21.png}
\figsetgrpnote{ \textbf{UGC 8638 Top Row} \hi moment maps from VLA observations. Left: \hi column density in 10$^{21}$ hydrogen atoms cm$^{-2}$, Center: \hi velocity map with isovelocity contours spaced every 5 km s$^{-1}$, Right: \hi velocity dispersion map with isovelocity contours at 1.5 km s$^{-1}$ spacing. The beam size (13.16''$\times$10.79'') of the \hi data cube used is shown in the bottom left of the left panel. \textbf{Middle Row} maps from observations with the SparsePak IFU on the WIYN 3.5m telescope, with H$\alpha$ line measurements from PAN.  Left: H$\alpha$ line flux on a log scale in units of 10$^{-16}$ erg s$^{-1}$ cm$^{-1}$, Center: H$\alpha$ line centers map, Right: H$\alpha$ velocity dispersion ($\sigma_{H\alpha}$) map. Each filled circle corresponds to a fiber's size and position on the sky. \textbf{Bottom Row} Left: Three color image from \textit{HST} F814W (red), average of F814W and F606W (green), and F606W (blue) observations with ACS, Center: \hi dispersion map from VLA observations with isovelocity contours in 1.5 km s$^{-1}$ step size, Right: $\sigma_{H\alpha}$ map from the SparsePak IFU on the WIYN 3.5m telescope. Overlaid on all three panels are the outlines of the regions used for the analysis. }
\figsetgrpend

\figsetgrpstart
\figsetgrpnum{13.22}
\figsetgrptitle{Maps of UGC8651}
\figsetplot{f16_22.png}
\figsetgrpnote{ \textbf{UGC 8651 Top Row} \hi moment maps from VLA observations. Left: \hi column density in 10$^{21}$ hydrogen atoms cm$^{-2}$, Center: \hi velocity map with isovelocity contours spaced every 5 km s$^{-1}$, Right: \hi velocity dispersion map with isovelocity contours at 1 km s$^{-1}$ spacing. The beam size (13.95''$\times$11.28'') of the \hi data cube used is shown in the bottom left of the left panel. \textbf{Middle Row} maps from observations with the SparsePak IFU on the WIYN 3.5m telescope, with H$\alpha$ line measurements from PAN.  Left: H$\alpha$ line flux on a log scale in units of 10$^{-16}$ erg s$^{-1}$ cm$^{-1}$, Center: H$\alpha$ line centers map, Right: H$\alpha$ velocity dispersion ($\sigma_{H\alpha}$) map. Each filled circle corresponds to a fiber's size and position on the sky. \textbf{Bottom Row} Left: Three color image from \textit{HST} F814W (red), average of F814W and F606W (green), and F606W (blue) observations with ACS, Center: \hi dispersion map from VLA observations with isovelocity contours in 1 km s$^{-1}$ step size, Right: $\sigma_{H\alpha}$ map from the SparsePak IFU on the WIYN 3.5m telescope. Overlaid on all three panels are the outlines of the regions used for the analysis. }
\figsetgrpend

\figsetgrpstart
\figsetgrpnum{13.23}
\figsetgrptitle{Maps of NGC5253}
\figsetplot{f16_23.png}
\figsetgrpnote{ \textbf{NGC 5253 Top Row} \hi moment maps from VLA observations. Left: \hi column density in 10$^{21}$ hydrogen atoms cm$^{-2}$, Center: \hi velocity map with isovelocity contours spaced every 10 km s$^{-1}$, Right: \hi velocity dispersion map with isovelocity contours at 2.5 km s$^{-1}$ spacing. The beam size (17.60''$\times$10.11'') of the \hi data cube used is shown in the bottom left of the left panel. \textbf{Bottom Row} Left: Three color image from \textit{HST} F814W (red), average of F814W and F555W (green), and F555W (blue) observations with ACS, Center: \hi dispersion map from VLA observations with isovelocity contours in 2.5 km s$^{-1}$ step size, Right: Ground based H$\alpha$ map published in \cite{Dale09}. Overlaid on all three panels are the outlines of the regions used for the analysis. }
\figsetgrpend

\figsetgrpstart
\figsetgrpnum{13.24}
\figsetgrptitle{Maps of UGC9128}
\figsetplot{f16_24.png}
\figsetgrpnote{ \textbf{UGC 9128 Top Row} \hi moment maps from VLA observations Left:   \hi column density in 10$^{21}$ hydrogen atoms cm$^{-2}$, Center: \hi velocity map with isovelocity contours spaces every 5 km s$^{-1}$, Right: \hi velocity dispersion map with isovelocity contours at 2.5 km s$^{-1}$ spacing.  The beam size (13.279''$\times$10.326'') of the \hi data cube used is shown in the bottom left of the left panel. \textbf{Middle Row} maps from observations with the SparsePak IFU on the WIYN 3.5m telescope, with H$\alpha$ line measurements from PAN.  Left: H$\alpha$ line flux on a log scale in units of 10$^{-16}$ erg s$^{-1}$ cm$^{-1}$, Center: H$\alpha$ line centers map, Right: H$\alpha$ velocity dispersion ($\sigma_{H\alpha}$) map. Each filled circle corresponds to a fiber's size and position on the sky.  \textbf{Bottom Row} Left: Three color image from \textit{HST} F814W (red), average of F814W and F606W (green), and F606W (blue) observations with ACS, Center: \hi dispersion map from VLA observations with isovelocity contours in 2.5 km s$^{-1}$ step size, Right: $H\alpha$ FWHM map from the SparsePak IFU on the WIYN 3.5m telescope. Overlaid on all three panels are the outlines of the regions used for the analysis.}
\figsetgrpend

\figsetgrpstart
\figsetgrpnum{13.25}
\figsetgrptitle{Maps of UGC09240}
\figsetplot{f16_25.png}
\figsetgrpnote{ \textbf{UGC 9240 Top Row} \hi moment maps from VLA observations Left:   \hi column density in 10$^{21}$ hydrogen atoms cm$^{-2}$, Center: \hi velocity map with isovelocity contours spaces every 5 km s$^{-1}$, Right: \hi velocity dispersion map with isovelocity contours at 2 km s$^{-1}$ spacing.  The beam size (16.11''$\times$14.29'') of the \hi data cube used is shown in the bottom left of the left panel. \textbf{Middle Row} maps from observations with the SparsePak IFU on the WIYN 3.5m telescope, with H$\alpha$ line measurements from PAN.  Left: H$\alpha$ line flux on a log scale in units of 10$^{-16}$ erg s$^{-1}$ cm$^{-1}$, Center: H$\alpha$ line centers map, Right: H$\alpha$ velocity dispersion ($\sigma_{H\alpha}$) map. Each filled circle corresponds to a fiber's size and position on the sky. \textbf{Bottom Row} Left: Three color image from \textit{HST} F814W (red), average of F814W and F606W (green), and F606W (blue) observations with ACS, Center: \hi dispersion map from VLA observations with isovelocity contours in 2 km s$^{-1}$ step size, Right: $\sigma_{H\alpha}$ map from the SparsePak IFU on the WIYN 3.5m telescope. Overlaid on all three panels are the outlines of the regions used for the analysis. }
\figsetgrpend

\figsetgrpstart
\figsetgrpnum{13.26}
\figsetgrptitle{Maps of NGC6789}
\figsetplot{f16_26.png}
\figsetgrpnote{ \textbf{NGC 6789 Top Row} \hi moment maps from VLA observations Left:   \hi column density in 10$^{21}$ hydrogen atoms cm$^{-2}$, Center: \hi velocity map with isovelocity contours spaces every 5 km s$^{-1}$, Right: \hi velocity dispersion map with isovelocity contours at 2.5 km s$^{-1}$ spacing. The beam size (11.92''$\times$10.52'') of the   \hi data cube used is shown in the bottom left of the left panel. \textbf{Middle Row} maps from observations with the SparsePak IFU on the WIYN 3.5m telescope, with H$\alpha$ line measurements from PAN.  Left: H$\alpha$ line flux on a log scale in units of 10$^{-16}$ erg s$^{-1}$ cm$^{-1}$, Center: H$\alpha$ line centers map, Right: H$\alpha$ velocity dispersion ($\sigma_{H\alpha}$) map.  Each filled circle corresponds to a fiber's size and position on the sky. \textbf{Bottom Row} Left: Three color image from \textit{HST} F814W (red), average of F814W and F555W (green),  and F555W (blue observations with WFPC2), Center: \hi dispersion map from VLA observations with isovelocity contours in 2.5 km s$^{-1}$ step size, Right: $H\alpha$ FWHM map from the SparsePak IFU on the WIYN 3.5m telescope. Overlaid on all three panels are the outlines of the regions used for the analysis.}
\figsetgrpend

\figsetend

\begin{figure*}
\figurenum{13}
\plotone{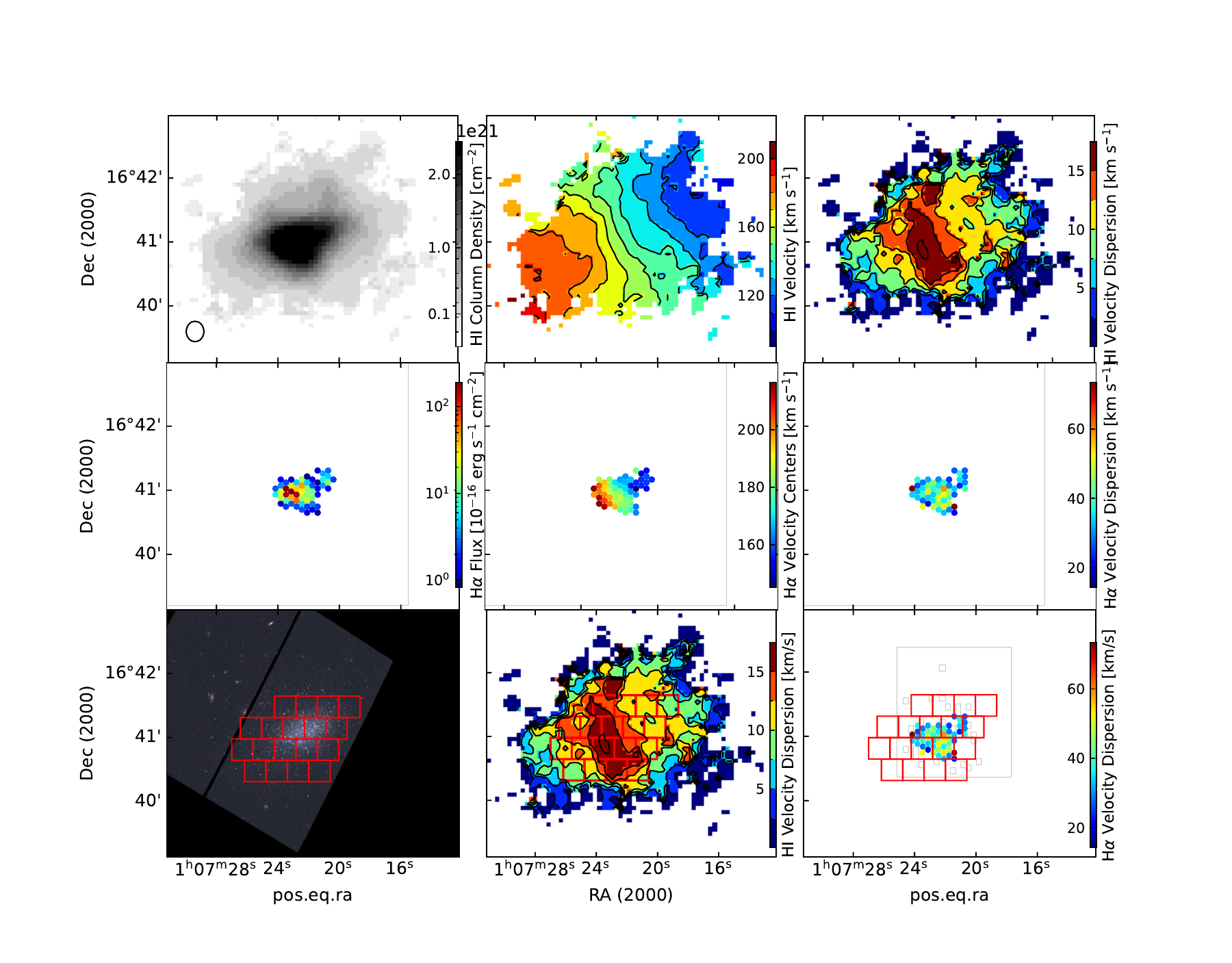}
\caption{\textbf{UGC 0685 Maps Top Row}  UGC 0685 \hi moment maps from VLA observations. Left: \hi column density in 10$^{21}$ hydrogen atoms cm$^{-2}$, Center: \hi velocity map with isovelocity contours spaced every 10 km s$^{-1}$, Right: \hi velocity dispersion map with isovelocity contours at 2.5 km s$^{-1}$ spacing. The beam size (19.25''$\times$16.53'') of the \hi data cube used is shown in the bottom left of the left panel. \textbf{Middle Row} UGC 0685 maps from observations with the SparsePak IFU on the WIYN 3.5m telescope, with H$\alpha$ line measurements from PAN.  Left: H$\alpha$ line flux on a log scale in units of 10$^{-16}$ erg s$^{-1}$ cm$^{-1}$, Center: H$\alpha$ line centers map, Right: H$\alpha$ velocity dispersion ($\sigma_{H\alpha}$) map. Each filled circle corresponds to a fiber's size and position on the sky. \textbf{Bottom Row} Left: Thre color image from \textit{HST} F814W (red), average of F814W and F606W (green), and F606W (blue) observations with ACS, Center: \hi dispersion map from VLA observations with isovelocity contours in 2.5 km s$^{-1}$ step size, Right: $\sigma_{H\alpha}$ map from the SparsePak IFU on the WIYN 3.5m telescope. Overlaid on all three panels are the outlines of the regions used for the analysis. \textbf{See digital version for full figure set} }\label{appendix}
\end{figure*}

\begin{figure*}
\figurenum{13}
\plotone{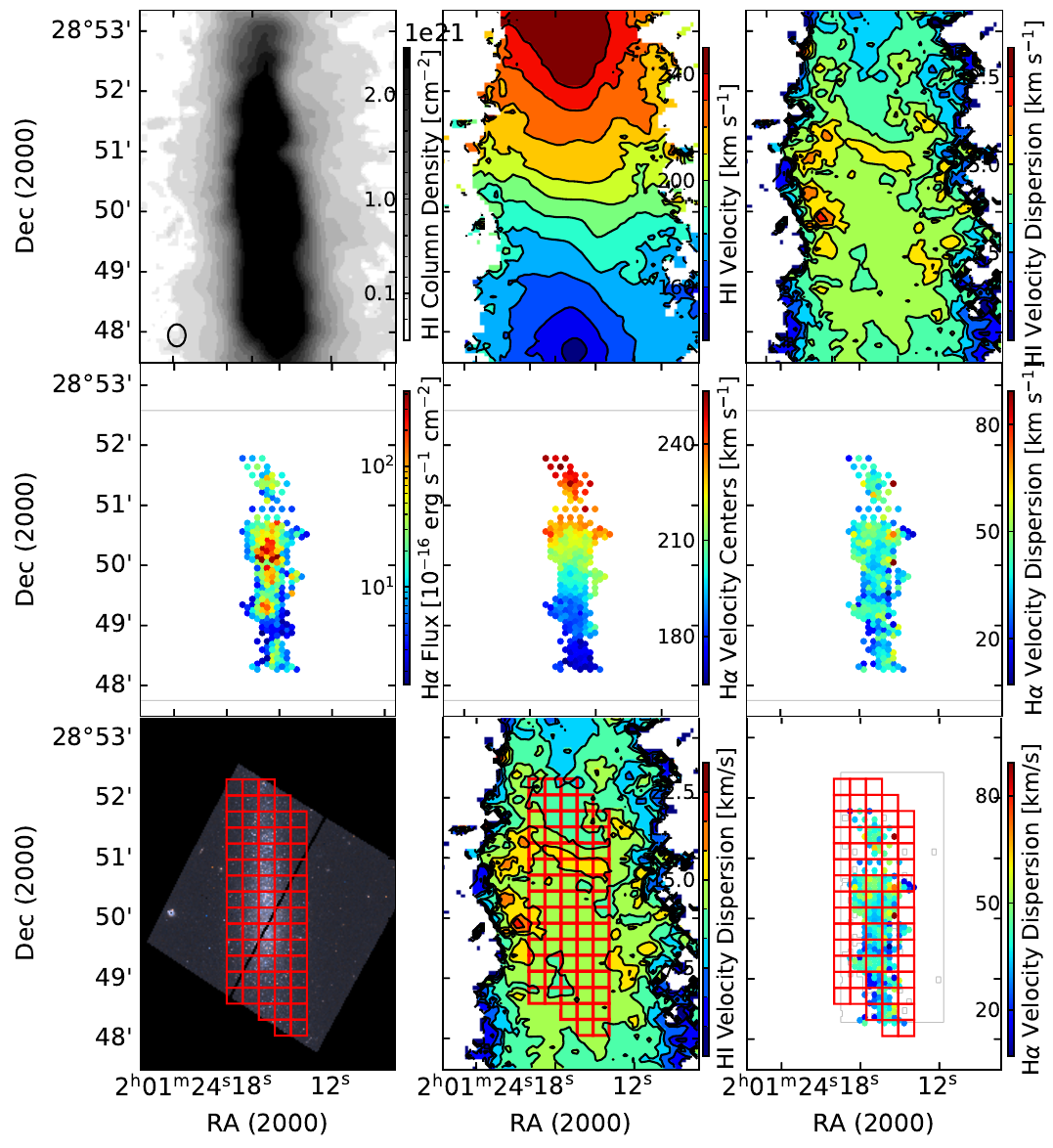}
\caption{\textbf{NGC 0784 Top Row} NGC0784 \hi moment maps from VLA observations. Left: \hi column density in 10$^{21}$ hydrogen atoms cm$^{-2}$, Center: \hi velocity map with isovelocity contours spaced every 10 km s$^{-1}$, Right: \hi velocity dispersion map with isovelocity contours at 2.5 km s$^{-1}$ spacing. The beam size (22.20''$\times$17.62'') of the \hi data cube used is shown in the bottom left of the left panel.  \textbf{ Middle Row} NGC0784 maps from observations with the SparsePak IFU on the WIYN 3.5m telescope, with H$\alpha$ line measurements from PAN.  Left: H$\alpha$ line flux on a log scale in units of 10$^{-16}$ erg s$^{-1}$ cm$^{-1}$, Center: H$\alpha$ line centers map, Right: H$\alpha$ velocity dispersion ($\sigma_{H\alpha}$) map. Each filled circle corresponds to a fiber's size and position on the sky.  \textbf{Bottom Row} Left: Three color image from \textit{HST} F814W (red), average of F814W and F606W (green), and F606W (blue) observations with ACS, Center: \hi dispersion map from VLA observations with isovelocity contours in 2.5 km s$^{-1}$ step size, Right: $\sigma_{H\alpha}$ map from the SparsePak IFU on the WIYN 3.5m telescope. Overlaid on all three panels are the outlines of the regions used for the analysis.  }
\end{figure*}

\begin{figure*}
\figurenum{13}
\plotone{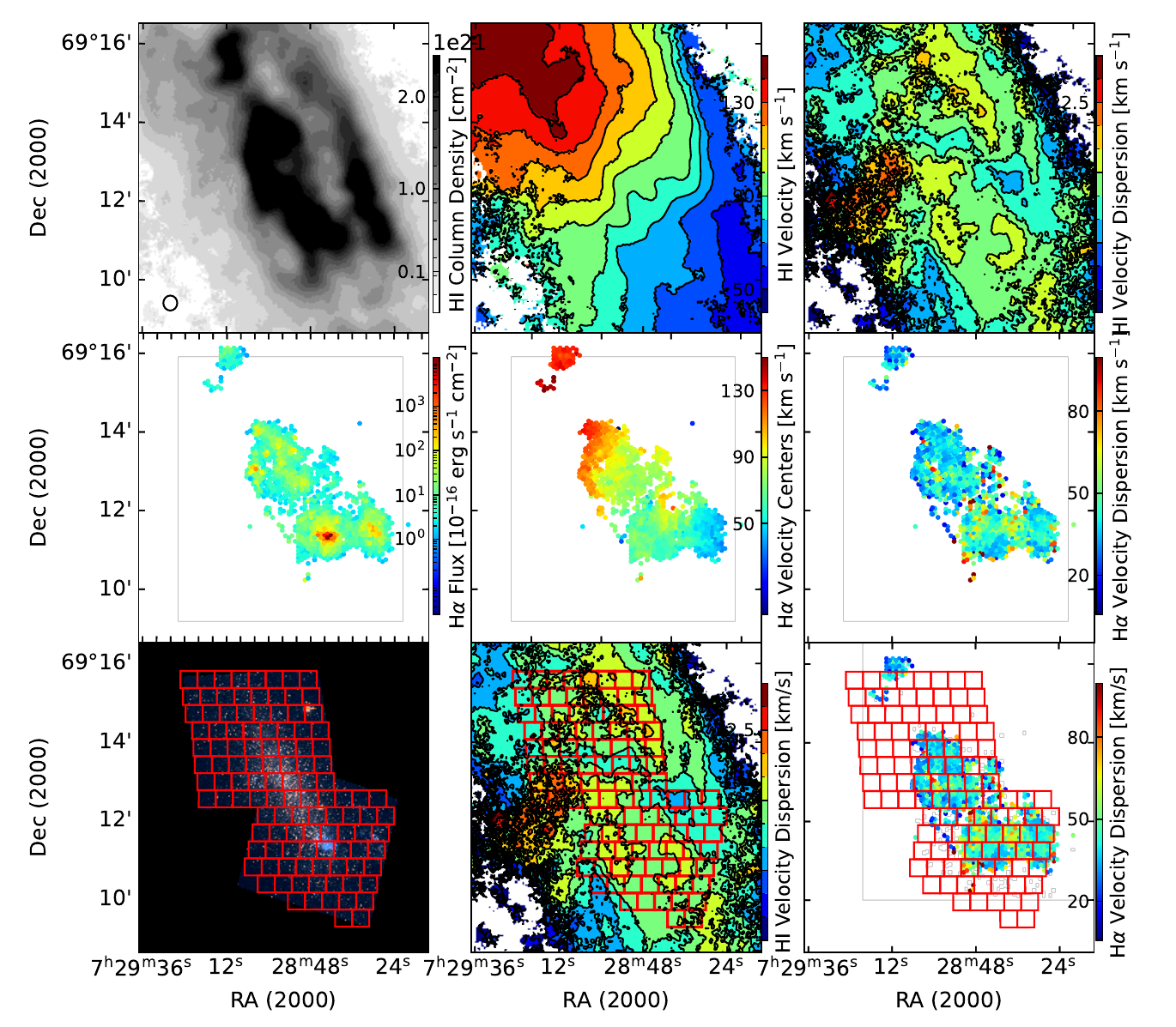}
\caption{\textbf{NGC 2366 Top Row} \hi moment maps from VLA observations Left: \hi column density in 10$^{21}$ hydrogen atoms cm$^{-2}$, Center: \hi velocity map with isovelocity contours spaces every 10 km s$^{-1}$, Right: \hi velocity dispersion map with isovelocity contours at 2.5 km s$^{-1}$ spacing. The beam size (22.89''$\times$21.25'') of the \hi data cube used is shown in the bottom left of the left panel. \textbf{Middle Row} maps from observations with the SparsePak IFU on the WIYN 3.5m telescope, with H$\alpha$ line measurements from PAN.  Left: H$\alpha$ line flux on a log scale in units of 10$^{-16}$ erg s$^{-1}$ cm$^{-1}$, Center: H$\alpha$ line centers map, Right: H$\alpha$ velocity dispersion ($\sigma_{H\alpha}$) map. Each filled circle corresponds to a fiber's size and position on the sky.  \textbf{Bottom Row} Left: Three color image from \textit{HST} F814W (red), average of F814W and F555W (green), and F555W (blue) observations with ACS, Center: \hi dispersion map from VLA observations with isovelocity contours in 2.5 km s$^{-1}$ step size, Right: $\sigma_{H\alpha}$ map from the SparsePak IFU on the WIYN 3.5m telescope. Overlaid on all three panels are the outlines of the regions used for the analysis. }
\end{figure*}

\begin{figure*}
\figurenum{13}
\plotone{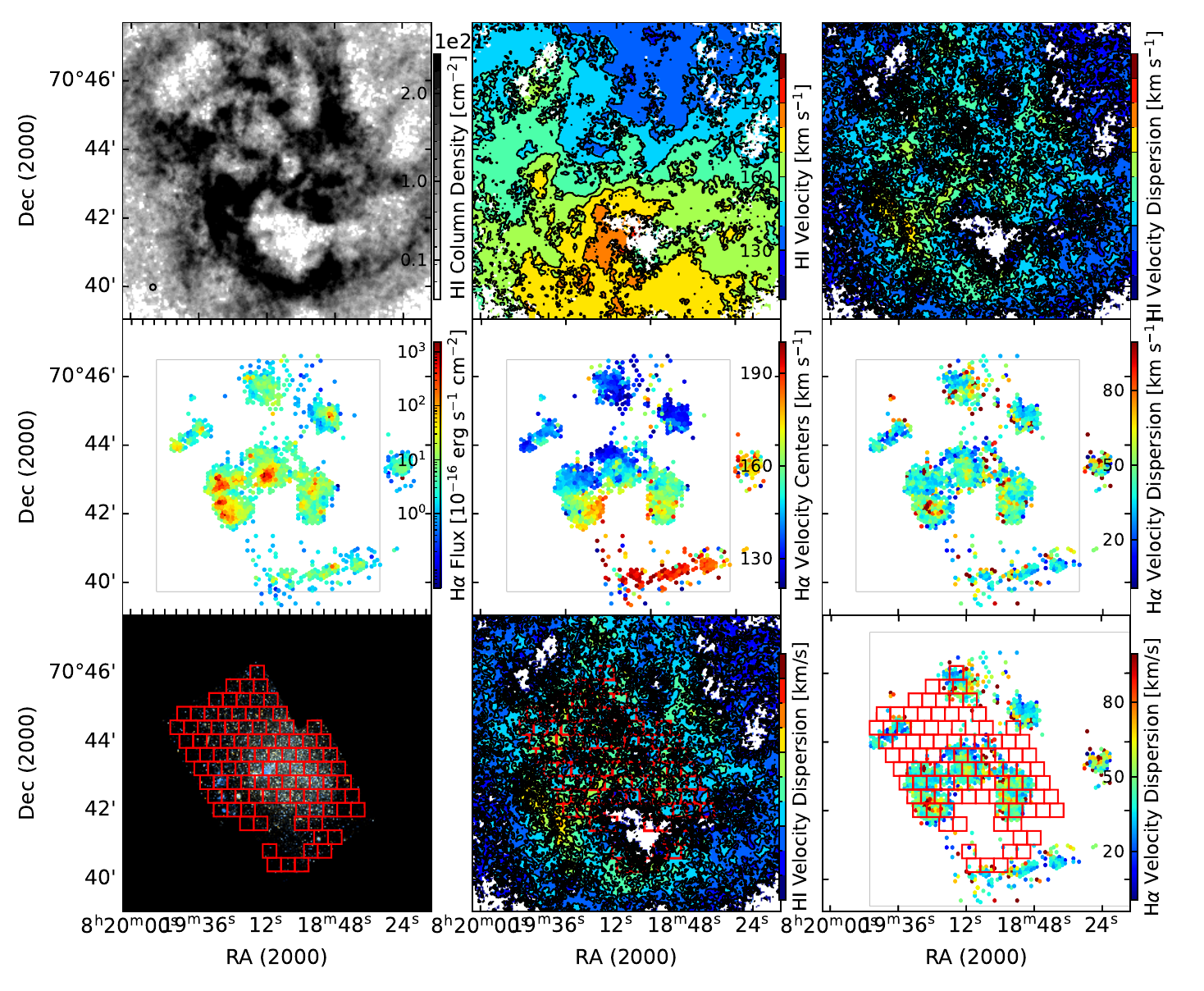}
\caption{\textbf{Holmberg II Top Row} \hi moment maps from VLA observations. Left: \hi column density in 10$^{21}$ hydrogen atoms cm$^{-2}$, Center: \hi velocity map with isovelocity contours spaced every 10 km s$^{-1}$, Right: \hi velocity dispersion map with isovelocity contours at 1.5 km s$^{-1}$ spacing. The beam size (10.73''$\times$10.40'') of the \hi data cube used is shown in the bottom left of the left panel. \textbf{Middle Row} maps from observations with the SparsePak IFU on the WIYN 3.5m telescope, with H$\alpha$ line measurements from PAN and FXCOR.  Left: H$\alpha$ flux on a log scale in units of 10$^{-16}$ erg s$^{-1}$ cm$^{-1}$, Center: H$\alpha$ recessional velocities, Right: H$\alpha$ velocity dispersion ($\sigma_{H\alpha}$) map. Each filled circle corresponds to a fiber's size and position on the sky.  \textbf{Bottom Row} Left: Three color image from \textit{HST} F814W (red), average of F814W and F555W (green), and F555W (blue) observations with ACS, Center: the VLA \hi dispersion map with isovelocity contours in 2.5 km s$^{-1}$ step size, Right: the WIYn 3.5m SparsePak IFU $\sigma_{H\alpha}$ map. Overlaid in red are the outlines of regions used for the analysis. }
\end{figure*}

\begin{figure*}
\figurenum{13}
\plotone{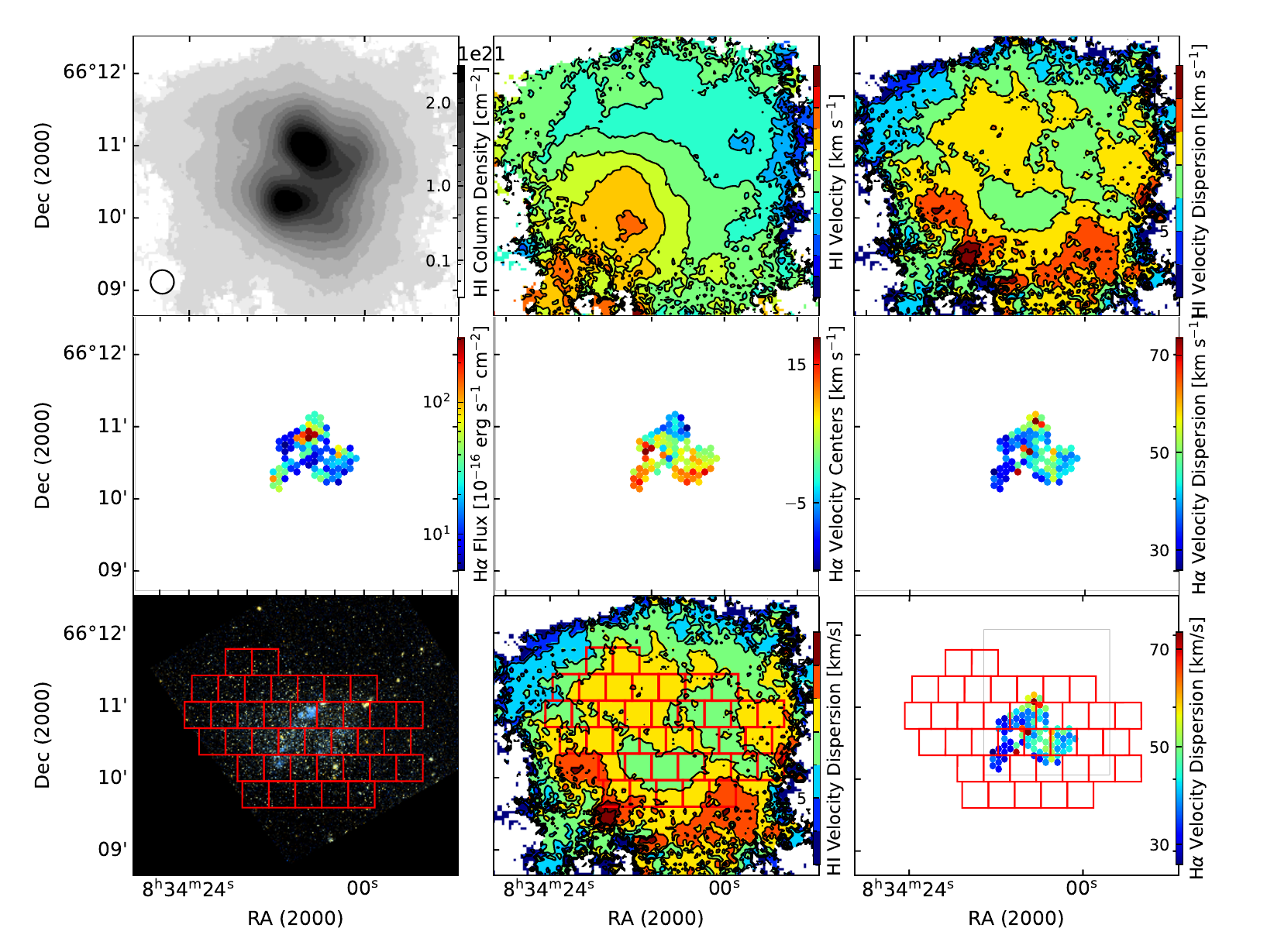}
\caption{\textbf{UGC 04459 Top Row} \hi moment maps from VLA observations. Left: \hi column density in 10$^{21}$ hydrogen atoms cm$^{-2}$, Center: \hi velocity map with isovelocity contours spaced every 5 km s$^{-1}$, Right: \hi velocity dispersion map with isovelocity contours at 2.5 km s$^{-1}$ spacing. The beam size (19.65''$\times$19.39'') of the \hi data cube used is shown in the bottom left of the left panel. \textbf{Middle Row} maps from observations with the SparsePak IFU on the WIYN 3.5m telescope, with H$\alpha$ line measurements from PAN.  Left: H$\alpha$ line flux on a log scale in units of 10$^{-16}$ erg s$^{-1}$ cm$^{-1}$, Center: H$\alpha$ line centers map, Right: H$\alpha$ velocity dispersion ($\sigma_{H\alpha}$) map. Each filled circle corresponds to a fiber's size and position on the sky. \textbf{Bottom Row} Left: Three color image from \textit{HST} F814W (red), average of F814W and F555W (green), and F555W (blue) observations with ACS, Center: \hi dispersion map from VLA observations with isovelocity contours in 2.5 km s$^{-1}$ step size, Right: $\sigma_{H\alpha}$ map from the SparsePak IFU on the WIYN 3.5m telescope. Overlaid on all three panels are the outlines of the regions used for the analysis. }
\end{figure*}

\begin{figure*}
\figurenum{13}
\plotone{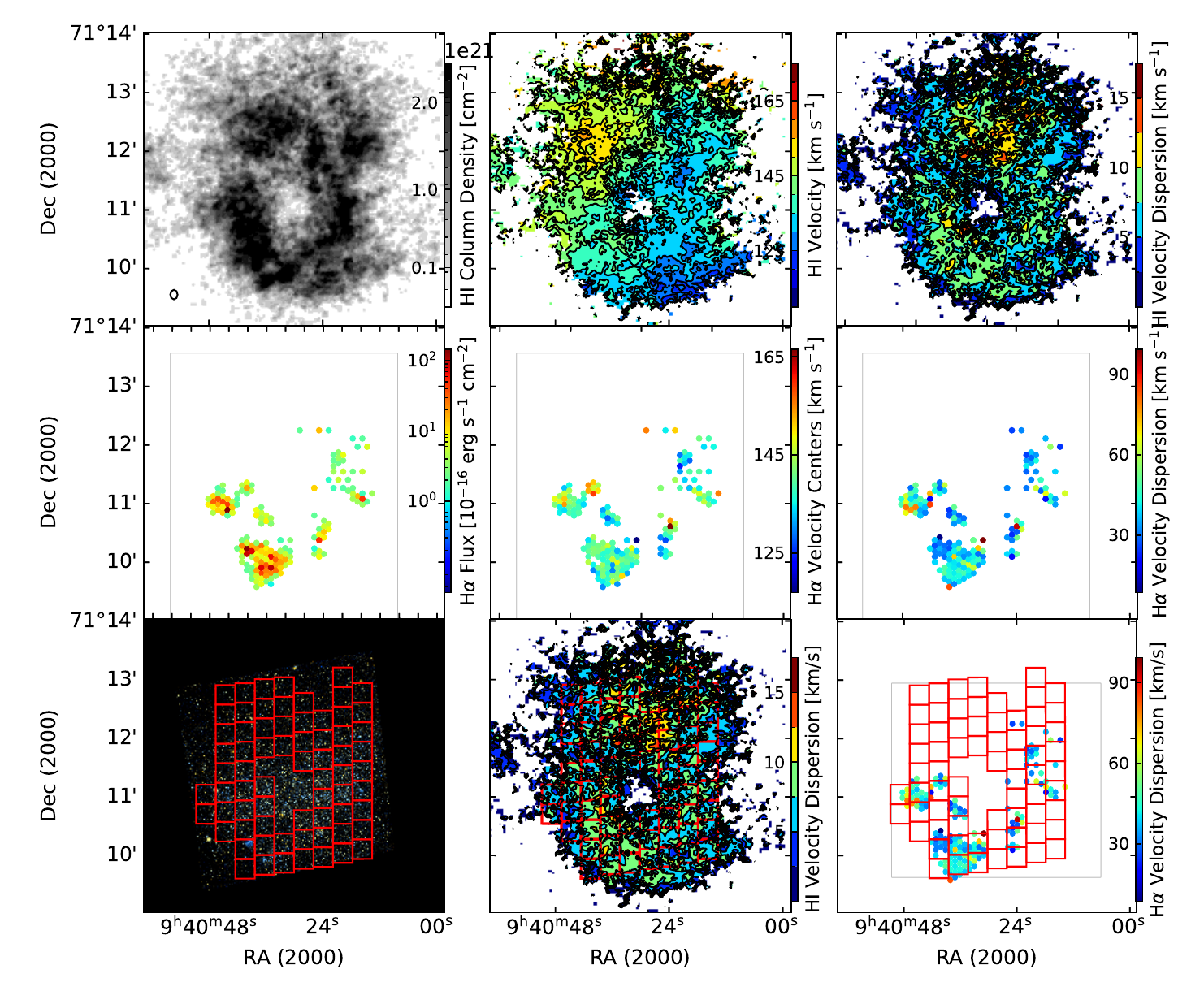}
\caption{\textbf{Holmberg I Top Row} \textbf{Holmberg I} \hi moment maps from VLA observations. Left: \hi column density in 10$^{21}$ hydrogen atoms cm$^{-2}$, Center: \hi velocity map with isovelocity contours spaced every 5 km s$^{-1}$, Right: \hi velocity dispersion map with isovelocity contours at 2.5 km s$^{-1}$ spacing. The beam size (9.77''$\times$7.50'') of the \hi data cube used is shown in the bottom left of the left panel. \textbf{Middle Row} maps from observations with the SparsePak IFU on the WIYN 3.5m telescope, with H$\alpha$ line measurements from PAN.  Left: H$\alpha$ line flux on a log scale in units of 10$^{-16}$ erg s$^{-1}$ cm$^{-1}$, Center: H$\alpha$ line centers map, Right: H$\alpha$ velocity dispersion ($\sigma_{H\alpha}$) map. Each filled circle corresponds to a fiber's size and position on the sky. \textbf{Bottom Row} Left: Three color image from \textit{HST} F814W (red), average of F814W and F555W (green), and F555W (blue) observations with ACS, Center: \hi dispersion map from VLA observations with isovelocity contours in 2.5 km s$^{-1}$ step size, Right: $\sigma_{H\alpha}$ map from the SparsePak IFU on the WIYN 3.5m telescope. Overlaid on all three panels are the outlines of the regions used for the analysis. }
\end{figure*}

\begin{figure*}
\figurenum{13}
\plotone{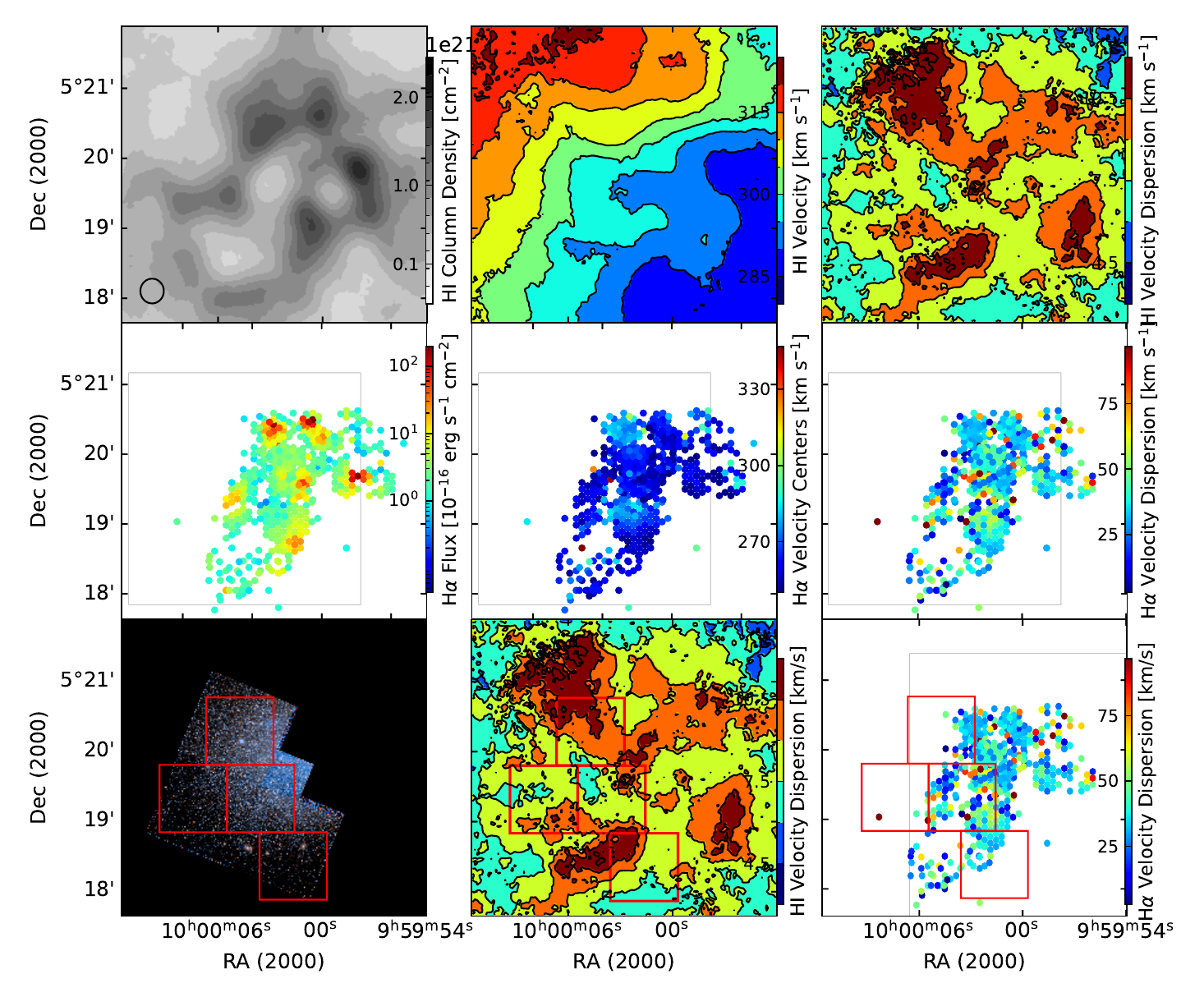}
\caption{\textbf{Sextans B Top Row} \hi moment maps from VLA observations. Left: \hi column density in 10$^{21}$ hydrogen atoms cm$^{-2}$, Center: \hi velocity map with isovelocity contours spaced every 5 km s$^{-1}$, Right: \hi velocity dispersion map with isovelocity contours at 1.5 km s$^{-1}$ spacing. The beam size (21.26''$\times$20.11'') of the \hi data cube used is shown in the bottom left of the left panel. \textbf{Middle Row} maps from observations with the SparsePak IFU on the WIYN 3.5m telescope, with H$\alpha$ line measurements from PAN.  Left: H$\alpha$ line flux on a log scale in units of 10$^{-16}$ erg s$^{-1}$ cm$^{-1}$, Center: H$\alpha$ line centers map, Right: H$\alpha$ velocity dispersion ($\sigma_{H\alpha}$) map. Each filled circle corresponds to a fiber's size and position on the sky.  \textbf{Top Row} Left: Two color image from \textit{HST} F814W (red), average of F814W and F606W (green), and F606W (blue) observations with WFPC2, Center: \hi dispersion map from VLA observations with isovelocity contours in 3 km s$^{-1}$ step size, Right: $\sigma_{H\alpha}$ map from the SparsePak IFU on the WIYN 3.5m telescope. Overlaid on all three panels are the outlines of the regions used for the analysis. }
\end{figure*}

\begin{figure*}
\figurenum{13}
\plotone{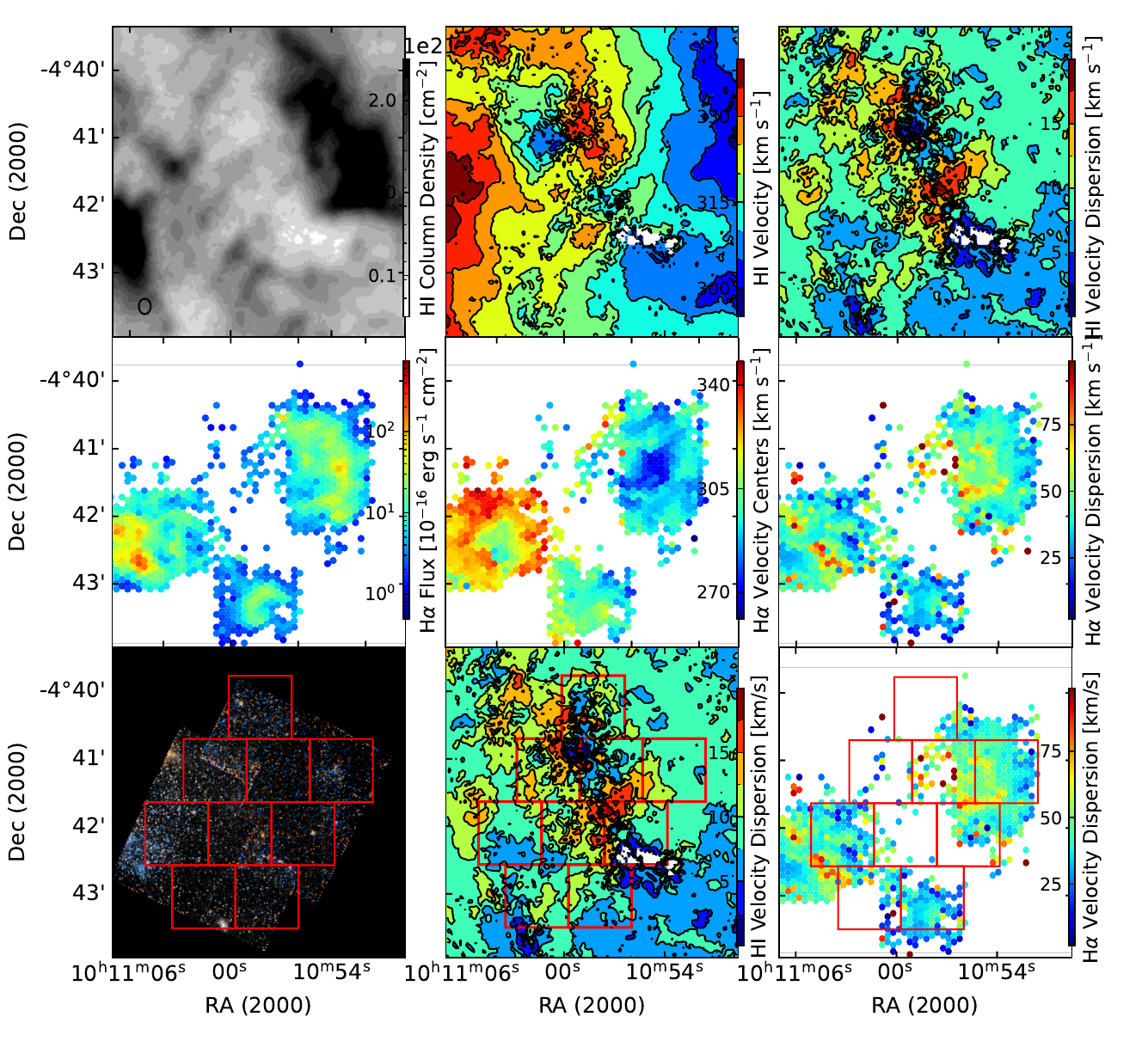}
\caption{\textbf{Sextans A Top Row} \hi moment maps from VLA observations. Left: \hi column density in 10$^{21}$ hydrogen atoms cm$^{-2}$, Center: \hi velocity map with isovelocity contours spaced every 5 km s$^{-1}$, Right: \hi velocity dispersion map with isovelocity contours at 2.5 km s$^{-1}$ spacing. The beam size (13.67''$\times$10.68'') of the \hi data cube used is shown in the bottom left of the left panel. \textbf{Middle Row} maps from observations with the SparsePak IFU on the WIYN 3.5m telescope, with H$\alpha$ line measurements from PAN.  Left: H$\alpha$ line flux on a log scale in units of 10$^{-16}$ erg s$^{-1}$ cm$^{-1}$, Center: H$\alpha$ line centers map, Right: H$\alpha$ velocity dispersion ($\sigma_{H\alpha}$) map. Each filled circle corresponds to a fiber's size and position on the sky. \textbf{Bottom Row} Left: Three color image from \textit{HST} F814W (red), average of F814W and F555W (green), and F555W (blue) observations with WFPC2, Center: \hi dispersion map from VLA observations with isovelocity contours in 2.5 km s$^{-1}$ step size, Right: $\sigma_{H\alpha}$ map from the SparsePak IFU on the WIYN 3.5m telescope. Overlaid on all three panels are the outlines of the regions used for the analysis. }
\end{figure*}

\begin{figure*}
\figurenum{13}
\plotone{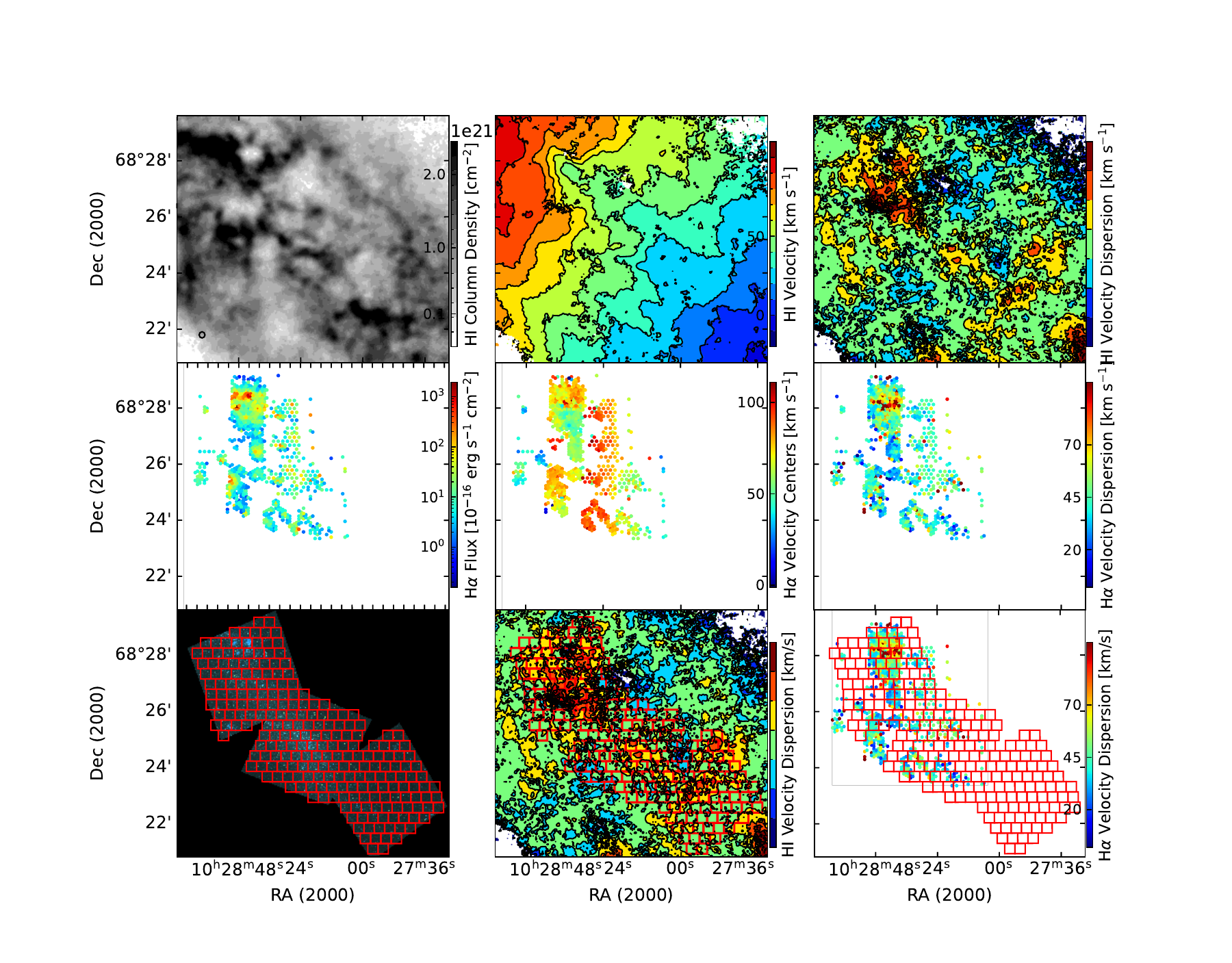}
\caption{\textbf{IC 2574 Top Row} \hi moment maps from VLA observations Left: \hi column density in 10$^{21}$ hydrogen atoms cm$^{-2}$, Center: \hi velocity map with isovelocity contours spaces every 10 km s$^{-1}$, Right: \hi velocity dispersion map with isovelocity contours at 2.5 km s$^{-1}$ spacing. The beam size (13.18''$\times$12.45'') of the \hi data cube used is shown in the bottom left of the left panel. \textbf{Middle Row} maps from observations with the SparsePak IFU on the WIYN 3.5m telescope, with H$\alpha$ line measurements from PAN.  Left: H$\alpha$ line flux on a log scale in units of 10$^{-16}$ erg s$^{-1}$ cm$^{-1}$, Center: H$\alpha$ line centers map, Right: H$\alpha$ velocity dispersion ($\sigma_{H\alpha}$) map. Each filled circle corresponds to a fiber's size and position on the sky. \textbf{Bottom Row} Left: Three color image from \textit{HST} F814W (red), average of F814W and F555W (green), and F555W (blue) observations with ACS, Center: \hi dispersion map from VLA observations with isovelocity contours in 2.5 km s$^{-1}$ step size, Right: $\sigma_{H\alpha}$ map from the SparsePak IFU on the WIYN 3.5m telescope. Overlaid on all three panels are the outlines of the regions used for the analysis. }
\end{figure*}

\begin{figure*}
\figurenum{13}
\plotone{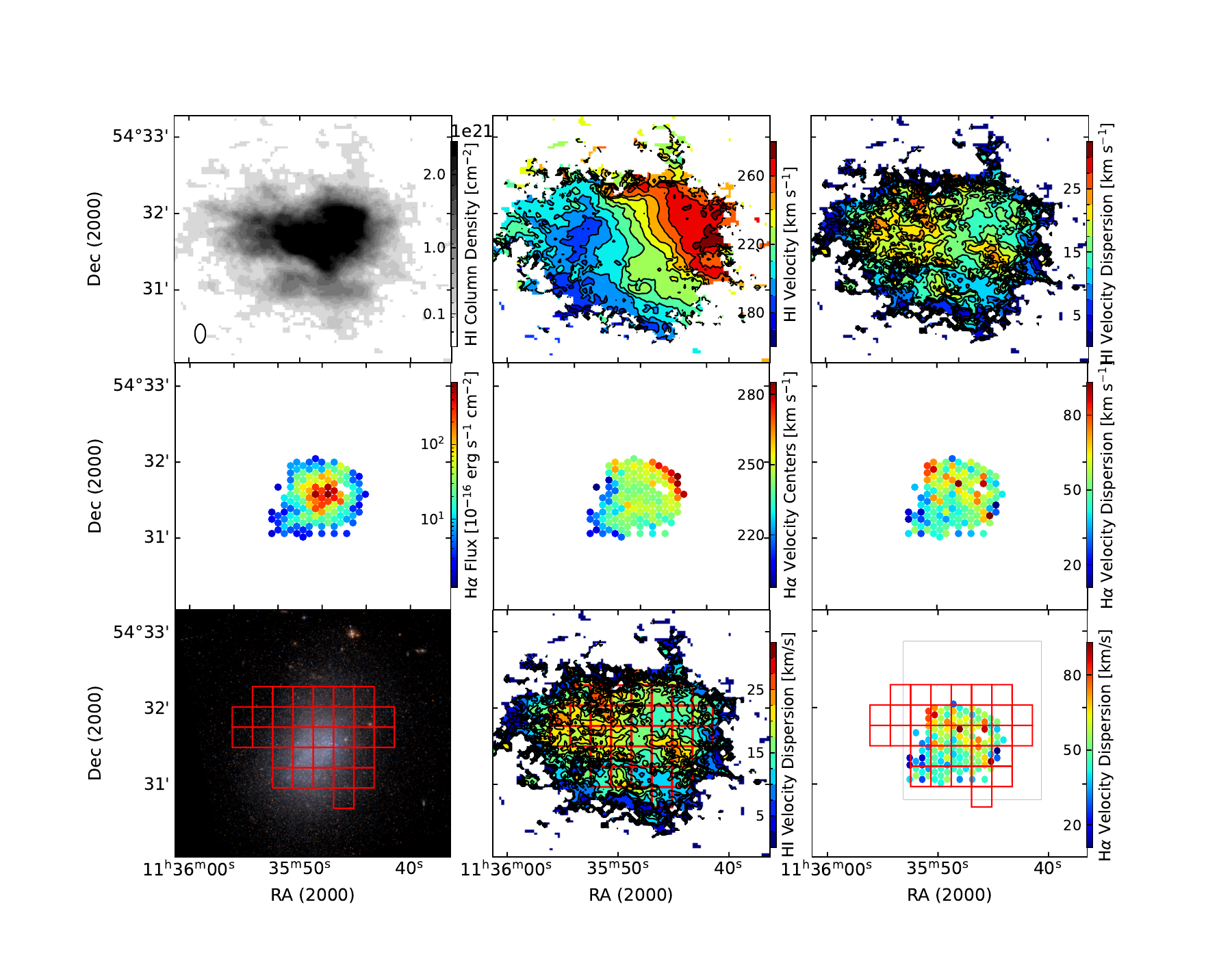}
\caption{\textbf{NGC 3738 Top Row} \hi moment maps from VLA observations. Left: \hi column density in 10$^{21}$ hydrogen atoms cm$^{-2}$, Center: \hi velocity map with isovelocity contours spaced every 10 km s$^{-1}$, Right: \hi velocity dispersion map with isovelocity contours at 5 km s$^{-1}$ spacing. The beam size (10.73''$\times$10.40'') of the \hi data cube used is shown in the bottom left of the left panel. \textbf{Middle Row} maps from observations with the SparsePak IFU on the WIYN 3.5m telescope, with H$\alpha$ line measurements from PAN.  Left: H$\alpha$ line flux on a log scale in units of 10$^{-16}$ erg s$^{-1}$ cm$^{-1}$, Center: H$\alpha$ line centers map, Right: H$\alpha$ velocity dispersion ($\sigma_{H\alpha}$) map. Each filled circle corresponds to a fiber's size and position on the sky. \textbf{Bottom Row} Left: Three color image from \textit{HST} 814W (red), average of F814W and F606W (green), and F606W (blue) observations with ACS, Center: \hi dispersion map from VLA observations with isovelocity contours in 5 km s$^{-1}$ step size, Right: $\sigma_{H\alpha}$ map from the SparsePak IFU on the WIYN 3.5m telescope. Overlaid on all three panels are the outlines of the regions used for the analysis. }
\end{figure*}

\begin{figure*}
\figurenum{13}
\plotone{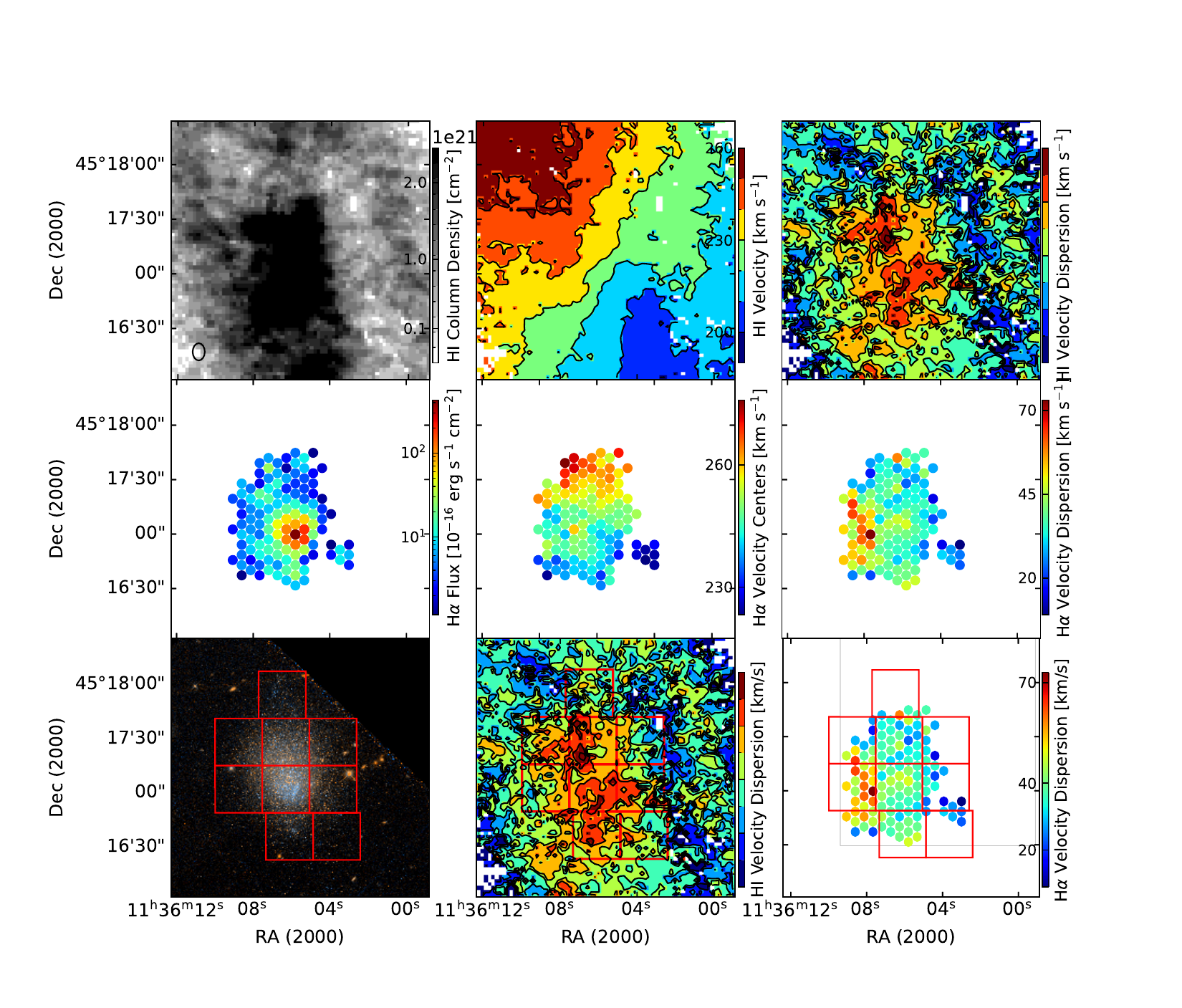}
\caption{\textbf{NGC 3741 Top Row} \hi moment maps from VLA observations. Left: \hi column density in 10$^{21}$ hydrogen atoms cm$^{-2}$, Center: \hi velocity map with isovelocity contours spaced every 10 km s$^{-1}$, Right: \hi velocity dispersion map with isovelocity contours at 1.5 km s$^{-1}$ spacing. The beam size (9.47''$\times$6.66'') of the \hi data cube used is shown in the bottom left of the left panel.  \textbf{Middle Row} maps from observations with the SparsePak IFU on the WIYN 3.5m telescope, with H$\alpha$ line measurements from PAN.  Left: H$\alpha$ line flux on a log scale in units of 10$^{-16}$ erg s$^{-1}$ cm$^{-1}$, Center: H$\alpha$ line centers map, Right: H$\alpha$ velocity dispersion ($\sigma_{H\alpha}$) map. Each filled circle corresponds to a fiber's size and position on the sky. \textbf{Bottom Row} Left: Three color image from \textit{HST} F814W (red), average of F814W and F475W (green), and F475W (blue) observations with ACS, Center: \hi dispersion map from VLA observations with isovelocity contours in 1.5 km s$^{-1}$ step size, Right: $\sigma_{H\alpha}$ map from the SparsePak IFU on the WIYN 3.5m telescope. Overlaid on all three panels are the outlines of the regions used for the analysis. }
\end{figure*}

\begin{figure*}
\figurenum{13}
\plotone{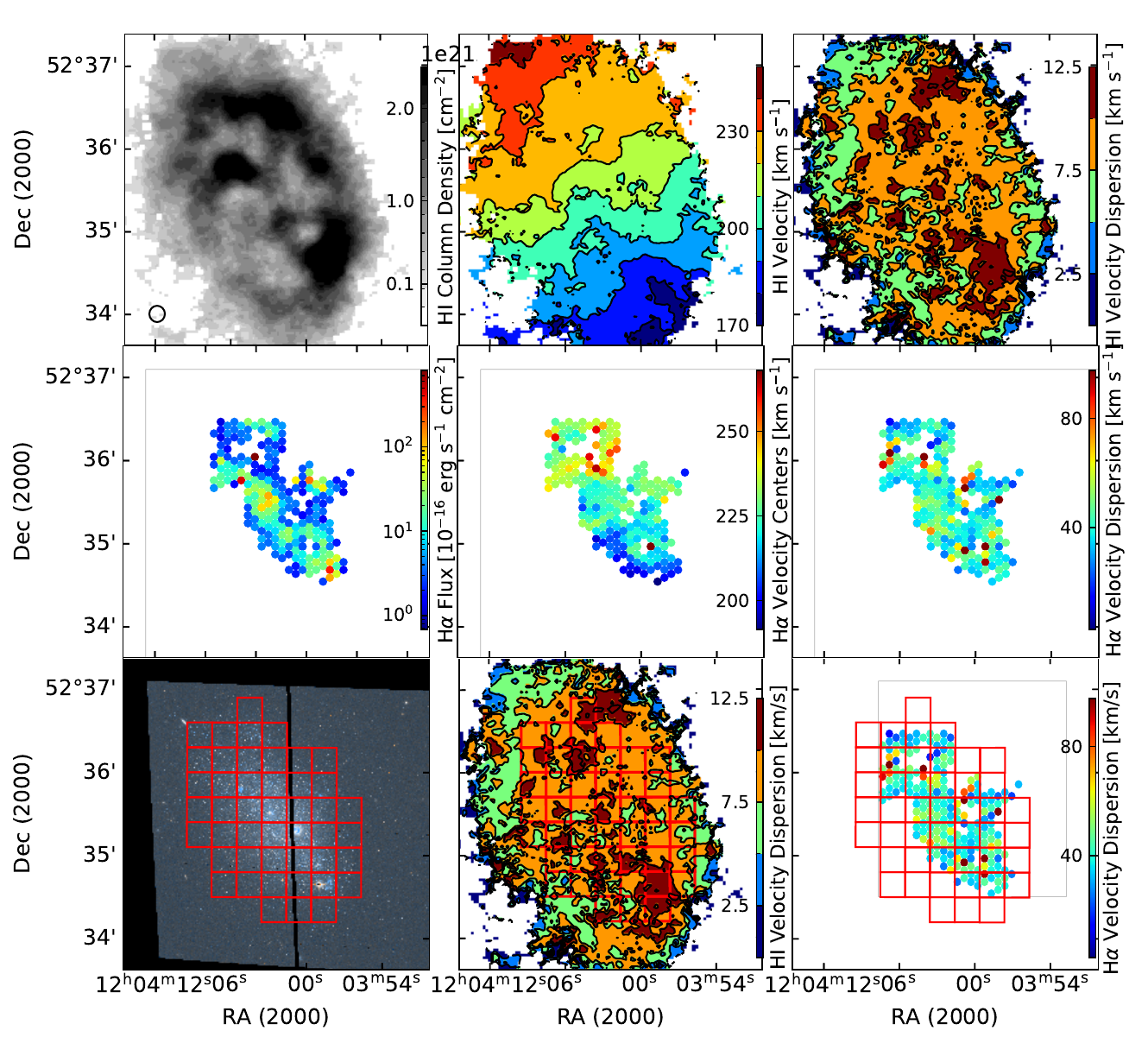}
\caption{\textbf{NGC 4068 Top Row} \hi moment maps from VLA observations Left: \hi column density in 10$^{21}$ hydrogen atoms cm$^{-2}$, Center: \hi velocity map with isovelocity contours spaces every 10 km s$^{-1}$, Right: \hi velocity dispersion map with isovelocity contours at 2.5 km s$^{-1}$ spacing. The beam size (11.83''$\times$11.29'') of the \hi data cube used is shown in the bottom left of the left panel. \textbf{Middle Row} maps from observations with the SparsePak IFU on the WIYN 3.5m telescope, with H$\alpha$ line measurements from PAN.  Left: H$\alpha$ line flux on a log scale in units of 10$^{-16}$ erg s$^{-1}$ cm$^{-1}$, Center: H$\alpha$ line centers map, Right: H$\alpha$ velocity dispersion ($\sigma_{H\alpha}$) map. Each filled circle corresponds to a fiber's size and position on the sky. \textbf{Bottom Row} Left: Three color image from \textit{HST} F814W (red), average of F814W and F606W (green), and F606W (blue) observations with ACS, Center: \hi dispersion map from VLA observations with isovelocity contours in 2.5 km s$^{-1}$ step size, Right: $H\alpha$ FWHM map from the SparsePak IFU on the WIYN 3.5m telescope. Overlaid on all three panels are the outlines of the regions used for the analysis.}
\end{figure*}

\begin{figure*}
\figurenum{13}
\plotone{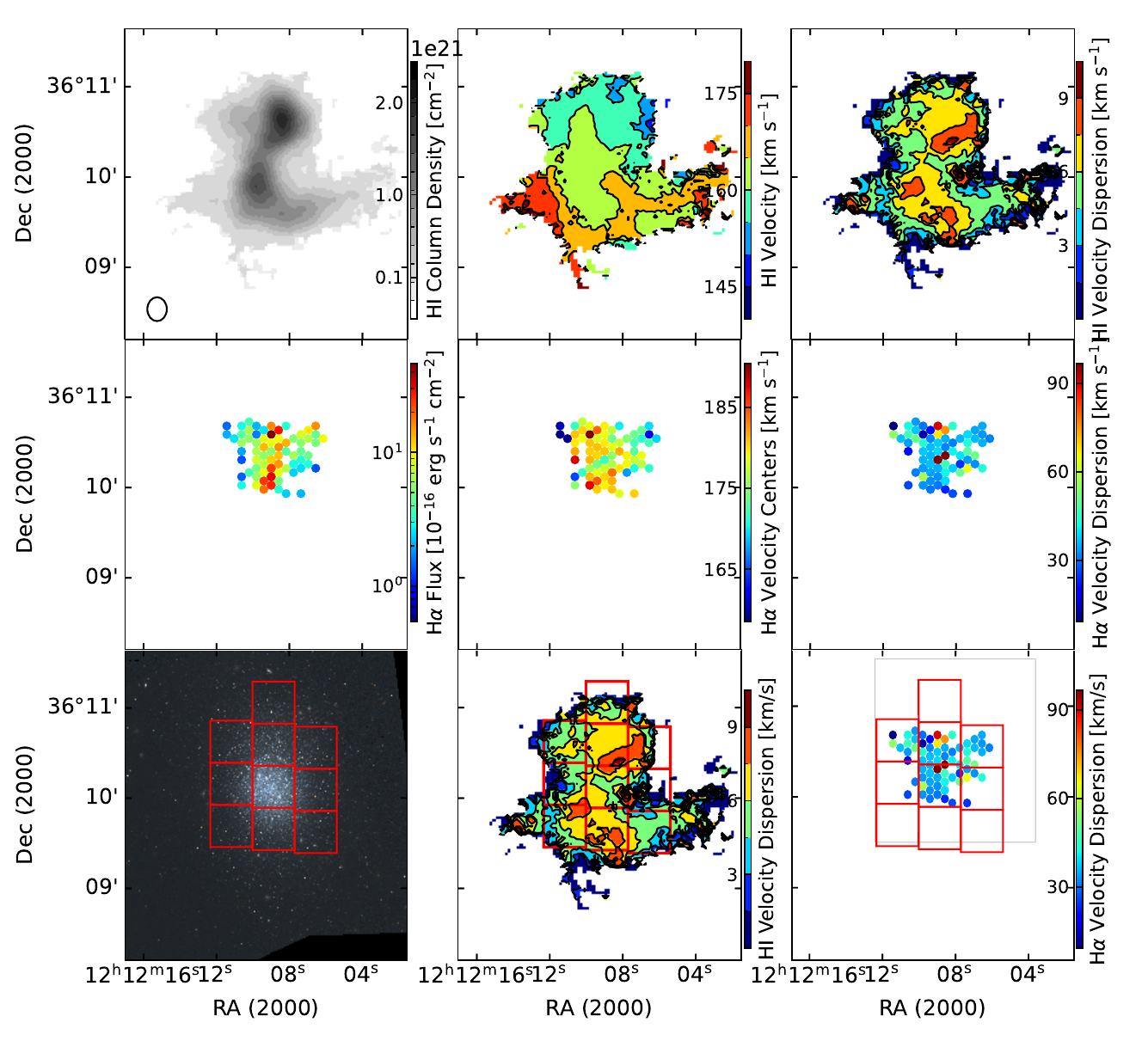}
\caption{\textbf{NGC 4163 Top Row} \hi moment maps from VLA observations Left: \hi column density in 10$^{21}$ hydrogen atoms cm$^{-2}$, Center: \hi velocity map with isovelocity contours spaces every 5 km s$^{-1}$, Right: \hi velocity dispersion map with isovelocity contours at 2 km s$^{-1}$ spacing. The beam size (15.911''$\times$12.941'') of the \hi data cube used is shown in the bottom left of the left panel. \textbf{Middle Row} maps from observations with the SparsePak IFU on the WIYN 3.5m telescope, with H$\alpha$ line measurements from PAN.  Left: H$\alpha$ line flux on a log scale in units of 10$^{-16}$ erg s$^{-1}$ cm$^{-1}$, Center: H$\alpha$ line centers map, Right: H$\alpha$ velocity dispersion ($\sigma_{H\alpha}$) map. Each filled circle corresponds to a fiber's size and position on the sky. \textbf{Bottom Row} Left: Three color image from \textit{HST} F814W (red), average of F814W and F606W (green), and F606W (blue) observations with ACS, Center: \hi dispersion map from VLA observations with isovelocity contours in 2 km s$^{-1}$ step size, Right: $H\alpha$ FWHM map from the SparsePak IFU on the WIYN 3.5m telescope. Overlaid on all three panels are the outlines of the regions used for the analysis. }
\end{figure*}

\begin{figure*}
\figurenum{13}
\plotone{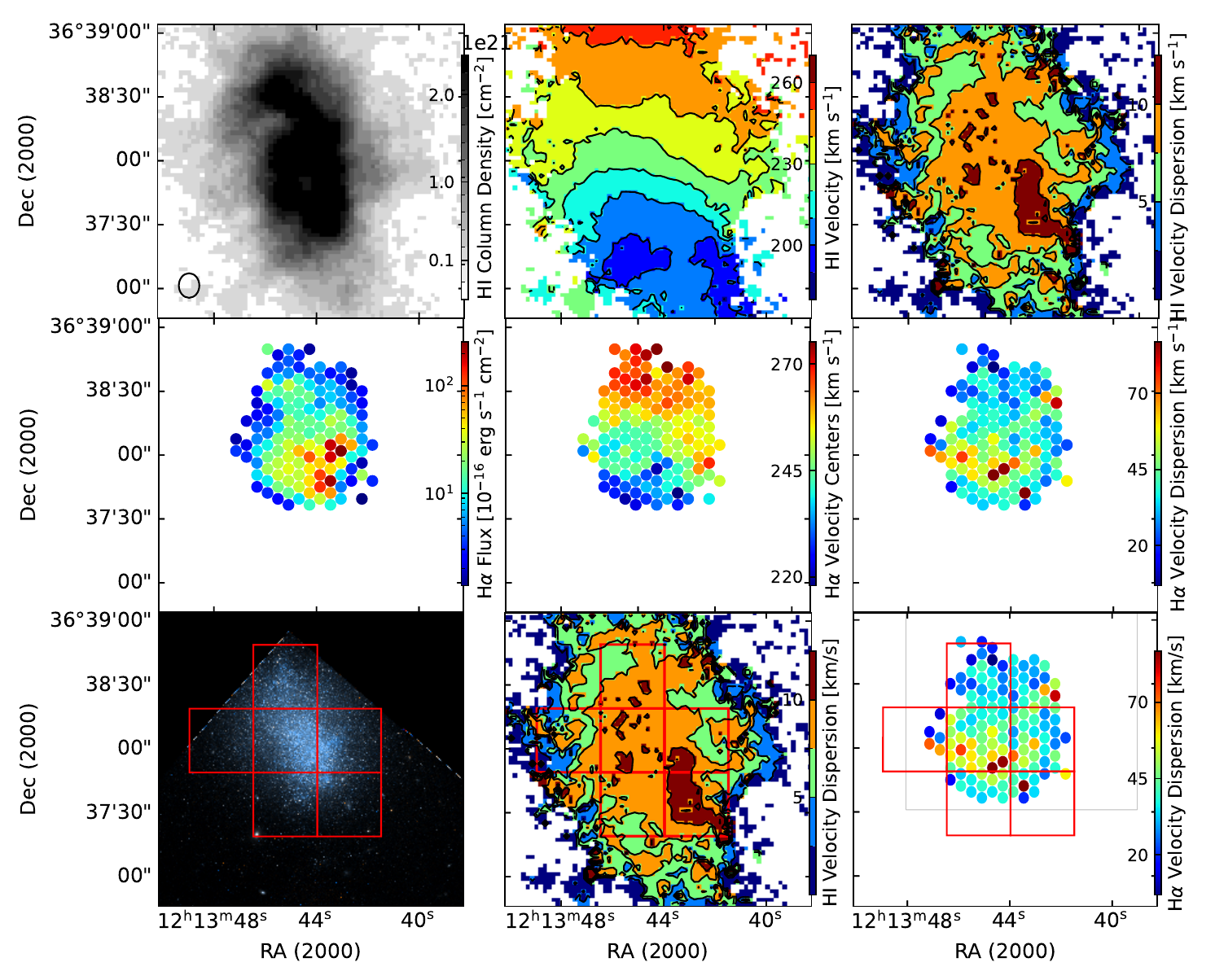}
\caption{\textbf{NGC 4190 Top Row} \hi moment maps from VLA observations. Left: \hi column density in 10$^{21}$ hydrogen atoms cm$^{-2}$, Center: \hi velocity map with isovelocity contours spaced every 10 km s$^{-1}$, Right: \hi velocity dispersion map with isovelocity contours at 2.5 km s$^{-1}$ spacing. The beam size (11.54''$\times$9.62'') of the \hi data cube used is shown in the bottom left of the left panel. \textbf{Middle Row} maps from observations with the SparsePak IFU on the WIYN 3.5m telescope, with H$\alpha$ line measurements from PAN.  Left: H$\alpha$ line flux on a log scale in units of 10$^{-16}$ erg s$^{-1}$ cm$^{-1}$, Center: H$\alpha$ line centers map, Right: H$\alpha$ velocity dispersion ($\sigma_{H\alpha}$) map. Each filled circle corresponds to a fiber's size and position on the sky. \textbf{Bottom Row} Left: Three color image from \textit{HST} F814W (red), average of F814W and F606W (green), and F606W (blue) observations with ACS, Center: \hi dispersion map from VLA observations with isovelocity contours in 2.5 km s$^{-1}$ step size, Right: $\sigma_{H\alpha}$ map from the SparsePak IFU on the WIYN 3.5m telescope. Overlaid on all three panels are the outlines of the regions used for the analysis.}
\end{figure*}

\begin{figure*}
\figurenum{13}
\plotone{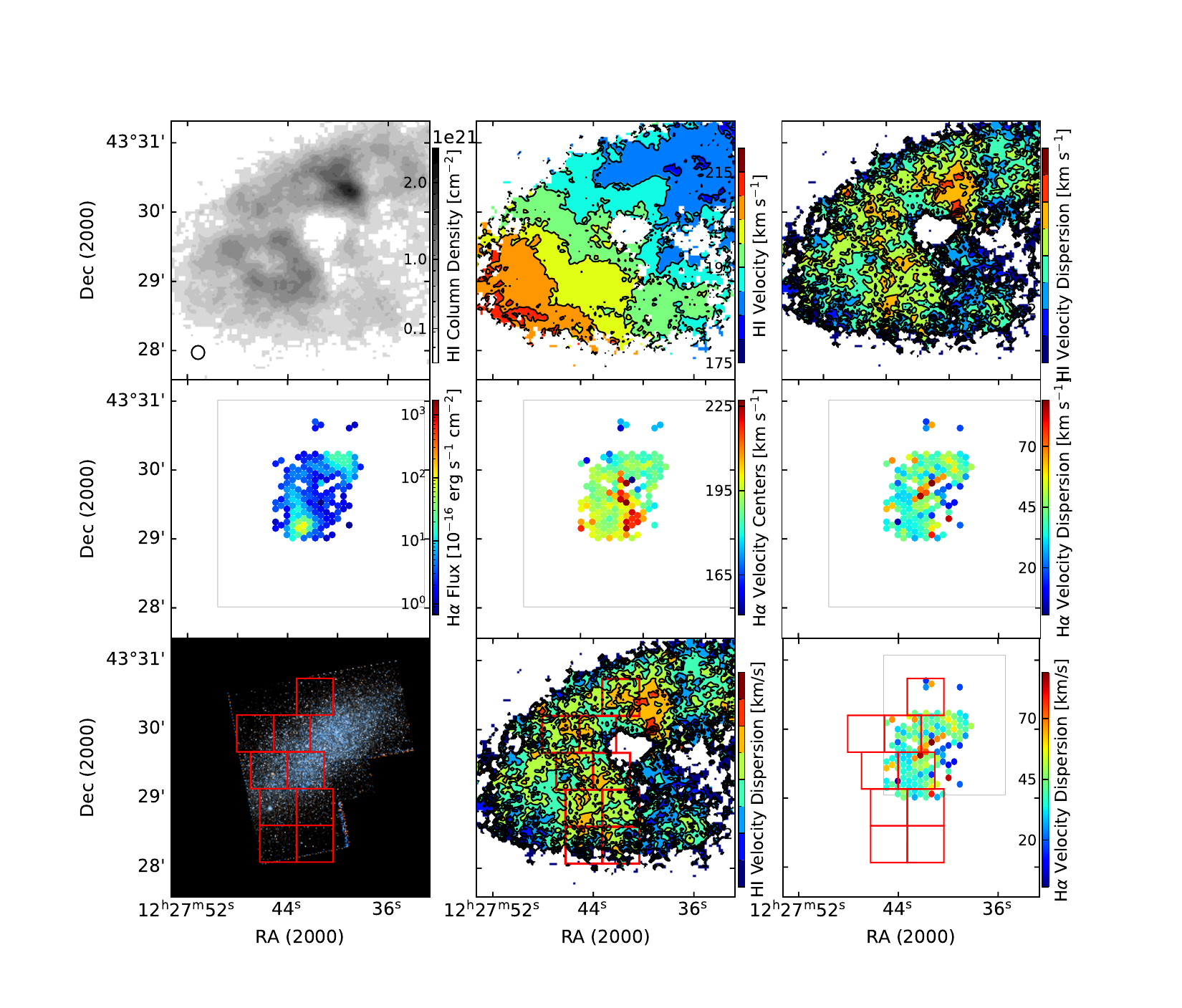}
\caption{\textbf{UGC 7577 Top Row} hi moment maps from VLA observations. Left: \hi column density in 10$^{21}$ hydrogen atoms cm$^{-2}$, Center: \hi velocity map with isovelocity contours spaced every 5 km s$^{-1}$, Right: \hi velocity dispersion map with isovelocity contours at 1 km s$^{-1}$ spacing. The beam size (11.84''$\times$11.30'') of the \hi data cube used is shown in the bottom left of the left panel. \textbf{Middle Row} maps from observations with the SparsePak IFU on the WIYN 3.5m telescope, with H$\alpha$ line measurements from PAN.  Left: H$\alpha$ line flux on a log scale in units of 10$^{-16}$ erg s$^{-1}$ cm$^{-1}$, Center: H$\alpha$ line centers map, Right: H$\alpha$ velocity dispersion ($\sigma_{H\alpha}$) map. Each filled circle corresponds to a fiber's size and position on the sky. \textbf{Bottom Row} Left: Three color image from \textit{HST} F814W (red), average of F814W and F606W (green), and F606W (blue) observations with WFPC2, Center: \hi dispersion map from VLA observations with isovelocity contours in 1 km s$^{-1}$ step size, Right: $\sigma_{H\alpha}$ map from the SparsePak IFU on the WIYN 3.5m telescope. Overlaid on all three panels are the outlines of the regions used for the analysis. }
\end{figure*}

\begin{figure*}
\figurenum{13}
\plotone{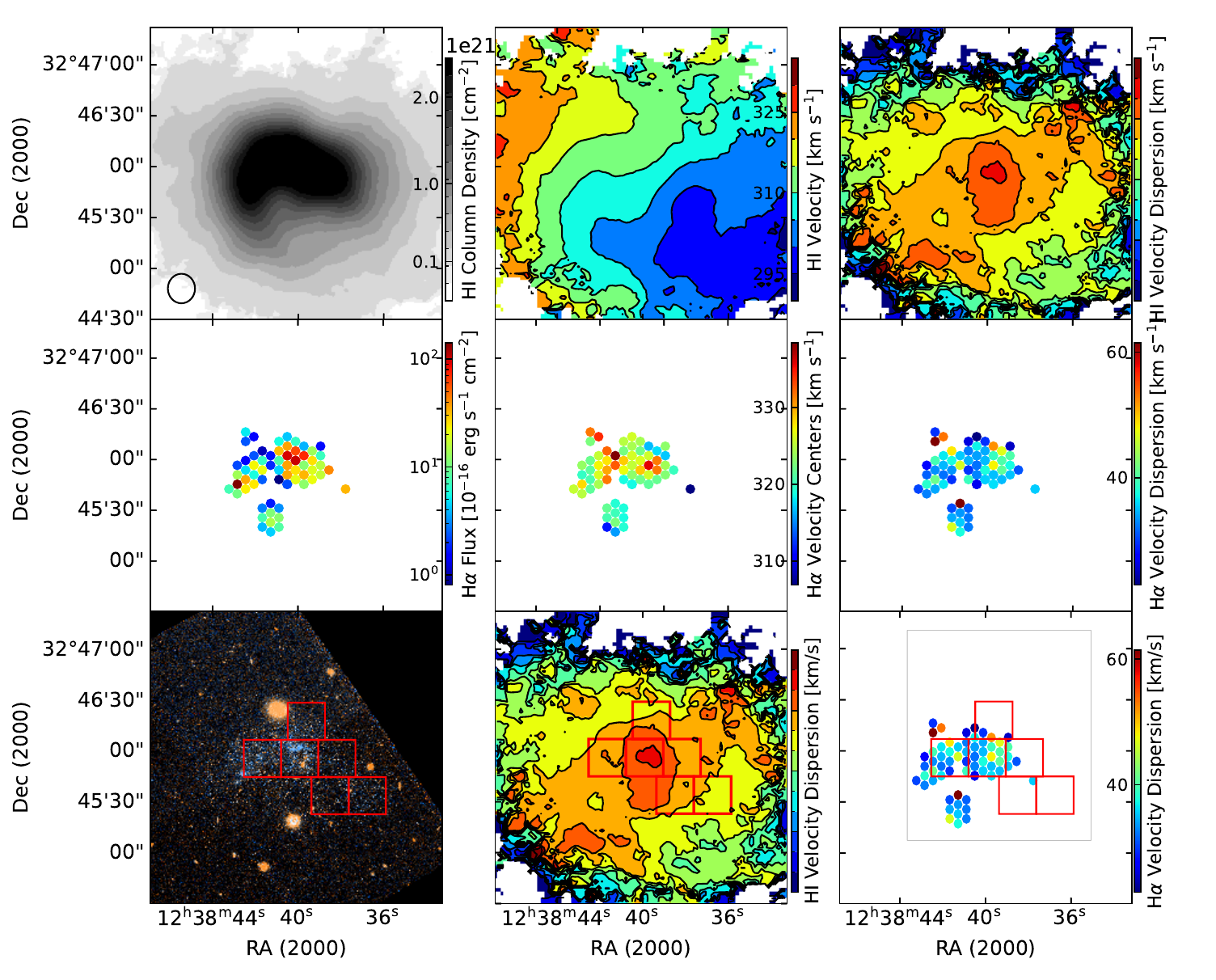}
\caption{\textbf{UGCA 292 Top Row} \hi moment maps from VLA observations. Left: \hi column density in 10$^{21}$ hydrogen atoms cm$^{-2}$, Center: \hi velocity map with isovelocity contours spaced every 5 km s$^{-1}$, Right: \hi velocity dispersion map with isovelocity contours at 1 km s$^{-1}$ spacing. The beam size (17.40''$\times$16.03'') of the \hi data cube used is shown in the bottom left of the left panel. \textbf{Middle Row} maps from observations with the SparsePak IFU on the WIYN 3.5m telescope, with H$\alpha$ line measurements from PAN.  Left: H$\alpha$ line flux on a log scale in units of 10$^{-16}$ erg s$^{-1}$ cm$^{-1}$, Center: H$\alpha$ line centers map, Right: H$\alpha$ velocity dispersion ($\sigma_{H\alpha}$) map. Each filled circle corresponds to a fiber's size and position on the sky. \textbf{Bottom Row} Left: Three color image from \textit{HST} F814W (red), average of F814W and F475W (green), and F475W (blue) observations with ACS, Center: \hi dispersion map from VLA observations with isovelocity contours in 1 km s$^{-1}$ step size, Right: $\sigma_{H\alpha}$ map from the SparsePak IFU on the WIYN 3.5m telescope. Overlaid on all three panels are the outlines of the regions used for the analysis. }
\end{figure*}

\begin{figure*}
\figurenum{13}
\plotone{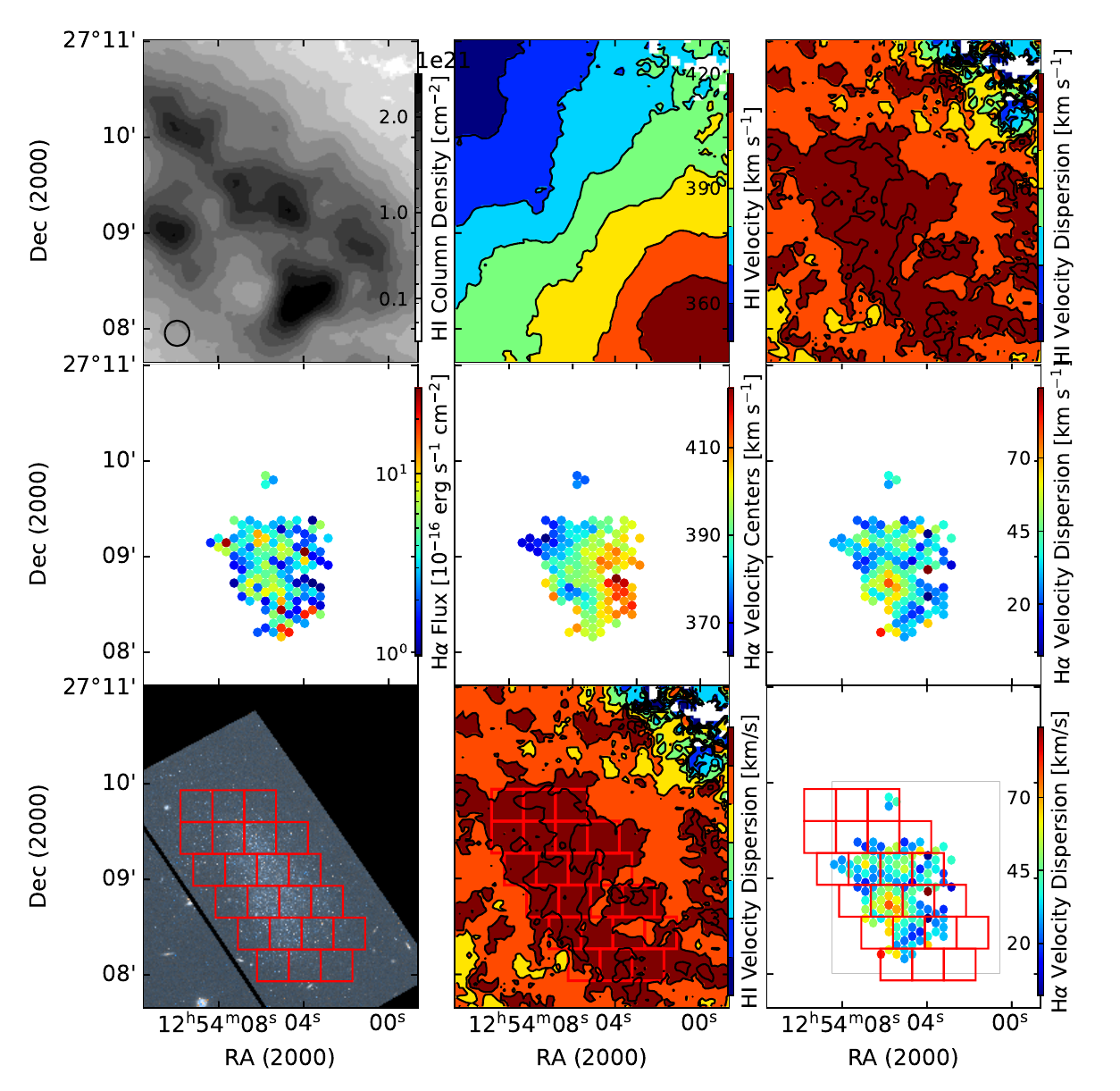}
\caption{\textbf{UGC 8024 Top Row} \hi moment maps from VLA observations. Left: \hi column density in 10$^{21}$ hydrogen atoms cm$^{-2}$, Center: \hi velocity map with isovelocity contours spaced every 10 km s$^{-1}$, Right: \hi velocity dispersion map with isovelocity contours at 1.5 km s$^{-1}$ spacing. The beam size (10.73''$\times$10.40'') of the \hi data cube used is shown in the bottom left of the left panel.  \textbf{Middle Row} maps from observations with the SparsePak IFU on the WIYN 3.5m telescope, with H$\alpha$ line measurements from PAN.  Left: H$\alpha$ line flux on a log scale in units of 10$^{-16}$ erg s$^{-1}$ cm$^{-1}$, Center: H$\alpha$ line centers map, Right: H$\alpha$ velocity dispersion ($\sigma_{H\alpha}$) map. Each filled circle corresponds to a fiber's size and position on the sky. \textbf{Bottom Row} Left: Three color image from \textit{HST} F814W (red), average of F814W and F606W (green), and F606W (blue) observations with ACS, Center: \hi dispersion map from VLA observations with isovelocity contours in 1.5 km s$^{-1}$ step size, Right: $\sigma_{H\alpha}$ map from the SparsePak IFU on the WIYN 3.5m telescope. Overlaid on all three panels are the outlines of the regions used for the analysis.}
\end{figure*}

\begin{figure*}
\figurenum{13}
\plotone{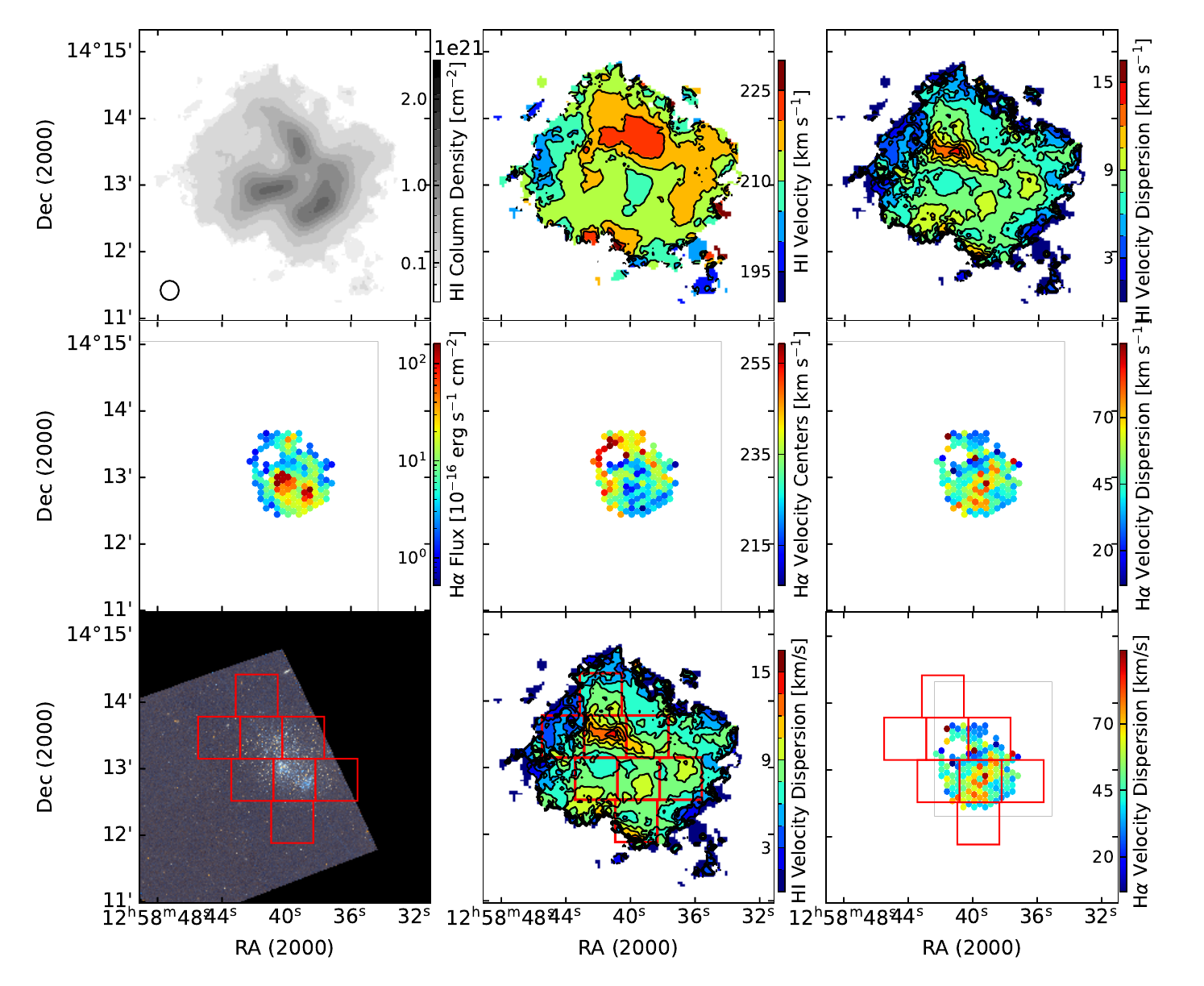}
\caption{\textbf{GR8 Top Row} \hi moment maps from VLA observations. Left: \hi column density in 10$^{21}$ hydrogen atoms cm$^{-2}$, Center: \hi velocity map with isovelocity contours spaced every 5 km s$^{-1}$, Right: \hi velocity dispersion map with isovelocity contours at 1.5 km s$^{-1}$ spacing. The beam size (17.43''$\times$16.50'') of the \hi data cube used is shown in the bottom left of the left panel. \textbf{Middle Row} maps from observations with the SparsePak IFU on the WIYN 3.5m telescope, with H$\alpha$ line measurements from PAN.  Left: H$\alpha$ line flux on a log scale in units of 10$^{-16}$ erg s$^{-1}$ cm$^{-1}$, Center: H$\alpha$ line centers map, Right: H$\alpha$ velocity dispersion ($\sigma_{H\alpha}$) map. Each filled circle corresponds to a fiber's size and position on the sky. \textbf{Bottom Row} Left: Three color image from \textit{HST} F814W (red), average of F814W and F475W (green), and F475W (blue) observations with ACS, Center: \hi dispersion map from VLA observations with isovelocity contours in 1.5 km s$^{-1}$ step size, Right: $\sigma_{H\alpha}$ map from the SparsePak IFU on the WIYN 3.5m telescope. Overlaid on all three panels are the outlines of the regions used for the analysis.}
\end{figure*}

\begin{figure*}
\figurenum{13}
\plotone{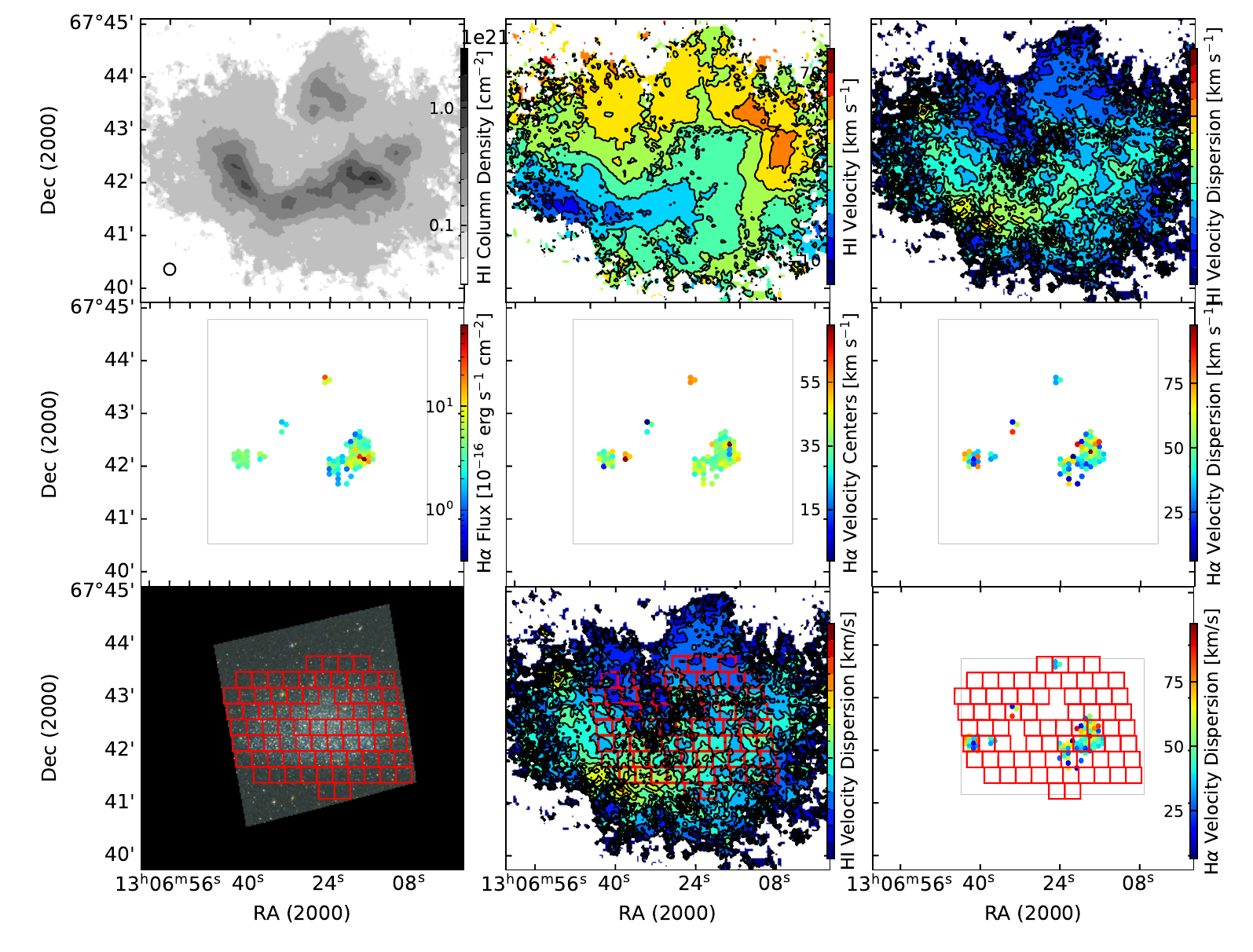}
\caption{\textbf{UGC 8201 Top Row} \hi moment maps from VLA observations. Left: \hi column density in 10$^{21}$ hydrogen atoms cm$^{-2}$, Center: \hi velocity map with isovelocity contours spaced every 10 km s$^{-1}$, Right: \hi velocity dispersion map with isovelocity contours at 2.5 km s$^{-1}$ spacing. The beam size (13.48''$\times$13.12'') of the \hi data cube used is shown in the bottom left of the left panel.  \textbf{Middle Row} maps from observations with the SparsePak IFU on the WIYN 3.5m telescope, with H$\alpha$ line measurements from PAN.  Left: H$\alpha$ line flux on a log scale in units of 10$^{-16}$ erg s$^{-1}$ cm$^{-1}$, Center: H$\alpha$ line centers map, Right: H$\alpha$ velocity dispersion ($\sigma_{H\alpha}$) map. Each filled circle corresponds to a fiber's size and position on the sky. \textbf{Bottom Row} Left: Three color image from \textit{HST} F814W (red), average of F814W and F555W (green), and F555W (blue) observations with ACS, Center: \hi dispersion map from VLA observations with isovelocity contours in 2.5 km s$^{-1}$ step size, Right: $\sigma_{H\alpha}$ map from the SparsePak IFU on the WIYN 3.5m telescope. Overlaid on all three panels are the outlines of the regions used for the analysis.}
\end{figure*}

\begin{figure*}
\figurenum{13}
\plotone{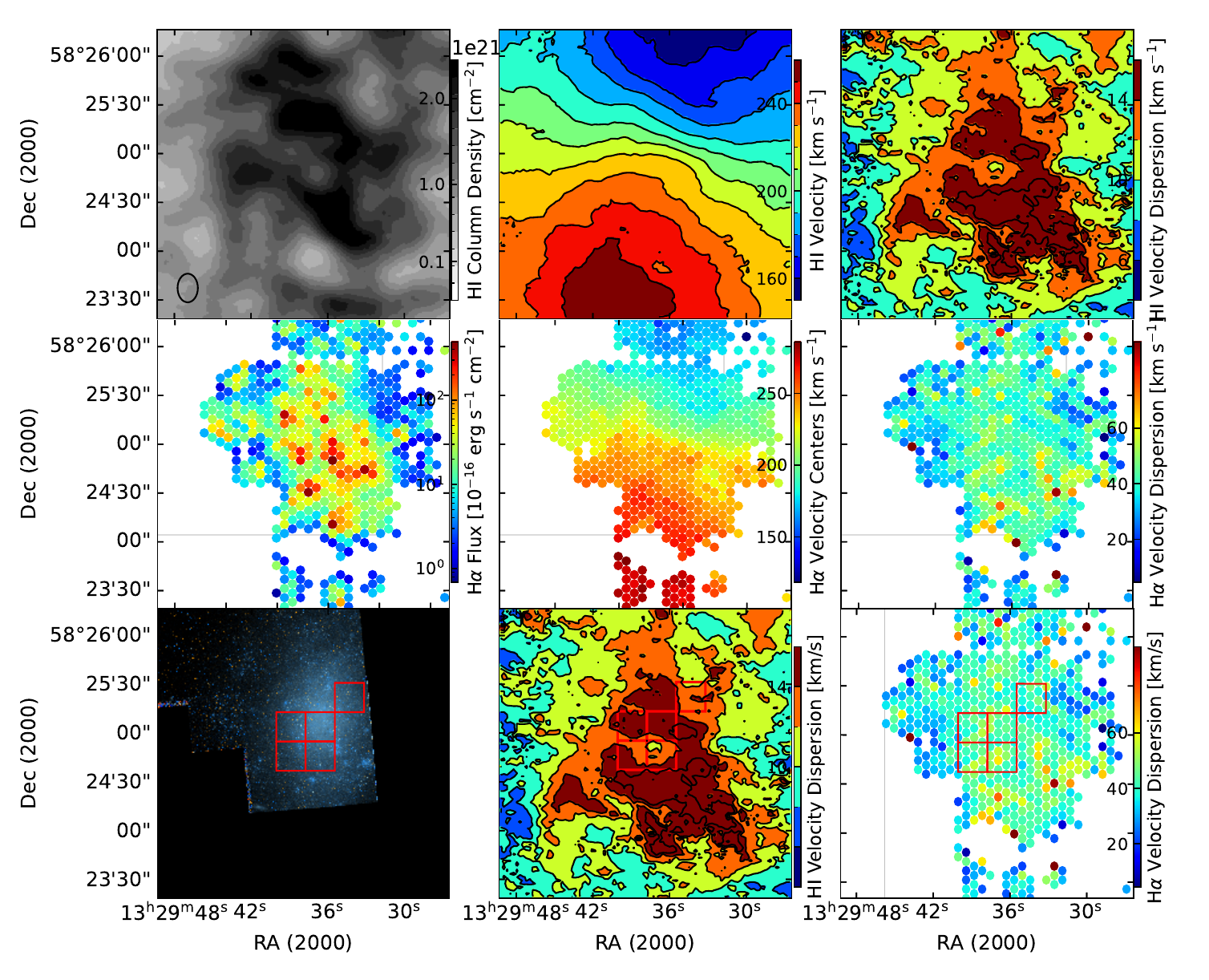}
\caption{\textbf{NGC 5204 Top Row} \hi moment maps from VLA observations. Left: \hi column density in 10$^{21}$ hydrogen atoms cm$^{-2}$, Center: \hi velocity map with isovelocity contours spaced every 10 km s$^{-1}$, Right: \hi velocity dispersion map with isovelocity contours at 2.5 km s$^{-1}$ spacing. The beam size (17.62''$\times$12.50'') of the \hi data cube used is shown in the bottom left of the left panel. \textbf{Middle Row} maps from observations with the SparsePak IFU on the WIYN 3.5m telescope, with H$\alpha$ line measurements from PAN.  Left: H$\alpha$ line flux on a log scale in units of 10$^{-16}$ erg s$^{-1}$ cm$^{-1}$, Center: H$\alpha$ line centers map, Right: H$\alpha$ velocity dispersion ($\sigma_{H\alpha}$) map. Each filled circle corresponds to a fiber's size and position on the sky. \textbf{Bottom Row} Left: Three color image from \textit{HST} F814W (red), average of F814W and F606W (green), and F606W (blue) observations with WFPC2, Center: \hi dispersion map from VLA observations with isovelocity contours in 2.5 km s$^{-1}$ step size, Right: $\sigma_{H\alpha}$ map from the SparsePak IFU on the WIYN 3.5m telescope. Overlaid on all three panels are the outlines of the regions used for the analysis.}
\end{figure*}

\begin{figure*}
\figurenum{13}
\plotone{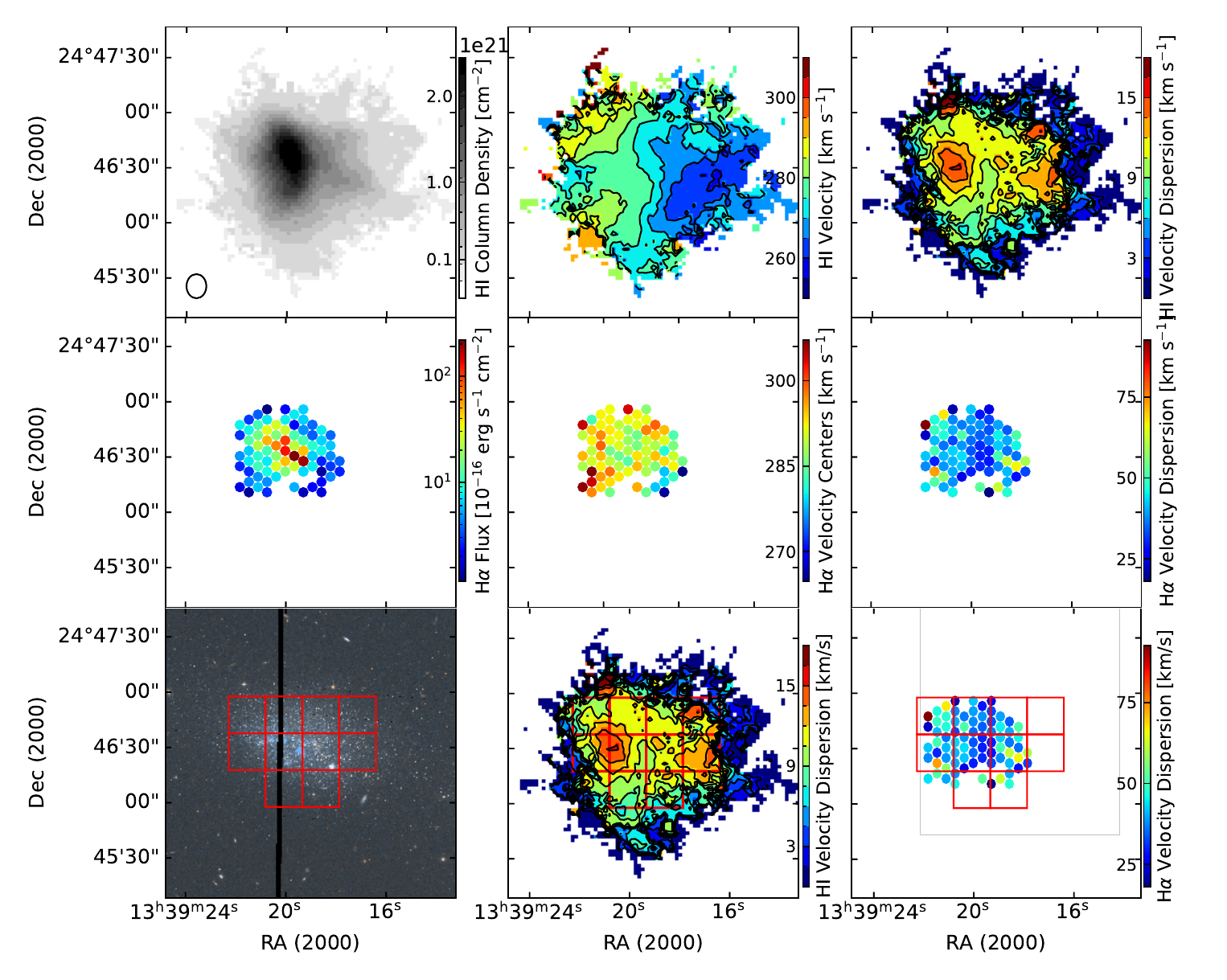}
\caption{\textbf{UGC 8638 Top Row} \hi moment maps from VLA observations. Left: \hi column density in 10$^{21}$ hydrogen atoms cm$^{-2}$, Center: \hi velocity map with isovelocity contours spaced every 5 km s$^{-1}$, Right: \hi velocity dispersion map with isovelocity contours at 1.5 km s$^{-1}$ spacing. The beam size (13.16''$\times$10.79'') of the \hi data cube used is shown in the bottom left of the left panel. \textbf{Middle Row} maps from observations with the SparsePak IFU on the WIYN 3.5m telescope, with H$\alpha$ line measurements from PAN.  Left: H$\alpha$ line flux on a log scale in units of 10$^{-16}$ erg s$^{-1}$ cm$^{-1}$, Center: H$\alpha$ line centers map, Right: H$\alpha$ velocity dispersion ($\sigma_{H\alpha}$) map. Each filled circle corresponds to a fiber's size and position on the sky. \textbf{Bottom Row} Left: Three color image from \textit{HST} F814W (red), average of F814W and F606W (green), and F606W (blue) observations with ACS, Center: \hi dispersion map from VLA observations with isovelocity contours in 1.5 km s$^{-1}$ step size, Right: $\sigma_{H\alpha}$ map from the SparsePak IFU on the WIYN 3.5m telescope. Overlaid on all three panels are the outlines of the regions used for the analysis. }
\end{figure*}

\begin{figure*}
\figurenum{13}
\plotone{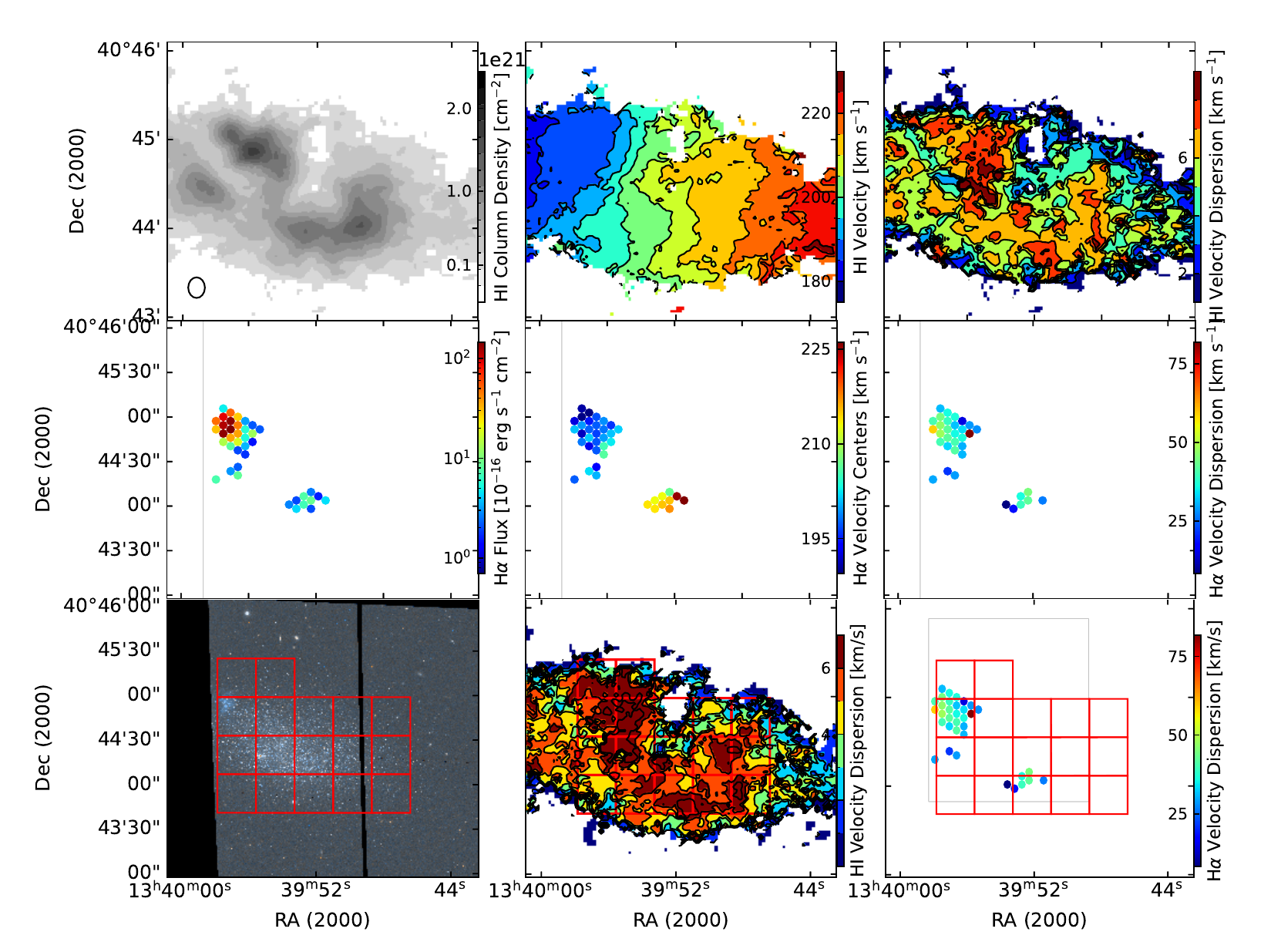}
\caption{\textbf{UGC 8651 Top Row} \hi moment maps from VLA observations. Left: \hi column density in 10$^{21}$ hydrogen atoms cm$^{-2}$, Center: \hi velocity map with isovelocity contours spaced every 5 km s$^{-1}$, Right: \hi velocity dispersion map with isovelocity contours at 1 km s$^{-1}$ spacing. The beam size (13.95''$\times$11.28'') of the \hi data cube used is shown in the bottom left of the left panel. \textbf{Middle Row} maps from observations with the SparsePak IFU on the WIYN 3.5m telescope, with H$\alpha$ line measurements from PAN.  Left: H$\alpha$ line flux on a log scale in units of 10$^{-16}$ erg s$^{-1}$ cm$^{-1}$, Center: H$\alpha$ line centers map, Right: H$\alpha$ velocity dispersion ($\sigma_{H\alpha}$) map. Each filled circle corresponds to a fiber's size and position on the sky. \textbf{Bottom Row} Left: Three color image from \textit{HST} F814W (red), average of F814W and F606W (green), and F606W (blue) observations with ACS, Center: \hi dispersion map from VLA observations with isovelocity contours in 1 km s$^{-1}$ step size, Right: $\sigma_{H\alpha}$ map from the SparsePak IFU on the WIYN 3.5m telescope. Overlaid on all three panels are the outlines of the regions used for the analysis. }
\end{figure*}

\begin{figure*}
\figurenum{13}
\plotone{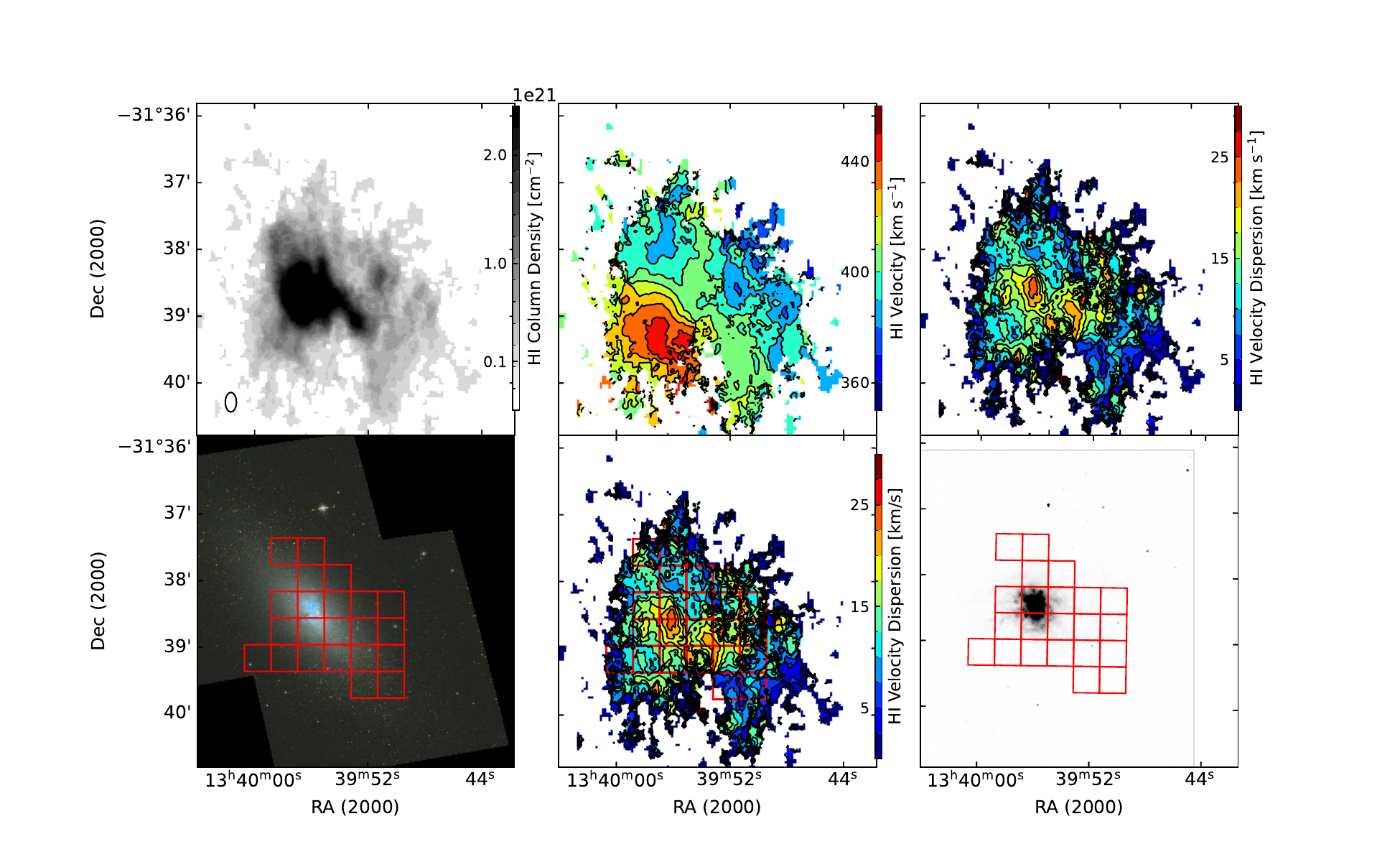}
\caption{\textbf{NGC 5253 Top Row} \hi moment maps from VLA observations. Left: \hi column density in 10$^{21}$ hydrogen atoms cm$^{-2}$, Center: \hi velocity map with isovelocity contours spaced every 10 km s$^{-1}$, Right: \hi velocity dispersion map with isovelocity contours at 2.5 km s$^{-1}$ spacing. The beam size (17.60''$\times$10.11'') of the \hi data cube used is shown in the bottom left of the left panel. \textbf{Bottom Row} Left: Three color image from \textit{HST} F814W (red), average of F814W and F555W (green), and F555W (blue) observations with ACS, Center: \hi dispersion map from VLA observations with isovelocity contours in 2.5 km s$^{-1}$ step size, Right: Ground based H$\alpha$ map published in \cite{Dale09}. Overlaid on all three panels are the outlines of the regions used for the analysis. }
\end{figure*}

\begin{figure*}
\figurenum{13}
\plotone{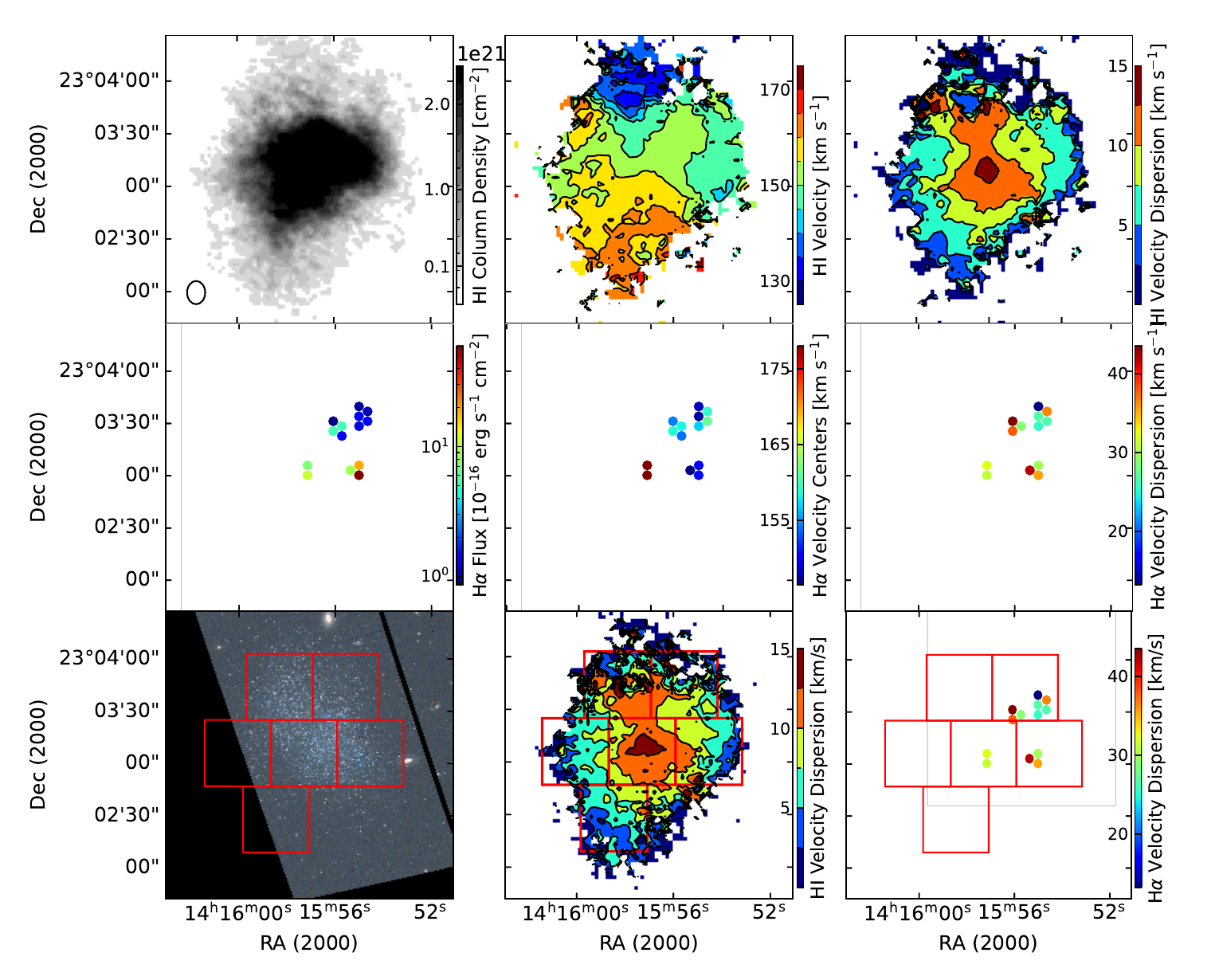}
\caption{\textbf{UGC 9128 Top Row} \hi moment maps from VLA observations Left:   \hi column density in 10$^{21}$ hydrogen atoms cm$^{-2}$, Center: \hi velocity map with isovelocity contours spaces every 5 km s$^{-1}$, Right: \hi velocity dispersion map with isovelocity contours at 2.5 km s$^{-1}$ spacing.  The beam size (13.279''$\times$10.326'') of the \hi data cube used is shown in the bottom left of the left panel. \textbf{Middle Row} maps from observations with the SparsePak IFU on the WIYN 3.5m telescope, with H$\alpha$ line measurements from PAN.  Left: H$\alpha$ line flux on a log scale in units of 10$^{-16}$ erg s$^{-1}$ cm$^{-1}$, Center: H$\alpha$ line centers map, Right: H$\alpha$ velocity dispersion ($\sigma_{H\alpha}$) map. Each filled circle corresponds to a fiber's size and position on the sky.  \textbf{Bottom Row} Left: Three color image from \textit{HST} F814W (red), average of F814W and F606W (green), and F606W (blue) observations with ACS, Center: \hi dispersion map from VLA observations with isovelocity contours in 2.5 km s$^{-1}$ step size, Right: $H\alpha$ FWHM map from the SparsePak IFU on the WIYN 3.5m telescope. Overlaid on all three panels are the outlines of the regions used for the analysis.}
\end{figure*}

\begin{figure*}
\figurenum{13}
\plotone{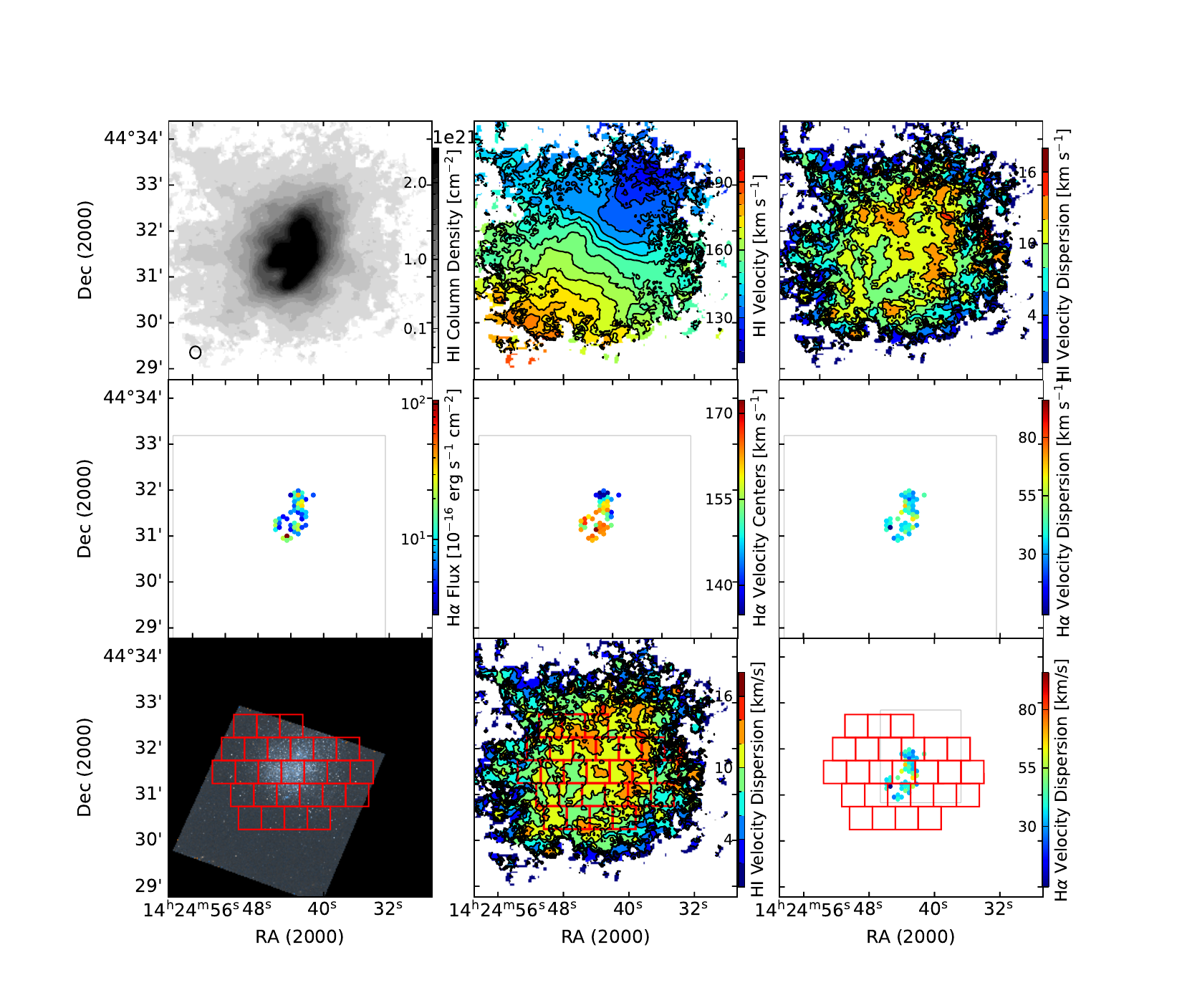}
\caption{ \textbf{UGC 9240 Top Row} \hi moment maps from VLA observations Left:   \hi column density in 10$^{21}$ hydrogen atoms cm$^{-2}$, Center: \hi velocity map with isovelocity contours spaces every 5 km s$^{-1}$, Right: \hi velocity dispersion map with isovelocity contours at 2 km s$^{-1}$ spacing.  The beam size (16.11''$\times$14.29'') of the \hi data cube used is shown in the bottom left of the left panel. \textbf{Middle Row} maps from observations with the SparsePak IFU on the WIYN 3.5m telescope, with H$\alpha$ line measurements from PAN.  Left: H$\alpha$ line flux on a log scale in units of 10$^{-16}$ erg s$^{-1}$ cm$^{-1}$, Center: H$\alpha$ line centers map, Right: H$\alpha$ velocity dispersion ($\sigma_{H\alpha}$) map. Each filled circle corresponds to a fiber's size and position on the sky. \textbf{Bottom Row} Left: Three color image from \textit{HST} F814W (red), average of F814W and F606W (green), and F606W (blue) observations with ACS, Center: \hi dispersion map from VLA observations with isovelocity contours in 2 km s$^{-1}$ step size, Right: $\sigma_{H\alpha}$ map from the SparsePak IFU on the WIYN 3.5m telescope. Overlaid on all three panels are the outlines of the regions used for the analysis. }
\end{figure*}

\begin{figure*}
\figurenum{13}
\plotone{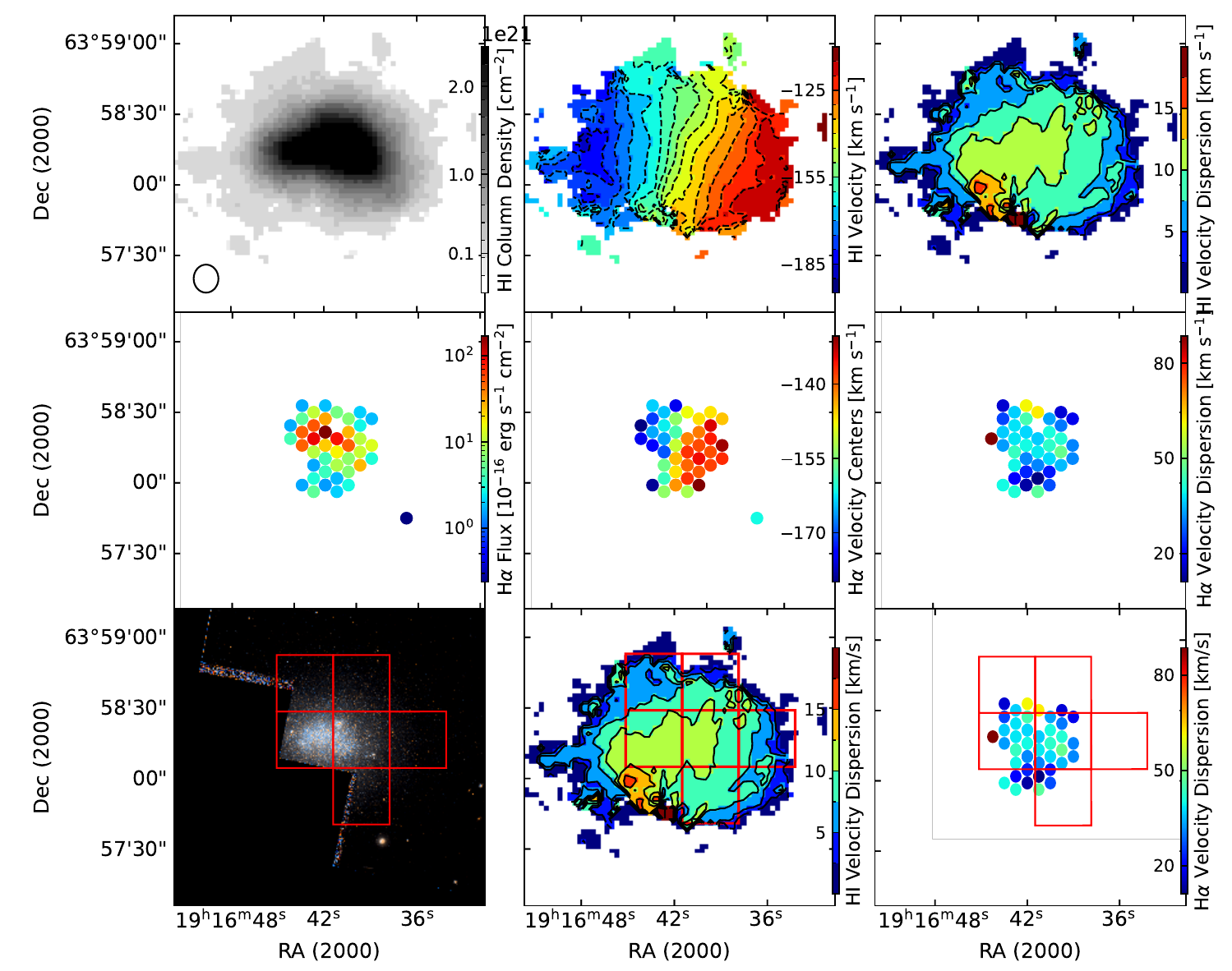}
\caption{\textbf{NGC 6789 Top Row} \hi moment maps from VLA observations Left:   \hi column density in 10$^{21}$ hydrogen atoms cm$^{-2}$, Center: \hi velocity map with isovelocity contours spaces every 5 km s$^{-1}$, Right: \hi velocity dispersion map with isovelocity contours at 2.5 km s$^{-1}$ spacing. The beam size (11.92''$\times$10.52'') of the   \hi data cube used is shown in the bottom left of the left panel. \textbf{Middle Row} maps from observations with the SparsePak IFU on the WIYN 3.5m telescope, with H$\alpha$ line measurements from PAN.  Left: H$\alpha$ line flux on a log scale in units of 10$^{-16}$ erg s$^{-1}$ cm$^{-1}$, Center: H$\alpha$ line centers map, Right: H$\alpha$ velocity dispersion ($\sigma_{H\alpha}$) map.  Each filled circle corresponds to a fiber's size and position on the sky. \textbf{Bottom Row} Left: Three color image from \textit{HST} F814W (red), average of F814W and F555W (green),  and F555W (blue observations with WFPC2), Center: \hi dispersion map from VLA observations with isovelocity contours in 2.5 km s$^{-1}$ step size, Right: $H\alpha$ FWHM map from the SparsePak IFU on the WIYN 3.5m telescope. Overlaid on all three panels are the outlines of the regions used for the analysis.}
\end{figure*}

\section{Appendix: Regional SFHs}\label{appendix_SFHs}

\textbf{Included in this appendix are data tables with the SFHs of the individual regions and the global SFHs.  Abridged form of these tables are included here and the full versions are available as machine readable tables online.}

\begin{table*}
\begin{rotatetable*}
\centering
\footnotesize
\caption{Regional Star Formation Histories} 
\begin{tabular}{l l l l l c H H c H H c H H c H H c H H c H H c H H}
\hline
Galaxy & Region & Size & RA & Dec & SFR$^{6.6}_{7.0}$ & SFR$^{6.6}_{7.0}$\_up & SFR$^{6.6}_{7.0}$\_down & SFR$^{7.0}_{7.4}$ & SFR$^{7.0}_{7.4}$\_up & SFR$^{7.0}_{7.4}$\_down & SFR$^{7.4}_{7.7}$ & SFR$^{7.4}_{7.7}$\_up & SFR$^{7.4}_{7.7}$\_down & SFR$^{7.7}_{8.0}$ & SFR$^{7.7}_{8.0}$\_up & SFR$^{7.7}_{8.0}$\_down & SFR$^{8.0}_{8.3}$ & SFR$^{8.0}_{8.3}$\_up & SFR$^{8.0}_{8.3}$\_down & SFR$^{8.3}_{8.7}$ & SFR$^{8.3}_{8.7}$\_up & SFR$^{8.3}_{8.7}$\_down & SFR$^{8.7}_{10.15}$ & SFR$^{8.7}_{10.15}$\_up & SFR$^{8.7}_{10.15}$\_down \\
 & & \arcsec & J2000.0 & J2000.0 & M$_\odot$ yr$^{-1}$ & M$_\odot$ yr$^{-1}$ & M$_\odot$ yr$^{-1}$ & M$_\odot$ yr$^{-1}$ & M$_\odot$ yr$^{-1}$ & M$_\odot$ yr$^{-1}$ & M$_\odot$ yr$^{-1}$ & M$_\odot$ yr$^{-1}$ & M$_\odot$ yr$^{-1}$ & M$_\odot$ yr$^{-1}$ & M$_\odot$ yr$^{-1}$ & M$_\odot$ yr$^{-1}$ & M$_\odot$ yr$^{-1}$ & M$_\odot$ yr$^{-1}$ & SM$_\odot$ yr$^{-1}$ & M$_\odot$ yr$^{-1}$ & M$_\odot$ yr$^{-1}$ & M$_\odot$ yr$^{-1}$ & M$_\odot$ yr$^{-1}$ &M$_\odot$ yr$^{-1}$ & M$_\odot$ yr$^{-1}$ \\
  (1) & (2) & (3) & (4) & (5) & (6) & (7) & (8) & (9) & (10) & (11) & (12) & (13) & (14) & (15) & (16) & (17) & (18) & (19) & (20) & (21) & (22) & (23) & (24) & (25) & (26)  \\
 \hline
 \hline
UGC0685 & reg\_3 & 20.0 & 16.85959510 & 16.68020527 & 0.00000 & 0.00013 & 0.00000 & 0.000059 & 0.000053 & 0.000059 & 0.000061 & 0.000055 & 0.000061 & 0.000041 & 0.000063 & 0.000041 & 0.0000850 & 0.000037 & 0.000035 & 0.000000 & 0.000013 & 0.000000 & 0.00023 & 0.00006 & 0.00011 \\
UGC0685 & reg\_6 & 20.0 & 16.85611498 & 16.67464996 & 0.000000 & 0.000058 & 0.000000 & 0.000000 & 0.000038 & 0.000000 & 0.000000 & 0.000042 & 0.000000 & 0.000025 & 0.000039 & 0.000025 & 0.000010 & 0.000015 & 0.000010 & 0.000006 & 0.000010 & 0.000006 & 0.00027 & 0.00005 & 0.00016 \\
UGC0685 & reg\_7 & 20.0 & 16.85379551 & 16.68020564 & 0.000000 & 0.00033 & 0.000000 & 0.00050 & 0.00025 & 0.00029 & 0.000000 & 0.00022 & 0.000000 & 0.000076 & 0.000074 & 0.000052 & 0.000064 & 0.000074 & 0.000027 & 0.000078 & 0.000079 & 0.000058 & 0.00057 & 0.00023 & 0.00014 \\
UGC0685 & reg\_8 & 20.0 & 16.85727567 & 16.685760993 & 0.000000 & 0.000066 & 0.000000 & 0.000000 & 0.000044 & 0.000000 & 0.000022 & 0.000025 & 0.000022 & 0.000004 & 0.000037 & 0.000004 & 0.000049 & 0.000026 & 0.000032 & 0.000000 & 0.000012 & 0.000000 & 0.000223 & 0.000064 & 0.000088 \\
UGC0685 & reg\_12 & 20.0 & 16.85031555 & 16.67465023 & 0.000000 & 0.000047 & 0.000000 & 0.000000 & 0.000028 & 0.000000 & 0.000000 & 0.000028 & 0.000000 & 0.000000 & 0.000020 & 0.000000 & 0.000000 & 0.000011 & 0.000000 & 0.000000 & 0.000006 & 0.000000 & 0.00033 & 0.00009 & 0.00011 \\
\hline 
\end{tabular}
\tablecomments{\small Column~1: Galaxy Name. Column~2: Region Identifier. Column~3: Angular size of the region. Column~4: the Right Ascension (J2000.0) of the center of the region. Column~5: the Declination (J2000.0) of the center of the region. Column~6: SFR between 10$^{6.6}$ and 10$^{7.0}$ years ago. Column~7: Error up on SFR between 10$^{6.6}$ and 10$^{7.0}$ years ago. Column~8: Error down on SFR between 10$^{6.6}$ and 10$^{7.0}$ years ago. Column~9: SFR between 10$^{7.0}$ and 10$^{7.4}$ years ago. Column~10: Error up on SFR between 10$^{7.0}$ and 10$^{7.4}$ years ago. Column~11: Error down on SFR between 10$^{7.0}$ and 10$^{7.4}$ years ago. Column~12: SFR between 10$^{7.4}$ and 10$^{7.7}$ years ago. Column~13: Error up on SFR between 10$^{7.4}$ and 10$^{7.7}$ years ago. Column~14: Error down on SFR between 10$^{7.4}$ and 10$^{7.7}$ years ago. Column~15: SFR between 10$^{7.7}$ and 10$^{8.0}$ years ago. Column~16: Error up on SFR between 10$^{7.7}$ and 10$^{8.0}$ years ago. Column~17: Error down on SFR between 10$^{7.7}$ and 10$^{8.0}$ years ago. Column~18: SFR between 10$^{8.0}$ and 10$^{8.3}$ years ago. Column~19: Error up on SFR between 10$^{8.0}$ and 10$^{8.3}$ years ago. Column~20: Error down on SFR between 10$^{8.0}$ and 10$^{8.3}$ years ago. Column~21: SFR between 10$^{8.3}$ and 10$^{8.7}$ years ago. Column~22: Error up on SFR between 10$^{8.3}$ and 10$^{8.7}$ years ago. Column~23: Error down on SFR between 10$^{8.3}$ and 10$^{8.7}$ years ago. Column~24: SFR between 10$^{8.7}$ and 10$^{10.15}$ years ago. Column~25: Error up on SFR between 10$^{8.7}$ and 10$^{10.15}$ years ago. Column~26: Error down on SFR between 10$^{8.7}$ and 10$^{10.15}$ years ago. See digital edition for machine readable version of the full table}
%\label{table:reg\_sfhs} 
\end{rotatetable*}
\end{table*}

\begin{table*}
\begin{rotatetable*}
\centering
\footnotesize
\caption{Global Star Formation Histories} 
\begin{tabular}{l c H H c H H c H H c H H c H H c H H c H H c H H c H H c H H c H H c H H}
\hline
Galaxy & SFR$^{6.6}_{7.0}$ & SFR$^{6.6}_{7.0}$\_up & SFR$^{6.6}_{7.0}$\_down & SFR$^{7.0}_{7.4}$ & SFR$^{7.0}_{7.4}$\_up & SFR$^{7.0}_{7.4}$\_down & SFR$^{7.4}_{7.6}$ & SFR$^{7.4}_{7.6}$\_up & SFR$^{7.4}_{7.6}$\_down & SFR$^{7.6}_{7.75}$ & SFR$^{7.6}_{7.75}$\_up & SFR$^{7.6}_{7.75}$\_down & SFR$^{7.75}_{7.85}$ & SFR$^{7.75}_{7.85}$\_up & SFR$^{7.75}_{7.85}$\_down & SFR$^{7.85}_{8.0}$ & SFR$^{7.85}_{8.0}$\_up & SFR$^{7.85}_{8.0}$\_down & SFR$^{8.0}_{8.15}$ & SFR$^{8.0}_{8.15}$\_up & SFR$^{8.0}_{8.15}$\_down & SFR$^{8.15}_{8.3}$ & SFR$^{8.15}_{8.3}$\_up & SFR$^{8.15}_{8.3}$\_down & SFR$^{8.3}_{8.45}$ & SFR$^{8.3}_{8.45}$\_up & SFR$^{8.3}_{8.45}$\_down & SFR$^{8.45}_{8.6}$ & SFR$^{8.45}_{8.6}$\_up & SFR$^{8.45}_{8.6}$\_down & SFR$^{8.6}_{8.75}$ & SFR$^{8.6}_{8.75}$\_up & SFR$^{8.6}_{8.75}$\_down & SFR$^{8.75}_{10.15}$ & SFR$^{8.75}_{10.15}$\_up & SFR$^{8.75}_{10.15}$\_down \\
 & M$_\odot$ yr$^{-1}$ & M$_\odot$ yr$^{-1}$ & M$_\odot$ yr$^{-1}$ & M$_\odot$ yr$^{-1}$ & M$_\odot$ yr$^{-1}$ & M$_\odot$ yr$^{-1}$ & M$_\odot$ yr$^{-1}$ & M$_\odot$ yr$^{-1}$ & M$_\odot$ yr$^{-1}$ & M$_\odot$ yr$^{-1}$ & M$_\odot$ yr$^{-1}$ & M$_\odot$ yr$^{-1}$ & M$_\odot$ yr$^{-1}$ & M$_\odot$ yr$^{-1}$ & SM$_\odot$ yr$^{-1}$ & M$_\odot$ yr$^{-1}$ & M$_\odot$ yr$^{-1}$ & M$_\odot$ yr$^{-1}$ & M$_\odot$ yr$^{-1}$ &M$_\odot$ yr$^{-1}$ & M$_\odot$ yr$^{-1}$ & M$_\odot$ yr$^{-1}$ & M$_\odot$ yr$^{-1}$ & M$_\odot$ yr$^{-1}$ & M$_\odot$ yr$^{-1}$ & M$_\odot$ yr$^{-1}$ & SM$_\odot$ yr$^{-1}$ & M$_\odot$ yr$^{-1}$ & M$_\odot$ yr$^{-1}$ & M$_\odot$ yr$^{-1}$ & M$_\odot$ yr$^{-1}$ & M$_\odot$ yr$^{-1}$ & M$_\odot$ yr$^{-1}$ & M$_\odot$ yr$^{-1}$ & M$_\odot$ yr$^{-1}$ & M$_\odot$ yr$^{-1}$ \\
 (1) & (2) & (3) & (4) & (5) & (6) & (7) & (8) & (9) & (10) & (11) & (12) & (13) & (14) & (15) & (16) & (17) & (18) & (19) & (20) & (21) & (22) & (23) & (24) & (25) & (26) & (27) & (28) & (29) & (30) & (31) & (32) & (33) & (34) & (35) & (36) & (37) \\
 \hline
 \hline
UGC~0685 & 0.019000 & 0.003000 & 0.011000 & 0.012200 & 0.001700 & 0.004500 & 0.010700 & 0.003700 & 0.002400 & 0.004400 & 0.004100 & 0.000800 & 0.005400 & 0.004100 & 0.002100 & 0.005800 & 0.002300 & 0.003500 & 0.011200 & 0.004100 & 0.004200 & 0.010500 & 0.003500 & 0.002200 & 0.011700 & 0.001700 & 0.004000 & 0.002100 & 0.002600 & 0.001300 & 0.073000 & 0.028000 & 0.069000 & 0.010700 & 0.000500 & 0.001400 \\
NGC~0784 & 0.182000 & 0.008000 & 0.109000 & 0.150000 & 0.016000 & 0.027000 & 0.099000 & 0.008000 & 0.021000 & 0.086000 & 0.017000 & 0.012000 & 0.085000 & 0.010000 & 0.029000 & 0.085000 & 0.006000 & 0.039000 & 0.116000 & 0.005000 & 0.064000 & 0.160000 & 0.006000 & 0.063000 & 0.174000 & 0.007000 & 0.106000 & 0.000000 & 0.023000 & 0.000000 & 0.984000 & 0.053000 & 0.539000 & 0.042000 & 0.013000 & 0.001000 \\
NGC~2366 & 0.154000 & 0.008000 & 0.061000 & 0.219000 & 0.007000 & 0.159000 & 0.046000 & 0.058000 & 0.008000 & 0.057000 & 0.036000 & 0.007000 & 0.048000 & 0.007000 & 0.031000 & 0.043000 & 0.003000 & 0.027000 & 0.056000 & 0.002000 & 0.031000 & 0.060000 & 0.007000 & 0.024000 & 0.063000 & 0.004000 & 0.025000 & 0.057000 & 0.006000 & 0.016000 & 0.080000 & 0.001000 & 0.038000 & 0.028400 & 0.002300 & 0.007300 \\
HolmbergII & 0.158000 & 0.008000 & 0.053000 & 0.265000 & 0.008000 & 0.170000 & 0.111000 & 0.060000 & 0.016000 & 0.079000 & 0.058000 & 0.006000 & 0.078000 & 0.008000 & 0.038000 & 0.048000 & 0.003000 & 0.019000 & 0.042000 & 0.002000 & 0.017000 & 0.054000 & 0.004000 & 0.022000 & 0.060000 & 0.004000 & 0.016000 & 0.047000 & 0.008000 & 0.012000 & 0.062000 & 0.007000 & 0.016000 & 0.027600 & 0.005300 & 0.005400 \\
UGC~4459 & 0.021400 & 0.003600 & 0.008400 & 0.009200 & 0.000800 & 0.005500 & 0.000900 & 0.003700 & 0.000400 & 0.001400 & 0.001800 & 0.000600 & 0.001900 & 0.000700 & 0.001600 & 0.000070 & 0.000620 & 0.000010 & 0.000680 & 0.000330 & 0.000620 & 0.001500 & 0.000840 & 0.000880 & 0.003700 & 0.000500 & 0.001700 & 0.003400 & 0.000200 & 0.001300 & 0.001300 & 0.001700 & 0.000500 & 0.003160 & 0.000440 & 0.000380 \\
HolmbergI & 0.029300 & 0.003500 & 0.009100 & 0.044000 & 0.003000 & 0.030000 & 0.012000 & 0.013000 & 0.004000 & 0.009500 & 0.008700 & 0.000800 & 0.016000 & 0.002000 & 0.011000 & 0.011900 & 0.001200 & 0.005800 & 0.012600 & 0.001600 & 0.005100 & 0.012600 & 0.000500 & 0.004200 & 0.008400 & 0.002600 & 0.002100 & 0.009700 & 0.003900 & 0.001400 & 0.009000 & 0.000700 & 0.005300 & 0.007200 & 0.000800 & 0.001500 \\
SextansB & 0.002200 & 0.002100 & 0.001200 & 0.006200 & 0.000200 & 0.002300 & 0.000600 & 0.001400 & 0.000200 & 0.000810 & 0.000690 & 0.000810 & 0.000700 & 0.001100 & 0.000600 & 0.000700 & 0.000430 & 0.000430 & 0.000650 & 0.000240 & 0.000290 & 0.001200 & 0.000220 & 0.000510 & 0.001200 & 0.000330 & 0.000220 & 0.001360 & 0.000340 & 0.000330 & 0.000990 & 0.000550 & 0.000350 & 0.001580 & 0.000230 & 0.000090 \\
SextansA & 0.014900 & 0.004000 & 0.004500 & 0.018900 & 0.001500 & 0.006800 & 0.008600 & 0.004600 & 0.002200 & 0.012700 & 0.003100 & 0.005100 & 0.012000 & 0.002600 & 0.005400 & 0.003200 & 0.001200 & 0.000600 & 0.001700 & 0.001100 & 0.000300 & 0.002310 & 0.000970 & 0.000350 & 0.003200 & 0.001300 & 0.000900 & 0.007100 & 0.000700 & 0.003700 & 0.001600 & 0.005000 & 0.000100 & 0.001850 & 0.000160 & 0.000520 \\
IC~2574 & 0.215000 & 0.010000 & 0.102000 & 0.310000 & 0.009000 & 0.190000 & 0.039000 & 0.087000 & 0.009000 & 0.076000 & 0.045000 & 0.008000 & 0.079000 & 0.008000 & 0.037000 & 0.062000 & 0.003000 & 0.025000 & 0.048000 & 0.008000 & 0.015000 & 0.052000 & 0.016000 & 0.017000 & 0.075000 & 0.021000 & 0.025000 & 0.078000 & 0.015000 & 0.016000 & 0.067000 & 0.020000 & 0.025000 & 0.062600 & 0.005400 & 0.009200 \\
NGC~3738 & 0.150000 & 0.007000 & 0.098000 & 0.061000 & 0.017000 & 0.006000 & 0.055000 & 0.009000 & 0.018000 & 0.072000 & 0.009000 & 0.033000 & 0.089000 & 0.017000 & 0.052000 & 0.073000 & 0.017000 & 0.023000 & 0.089000 & 0.016000 & 0.049000 & 0.157000 & 0.010000 & 0.098000 & 0.024000 & 0.010000 & 0.023000 & 0.002000 & 0.053000 & 0.002000 & 1.490000 & 0.147000 & 0.932000 & 0.057000 & 0.038000 & 0.005000 \\
NGC~3741 & 0.016600 & 0.003300 & 0.006700 & 0.010200 & 0.000600 & 0.004800 & 0.001700 & 0.003500 & 0.000400 & 0.001800 & 0.002500 & 0.000500 & 0.003300 & 0.000600 & 0.002600 & 0.002600 & 0.000700 & 0.001900 & 0.002710 & 0.000500 & 0.000510 & 0.002200 & 0.000830 & 0.000350 & 0.002540 & 0.000610 & 0.000760 & 0.002950 & 0.000950 & 0.000640 & 0.003300 & 0.000600 & 0.001000 & 0.001990 & 0.000230 & 0.000270 \\
NGC~4068 & 0.080000 & 0.007000 & 0.041000 & 0.075000 & 0.003000 & 0.021000 & 0.001000 & 0.028000 & 0.000000 & 0.019700 & 0.009900 & 0.004900 & 0.015000 & 0.014000 & 0.005000 & 0.019800 & 0.008400 & 0.005300 & 0.020000 & 0.011000 & 0.005000 & 0.019400 & 0.005600 & 0.005500 & 0.020000 & 0.012000 & 0.009000 & 0.021100 & 0.002700 & 0.009800 & 0.032000 & 0.067000 & 0.014000 & 0.017100 & 0.000100 & 0.004000 \\
NGC~4163 & 0.013800 & 0.002900 & 0.008500 & 0.008500 & 0.002100 & 0.002600 & 0.001300 & 0.002600 & 0.000600 & 0.004900 & 0.001700 & 0.002000 & 0.001900 & 0.002400 & 0.000900 & 0.001900 & 0.000800 & 0.001000 & 0.003200 & 0.000700 & 0.001000 & 0.002120 & 0.000560 & 0.000470 & 0.002400 & 0.000900 & 0.001100 & 0.003700 & 0.001200 & 0.001200 & 0.001600 & 0.001200 & 0.000800 & 0.007000 & 0.001300 & 0.000800 \\
NGC~4190 & 0.043000 & 0.005000 & 0.016000 & 0.013200 & 0.002700 & 0.007500 & 0.018000 & 0.011000 & 0.004000 & 0.034600 & 0.005800 & 0.008600 & 0.014000 & 0.011000 & 0.003000 & 0.025400 & 0.005500 & 0.007800 & 0.020000 & 0.003600 & 0.004700 & 0.014100 & 0.006800 & 0.005700 & 0.005000 & 0.011000 & 0.002000 & 0.004000 & 0.020000 & 0.004000 & 0.208000 & 0.006000 & 0.090000 & 0.026000 & 0.004000 & 0.021000 \\
UGC~7577 & 0.016000 & 0.003000 & 0.012000 & 0.008300 & 0.004000 & 0.002200 & 0.001800 & 0.002300 & 0.001300 & 0.006300 & 0.001000 & 0.004300 & 0.003400 & 0.003400 & 0.002100 & 0.006000 & 0.000900 & 0.002400 & 0.002100 & 0.001200 & 0.000600 & 0.002400 & 0.000800 & 0.001400 & 0.005300 & 0.000900 & 0.003000 & 0.002100 & 0.001400 & 0.000400 & 0.031000 & 0.002000 & 0.031000 & 0.005290 & 0.000780 & 0.000610 \\
UGCA292 & 0.007600 & 0.002100 & 0.001500 & 0.001800 & 0.000600 & 0.001300 & 0.002300 & 0.001100 & 0.001500 & 0.001600 & 0.001800 & 0.000700 & 0.001400 & 0.001100 & 0.001100 & 0.000460 & 0.000590 & 0.000450 & 0.000730 & 0.000150 & 0.000430 & 0.000850 & 0.000410 & 0.000400 & 0.002300 & 0.000200 & 0.001500 & 0.001550 & 0.000400 & 0.000190 & 0.001200 & 0.000510 & 0.000520 & 0.000479 & 0.000048 & 0.000078 \\
UGC~8024 & 0.021000 & 0.009000 & 0.011000 & 0.043200 & 0.002600 & 0.008700 & 0.003600 & 0.008500 & 0.000800 & 0.016200 & 0.003700 & 0.004300 & 0.008800 & 0.003500 & 0.004500 & 0.005600 & 0.003900 & 0.001000 & 0.007800 & 0.005900 & 0.000500 & 0.010500 & 0.005400 & 0.001200 & 0.009800 & 0.005700 & 0.000500 & 0.007900 & 0.003700 & 0.000600 & 0.011200 & 0.009100 & 0.003400 & 0.004500 & 0.000200 & 0.001200 \\
GR8 & 0.010300 & 0.001900 & 0.001600 & 0.003800 & 0.000800 & 0.001500 & 0.000200 & 0.001400 & 0.000000 & 0.002400 & 0.000000 & 0.001700 & 0.000010 & 0.000700 & 0.000010 & 0.000500 & 0.000260 & 0.000250 & 0.000490 & 0.000150 & 0.000160 & 0.000760 & 0.000130 & 0.000310 & 0.000850 & 0.000090 & 0.000150 & 0.000420 & 0.000140 & 0.000070 & 0.000420 & 0.000220 & 0.000110 & 0.000669 & 0.000023 & 0.000020 \\
UGC~8201 & 0.023000 & 0.004000 & 0.012000 & 0.104000 & 0.005000 & 0.046000 & 0.038000 & 0.029000 & 0.007000 & 0.091000 & 0.014000 & 0.014000 & 0.072000 & 0.007000 & 0.021000 & 0.049000 & 0.002000 & 0.014000 & 0.019900 & 0.006900 & 0.001100 & 0.018200 & 0.004100 & 0.002200 & 0.018500 & 0.005300 & 0.003800 & 0.027000 & 0.009000 & 0.015000 & 0.073000 & 0.003000 & 0.021000 & 0.014500 & 0.003600 & 0.001700 \\
NGC~5204 & 0.128000 & 0.016000 & 0.032000 & 0.056000 & 0.005000 & 0.023000 & 0.056000 & 0.005000 & 0.013000 & 0.046000 & 0.013000 & 0.012000 & 0.052000 & 0.008000 & 0.035000 & 0.036000 & 0.013000 & 0.015000 & 0.049000 & 0.011000 & 0.029000 & 0.076000 & 0.022000 & 0.049000 & 0.046000 & 0.017000 & 0.046000 & 0.000000 & 0.062000 & 0.000000 & 0.633000 & 0.025000 & 0.434000 & 0.065000 & 0.045000 & 0.022000 \\
UGC~8638 & 0.018000 & 0.004000 & 0.011000 & 0.006000 & 0.002300 & 0.002300 & 0.000000 & 0.001800 & 0.000000 & 0.001600 & 0.001400 & 0.000500 & 0.000300 & 0.001600 & 0.000300 & 0.000790 & 0.000370 & 0.000790 & 0.002200 & 0.001200 & 0.001300 & 0.003900 & 0.000500 & 0.001400 & 0.002270 & 0.000440 & 0.000510 & 0.003100 & 0.000500 & 0.001300 & 0.000970 & 0.000370 & 0.000890 & 0.005900 & 0.000100 & 0.001400 \\
UGC~8651 & 0.007100 & 0.001700 & 0.003500 & 0.005600 & 0.000800 & 0.002300 & 0.000600 & 0.003700 & 0.000000 & 0.004300 & 0.000100 & 0.003200 & 0.001500 & 0.002000 & 0.001000 & 0.001900 & 0.000500 & 0.001100 & 0.001320 & 0.000680 & 0.000460 & 0.002040 & 0.000860 & 0.000620 & 0.001800 & 0.001400 & 0.000300 & 0.002520 & 0.000620 & 0.000470 & 0.002500 & 0.001600 & 0.000900 & 0.002670 & 0.000260 & 0.000600 \\
NGC~5253 & 0.190000 & 0.006000 & 0.115000 & 0.026000 & 0.016000 & 0.005000 & 0.028000 & 0.006000 & 0.017000 & 0.053000 & 0.008000 & 0.031000 & 0.014000 & 0.020000 & 0.010000 & 0.049000 & 0.008000 & 0.030000 & 0.047000 & 0.008000 & 0.026000 & 0.080000 & 0.009000 & 0.040000 & 0.112000 & 0.008000 & 0.073000 & 0.004000 & 0.026000 & 0.002000 & 1.556000 & 0.026000 & 0.948000 & 0.069000 & 0.014000 & 0.017000 \\
UGC~9128 & 0.000610 & 0.000490 & 0.000350 & 0.000480 & 0.000840 & 0.000390 & 0.004800 & 0.000000 & 0.003400 & 0.000300 & 0.003100 & 0.000000 & 0.000800 & 0.001100 & 0.000800 & 0.000690 & 0.000530 & 0.000550 & 0.002100 & 0.000400 & 0.001300 & 0.004100 & 0.000300 & 0.001400 & 0.001520 & 0.000650 & 0.000170 & 0.001020 & 0.000260 & 0.000300 & 0.002330 & 0.000110 & 0.000800 & 0.000945 & 0.000033 & 0.000024 \\
UGC~9240 & 0.005200 & 0.002100 & 0.004000 & 0.025000 & 0.002000 & 0.015000 & 0.004900 & 0.007000 & 0.001800 & 0.004800 & 0.003600 & 0.003400 & 0.008800 & 0.002600 & 0.005300 & 0.007400 & 0.001000 & 0.003400 & 0.006400 & 0.000600 & 0.003000 & 0.005200 & 0.001400 & 0.000700 & 0.006300 & 0.001200 & 0.002800 & 0.007900 & 0.001800 & 0.002200 & 0.007400 & 0.000300 & 0.002800 & 0.005500 & 0.000800 & 0.001100 \\
NGC~6789 & 0.008700 & 0.001900 & 0.004100 & 0.004200 & 0.001200 & 0.002800 & 0.004900 & 0.002200 & 0.001400 & 0.003300 & 0.001700 & 0.001400 & 0.004900 & 0.003400 & 0.004400 & 0.004100 & 0.001200 & 0.002100 & 0.005800 & 0.001800 & 0.004700 & 0.003900 & 0.002000 & 0.002400 & 0.001300 & 0.002500 & 0.001300 & 0.000100 & 0.005100 & 0.000100 & 0.074000 & 0.009000 & 0.046000 & 0.006700 & 0.003500 & 0.002800 \\
\hline 
\end{tabular}
\tablecomments{\small Column~1: Galaxy Name. Column~2-4: SFR between 10$^{6.6}$ and 10$^{7.0}$ years ago and the error up and down. Column~5-7: SFR between 10$^{7.0}$ and 10$^{7.4}$ years ago and the error up and down. Column~8-10: SFR between 10$^{7.4}$ and 10$^{7.6}$ years ago and the error up and down. Column~11-13: SFR between 10$^{7.6}$ and 10$^{7.75}$ years ago and the error up and down. Column~14-16: SFR between 10$^{7.75}$ and 10$^{7.85}$ years ago and the error up and down. Column~17-19: SFR between 10$^{7.85}$ and 10$^{8.0}$ years ago and the error up and down. Column~20-22: SFR between 10$^{8.0}$ and 10$^{8.15}$ years ago and the error up and down. Column~23-25: SFR between 10$^{8.15}$ and 10$^{8.3}$ years ago and the error up and down. Column~26-28: SFR between 10$^{8.3}$ and 10$^{8.45}$ years ago and the error up and down. Column~29-31: SFR between 10$^{8.45}$ and 10$^{8.6}$ years ago and the error up and down. Column~32-34: SFR between 10$^{8.6}$ and 10$^{8.75}$ years ago and the error up and down. Column~35-37: SFR between 10$^{8.75}$ and 10$^{10.15}$ years ag and the error up and down.. \textbf{See digital edition for machine readable version of the full table.}}
%\label{table:global\_sfhs} 
\end{rotatetable*}
\end{table*}

{\bibliography{ref}}

\end{document}